\documentclass[aps,rmp,reprint,unsortedaddress,floatfix]{revtex4-2}
\usepackage{braket}
\usepackage{bbm}
\usepackage{tikz}
\usetikzlibrary{arrows.meta,decorations.pathmorphing, decorations.shapes, backgrounds,positioning,fit,decorations.pathreplacing, hobby}
\usetikzlibrary{shapes.geometric, calc, trees}
\usepackage{amsmath}
\usepackage{amsfonts}
\usepackage{graphicx}
\usepackage{tabularx}
\usepackage{makecell}
\usepackage[table]{xcolor}

\usepackage{wasysym}
\definecolor{OliveGreen}{rgb}{0,0.6,0}
\usepackage[normalem]{ulem}
\usepackage[hidelinks]{hyperref} 
\hypersetup{colorlinks=true, linkcolor=blue, citecolor=blue, urlcolor=blue,pdfborder={0 0 0}}
\usepackage{comment}
\usepackage{physics}
\usepackage{soul}
\usepackage{ifthen}
\usepackage{todonotes}

\usepackage{orcidlink}

\newcommand{\e}{e}
\renewcommand{\i}{i}
\newcommand{\h}{{H}}
\renewcommand{\eqref}[1]{Eq.~(\ref{#1})}

\renewcommand{\d}{\mathrm{d}}

\newcommand{\aw}{{a}_\omega}
\newcommand{\awd}{{a}_\omega^{\dagger}}
\newcommand{\bn}{{b}_n}
\newcommand{\bnd}{{b}_n^{\dagger}}

\newcommand{\hc}{\text{h.c.}}

\renewcommand{\Re}{\mathrm{Re}}
\renewcommand{\Im}{\mathrm{Im}}

\newcommand{\add}[1]{{\color{violet}\textbf{#1}}}

\newcommand{\N}{3} 

\newcommand{\maketree}[2]{%
  \ifnum#1>0
    \foreach \i in {1,...,\N}{%
      child { node {$\rho_{#2\i}$}
        \maketree{\numexpr#1-1\relax}{#2\i} 
      }
    }%
  \fi
}

\newcommand{\jw}[1]{\add{#1}}

\definecolor{paletteRed}{RGB}{224, 102, 62}
\definecolor{paletteOrange}{RGB}{253, 145, 58}
\definecolor{paletteYellow}{RGB}{252, 189, 59}
\definecolor{paletteGreen}{RGB}{130, 201, 164}
\definecolor{paletteBlue}{RGB}{30, 117, 179}

\colorlet{paletteRed_Light1}{paletteRed!70}
\colorlet{paletteOrange_Light1}{paletteOrange!75}
\colorlet{paletteYellow_Light1}{paletteYellow!75}
\colorlet{paletteGreen_Light1}{paletteGreen!75}
\colorlet{paletteBlue_Light1}{paletteBlue!70}

\colorlet{paletteRed_Light2}{paletteRed!50}
\colorlet{paletteOrange_Light2}{paletteOrange!55}
\colorlet{paletteYellow_Light2}{paletteYellow!55}
\colorlet{paletteGreen_Light2}{paletteGreen!55}
\colorlet{paletteBlue_Light2}{paletteBlue!50}

\colorlet{paletteRed_Light3}{paletteRed!30}
\colorlet{paletteOrange_Light3}{paletteOrange!35}
\colorlet{paletteYellow_Light3}{paletteYellow!35}
\colorlet{paletteGreen_Light3}{paletteGreen!35}
\colorlet{paletteBlue_Light3}{paletteBlue!30}

\begin{document}
\title{Tensor network methods for non-perturbative dynamics of open quantum systems}

\author{Thibaut Lacroix\ \orcidlink{0000-0002-5190-040X}}
\thanks{These authors contributed equally.}
\affiliation{Institut f{\"u}r Theoretische Physik und IQST, Albert-Einstein-Allee 11, Universit{\"a}t Ulm, D-89081 Ulm, Germany}

\author{Adam Burgess\ \orcidlink{0000-0001-7544-8940}}
\thanks{These authors contributed equally.}
\affiliation{SUPA, Institute of Photonics and Quantum Sciences, Heriot-Watt University, Edinburgh EH14 4AS, United Kingdom}

\author{Nicola Lorenzoni\ \orcidlink{0000-0002-1372-1283}}
\thanks{These authors contributed equally.}
\affiliation{Institut f{\"u}r Theoretische Physik und IQST, Albert-Einstein-Allee 11, Universit{\"a}t Ulm, D-89081 Ulm, Germany}

\author{Julian Wiercinski\ \orcidlink{0009-0004-3391-8093}}
\thanks{These authors contributed equally.}
\author{Kian Damezin\ \orcidlink{0009-0000-9624-9828}}
\thanks{These authors contributed equally.}
\affiliation{SUPA, Institute of Photonics and Quantum Sciences, Heriot-Watt University, Edinburgh EH14 4AS, United Kingdom}

 \author{James Lim\ \orcidlink{0000-0002-1511-5834}}
 \affiliation{Institut f{\"u}r Theoretische Physik und IQST, Albert-Einstein-Allee 11, Universit{\"a}t Ulm, D-89081 Ulm, Germany}

  \author{Dario Tamascelli\ \orcidlink{0000-0001-6575-4469}}
 \affiliation{Universit{\`a} degli Studi di Milano, Dipartimento di Fisica ``Aldo Pontremoli'', Via Celoria 16, I-20133 Milano, Italy}

 \author{Alex W. Chin\ \orcidlink{0000-0001-7741-5915}}
 \affiliation{Sorbonne Universit{\'e}, CNRS, Institut des NanoSciences de Paris, 4 place Jussieu, 75005 Paris, France}

 \author{Moritz Cygorek\ \orcidlink{0000-0003-4406-851X}}
 \affiliation{Condensed Matter Theory, Department of Physics, TU Dortmund, 44221 Dortmund, Germany}

 \author{Brendon W. Lovett\ \orcidlink{0000-0001-5142-9585}}
 \author{Jonathan Keeling\ \orcidlink{0000-0002-4283-552X}}
 \affiliation{SUPA, School of Physics and Astronomy, University of St Andrews, St Andrews KY16 9SS, United Kingdom}

 \author{Susana F. Huelga\ \orcidlink{0000-0003-1277-8154}}
 \thanks{S.F.H., M.B.P., and E.M.G. contributed equally to conception and leadership of this work.}

 \author{Martin B. Plenio\ \orcidlink{0000-0003-4238-8843}}
 \thanks{S.F.H., M.B.P., and E.M.G. contributed equally to conception and leadership of this work.}
 \affiliation{Institut f{\"u}r Theoretische Physik und IQST, Albert-Einstein-Allee 11, Universit{\"a}t Ulm, D-89081 Ulm, Germany}

 \author{Erik M. Gauger\ \orcidlink{0000-0003-1232-9885}}
 \thanks{S.F.H., M.B.P., and E.M.G. contributed equally to conception and leadership of this work.}
 \affiliation{SUPA, Institute of Photonics and Quantum Sciences, Heriot-Watt University, Edinburgh EH14 4AS, United Kingdom}

\begin{abstract}
    The description of open quantum system dynamics beyond the perturbative treatment (usually associated with Markovian master equations) is a computationally challenging task due to the unfavorable exponential scaling of memory kernels.
    Developed over recent decades in the context of quantum information and condensed matter, tensor networks provide both a new formalism and a toolbox for overcoming previous computational bottlenecks.
    This framework enables the formulation of non-perturbative, numerically exact methods for describing the dynamics of open quantum systems to controllable numerical accuracy.
    In this review, we present these methods and discuss their commonalities and differences to paint a comprehensive view of the field.
\end{abstract}

\date{\today}
\maketitle

\tableofcontents
\clearpage


\section{INTRODUCTION}\label{sec:Intro}

The advent of the second quantum revolution has driven growing research interest in high-accuracy theoretical descriptions of real-world quantum systems across both natural and artificial platforms.
Although one is typically only interested in a subset of the total degrees of freedom, hereafter referred to as the \emph{system}---for example, a quantum device's operational states or the excitations of particular sites in a quantum transport system---this system inevitably couples to many additional degrees of freedom, collectively referred to as the \emph{environment}, which can significantly influence its dynamics.

Since the environment typically comprises a large number of degrees of freedom, treating them on the same level as the system inherently leads to the quantum many-body problem, namely the exponential growth of the Hilbert space dimension with the number of degrees of freedom~\cite{vonNeumann1955}. Consequently, parameterizing quantum states as vectors spanning the entire Hilbert space, a standard approach in quantum mechanics, becomes numerically prohibitive when more than a few subsystems interact.

This exponential growth of complexity has driven the development of the open quantum systems (OQS) framework, centered on methods that aim to circumvent the explicit treatment of the entire environment by modeling only its influence on the system dynamics. Ideally, the goal would be to determine \textit{master equations}, i.e., equations of motion solely for the system degrees of freedom, in which the influence of the environment is incorporated through effective system operators. However, deriving master equations based on an underlying microscopic Hamiltonian is feasible only in a limited number of cases. In general, this is possible only through a sequence of approximations, typically valid when approaching the so-called \textit{Markovian} regime, where the system--environment coupling is weak, and the environment is only weakly perturbed by the system dynamics. This means that any perturbation induced by the system on the environment is quickly dissipated in the environment, propagating away from the system, preventing any back action on the system and thereby suppressing environmental \textit{memory effects}~\cite{weiss_quantum_2012,rivas_open_2011}.

The Markovian framework, rigorously established in the 1970s~\cite{lindblad_generators_1976, GoriniJMathPhys1976}, underpins many quantitative and numerical studies of open system dynamics and has become deeply rooted in the OQS community. Beyond this latter approach, a wide range of approximate techniques have been developed that partially extend beyond the Markovian regime, while still commonly relying on the assumption of small system-environment coupling and memory effects, and hence are referred to as \emph{perturbative} approaches~\cite{breuer_theory_2009,rivas_open_2011,gardiner_quantum_2004}.

However, over recent decades, experimental and theoretical advances have consistently expanded the classes of dynamical processes shown to lie beyond the perturbative regime. In such cases, the use of the aforementioned approximations is not justified, and non-perturbative approaches become essential even for a qualitative understanding of the dynamics~\cite{deVega2017}. 
This motivates the development of methodologies capable of accurately simulating non-perturbative dynamics, providing benchmarks for approximate methods, and simultaneously addressing the exponential growth of the environmental Hilbert space.

To this end, one seeks methods that yield results whose errors, with respect to the exact dynamics, can be bounded and made arbitrarily small by systematically refining well-defined convergence parameters, without relying on uncontrolled approximations. We refer to methods satisfying these criteria as \emph{numerically exact}.
According to this definition, approaches such as mean-field methods or non-equilibrium Green’s function techniques do not qualify as numerically exact, as they rely on effective or phenomenological representations of the environment and/or uncontrolled approximations, without a guaranteed route to systematically recover the exact underlying dynamics. 
The term ``numerically exact'' is, however, often used more loosely in the literature.

Generally, non-Markovian behavior can be expected to arise in the presence of environmental memory effects, such as information backflow and system--environment correlations~\cite{Breuer2009, huelga_non-markovianity-assisted_2012, breuer_colloquium_2016, li_concepts_2018}. These may have a variety of origins, including strong system--environment coupling, environmental dynamics that are slower than or comparable to those of the system, and the finite extent of the environment.

In the following, we provide a brief, non-exhaustive list of archetypal systems characterized by non-Markovian dynamics.

First, a wide range of quantum optical and solid-state systems, which act as the sources of non-classical light, necessitate an OQS description where the environment consists of a quantized electromagnetic field. In such cases, the Markov approximation typically applies when the environment is the electromagnetic vacuum in free space, where excitations emitted by the system may couple to freely propagating field modes that quickly travel away, irreversibly carrying information into the environment and thus suppressing memory effects~\cite{breuer_theory_2009, rivas_open_2011}.  However, when the electromagnetic environment is considered inside structured materials, photons emitted by the system may be reflected or reabsorbed at later times, leading to a backflow of information from the environment to the system and thus to non-Markovian dynamics. 
This occurs, for instance, when the system is embedded within photonic crystals~\cite{john_spontaneous_1994, HoeppePRL2012}, dielectric microcavities~\cite{WuOptica2010, MedinaPRL2021}, or waveguides with time-delayed feedback~\cite{PichlerPNAS2017, RoccatiPRL2024}.

In solid-state settings, the environment may instead consist of spin degrees of freedom, whose intrinsic dynamical timescales, or the finite size of the spin ensemble, have also been shown to induce non-Markovian effects~\cite{PhysRevA.97.062126, wu_diamond_2016,Coish_hyperfine_2004,Chekhovich2013}.

In other quantum optical and solid-state systems, where the system is embedded in a crystal host, the environment consists of an ensemble of phonons. The slower propagation and the presence of often strong couplings are known to induce non-Markovian behavior~\cite{ForstnerPRL2003, GallandPRL2008, KrummheuerPRB2005, lacroix_non-markovian_2024}.

Another example where phonon environments play an important role is the field of quantum effects in biological systems~\cite{Huelga2013}. A central example is the analysis of ultrafast processes occurring in biomolecular systems, where the presence and functional relevance of quantum effects remain open questions~\cite{EngelNat2007,cao_quantum_2020, lorenzoni_full_2025,Gustin_2023,Kaufman_2023,Matselyukh_2022}.
In this domain, processes of interest include electronic charge and excitation energy dynamics within molecular aggregates, where the environment is represented by phonon modes arising from nuclear motion~\cite{May2011}. Since these environments include vibrational modes with lifetimes comparable to or larger than the characteristic timescales of the system dynamics and couple strongly to the system, memory effects naturally emerge~\cite{Chin2013, RodenJChemPhys2011}. Additionally, the regime of intermediate environment coupling strengths, where traditional master equation approaches fail, has been shown to provide higher transfer performance~\cite{CarusoJChemPhys2009, prior_efficient_2010, chin_noise-assisted_2010, Plenio_2008, mohseni_environment-assisted_2008}. More examples of systems displaying non-Markovian dynamics can be found in the review by \textcite{deVega2017}.

The widespread relevance of non-Markovian effects motivates the use of approaches that accurately account for the environment's influence while reducing the exponentially large Hilbert space to a manageable size. In many-body theory, this challenge has led to a deeper analysis of the physically relevant subspaces in which the dynamics of a quantum state occur due to the constraints imposed by the nature of interactions in the Hamiltonian.
Over the past decades, significant progress has been achieved, and ongoing research shows that these constraints dramatically reduce the accessible portion of the total Hilbert space a system can explore. Among the most relevant physical limitations is the decay of interactions with distance, which constrains the evolution of their correlations in many-body systems. In this regard, a more general result shows that in polynomial time the dynamics of a many-body state can explore only an exponentially small fraction of the total Hilbert space~\cite{poulin_quantum_2011}. Under additional assumptions, this exponentially small subspace can be identified with the space of states satisfying an \textit{area law}~\cite{EisertRevModPhys2010}. In such states, correlations between particles are predominantly mediated by nearest neighbor interactions, whereas correlations with distant particles are exponentially suppressed.

After the reformulation of the celebrated \emph{density matrix renormalization group} (DMRG)~\cite{WhitePRL1992} in terms of \emph{matrix product states} (MPS)~\cite{OstlundPRL1995, OstlundPRB1997, fannes_finitely_1992}, the formalism of \textit{tensor networks} (TN) emerged as a powerful framework for parameterizing quantum states with a tunable level of correlations~\cite{orus_practical_2014, schollwock_density-matrix_2011}. In principle, TNs can represent any state in the Hilbert space at the cost of an exponential increase in the number of parameters.
Imposing a cutoff on the size of the tensors used in the representation of states, the TN framework achieves good approximations in many physically relevant settings, notably ground and thermal states that satisfy an area law, while only using a polynomially scaling number of parameters~\cite{molnar_approximating_2015}.

Tensor networks therefore offer an efficient representation of environmental degrees of freedom and enable non-perturbative numerical methods with controlled convergence, thereby falling under the definition of numerically exact methods given above.

Since the deployment of the TN framework, the field of many-body physics has witnessed a rapid emergence of a vast and diverse toolbox of methods~\cite{orus_practical_2014, schollwock_density-matrix_2011, paeckel_time-evolution_2019, simone_montangero_introduction_2018}, greatly enriching the repertoire of numerically exact approaches. 
However, this rapid growth has also made the field increasingly difficult to navigate. The study of open quantum systems has similarly experienced a proliferation of approaches, making it challenging to maintain a clear overview of the landscape.
\begin{figure*}
    \centering
    \includegraphics[width=\textwidth]{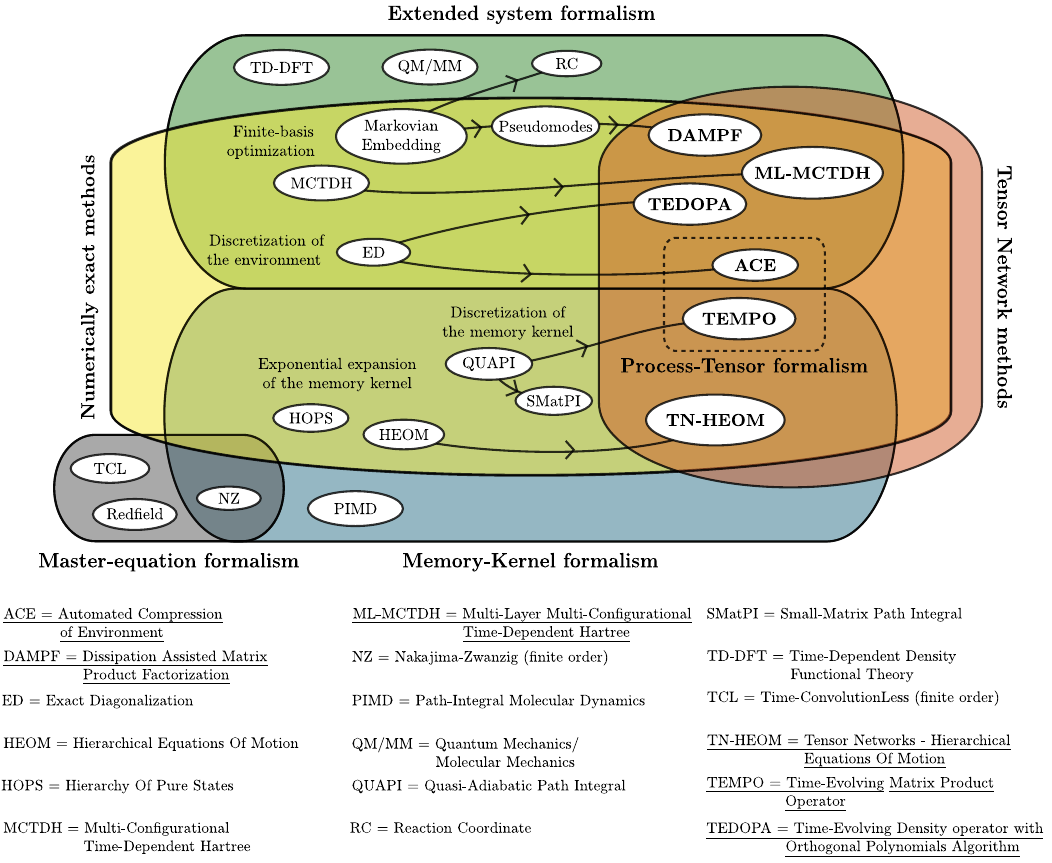}
    \caption{\textbf{A classification of numerical methods for open quantum system dynamics.} A Venn diagram illustrating the categorization of numerical tools for open quantum system dynamics is shown. The groups are defined according to the underlying formalism of each method, such as those based on an explicit representation of the environment or on memory kernel-based approaches, which include path integral representations. Methodologies that employ a tensor network formalism, as well as those identified as numerically exact, are also indicated.}
    \label{fig:methods}
\end{figure*}

A first broad classification of numerically exact approaches for open quantum systems, illustrated in Fig.~\ref{fig:methods}, can be made according to the representation of the equation of motion they aim to solve~\cite{xu_colloquium_2026}. 
More details on the underlying formalisms will be given in Sec.~\ref{sec:OQS}.
Two major classes defined in this way are the memory-kernel and path-integral methodologies, in which the influence of the environment on the system dynamics is encoded in a \emph{memory kernel} (see Sec.~\ref{sec:memory-kernel}) or an \emph{influence functional} (see Sec.~\ref{sec:IF}), respectively. Among these, a pioneering approach is the well-established Hierarchical Equations of Motion (HEOM) method~\cite{Tanimura2020}, which relies on repeated time differentiation of the influence functional. Over time, several descendant formulations of HEOM have emerged, including tensor-network-based implementations that natively employ the TN formalism~\cite{Shi2018, Ke2023,Chen2025}. 
A stochastic unraveling of the hierarchical equations of motion leads to the Hierarchy of Pure States (HOPS) method~\cite{SuessPRL2014}.
An equally seminal path-integral-based method is the Quasi-Adiabatic Path Integral (QUAPI) approach~\cite{makri_tensor_1995,Makri1995_2,makri_tensor_1995-1}, which instead relies on a discrete representation of the influence functional. QUAPI has, in turn, inspired several successor formulations, such as the Small Matrix Path Integral (SMatPI) method~\cite{makri_small_2020} and the tensor-network-based Time-Evolving Matrix Product Operator (TEMPO) algorithm~\cite{strathearn_efficient_2018}.

Another major class comprises methods that rely on an explicit parametrization of the physical environment, thus defining an \emph{extended system} (see Sec.~\ref{sec:extended-sys}). This family includes several approaches that have been formulated within the tensor-network framework, such as the \textit{Multi-Layer Multi-Configuration Time-Dependent Hartree} (ML-MCTDH) method~\cite{wang_multilayer_2003, manthe_multilayer_2008, vendrell_multilayer_2011}, which expands the environmental degrees of freedom on a finite optimized basis; the \textit{Time-Evolving Density operator with Orthogonal Polynomials Algorithm} (TEDOPA)~\cite{chin_exact_2010, prior_efficient_2010, tamascelli_efficient_2019}, which employs a unitary mapping of continuous environments onto semi-infinite discrete chains; the \textit{Dissipation-Assisted Matrix Product Factorization} (DAMPF) approach~\cite{SomozaPRL2019}, which exploits the pseudomode theory; and the \textit{Automated Compression of Environments} (ACE) method~\cite{mortiz_2022_ACE}, which iteratively adds and compresses the degrees of freedom of a discretized environment.

While many of these methods can address similar problems, they lack a unified classification framework and a comprehensive comparison. As a result, it remains unclear which methodology is better suited to which open-system scenarios.

In light of the wide range of numerically exact approaches available for simulating open quantum systems, this review aims to organize and present these methods within the common formalism of tensor networks, introduce them coherently and systematically, and make these state-of-the-art methods more accessible.\\

This review is organized as follows: Sec.~\ref{sec:OQS} introduces the framework of open quantum systems and the associated concepts that will be central to the derivations of the methods; in Sec.~\ref{sec:TN}, the formalism of tensor networks is presented, and we show how, and under which conditions, it enables the exponential dimensionality problem to be overcome. Finally, in Sec.~\ref{sec:numerical-methods}, we present each numerical method, covering its derivation, applications, strengths, and limitations.
For each method, a list of its open-source implementations is provided.

\section{BACKGROUND}

\subsection{Open quantum system dynamics}\label{sec:OQS}

The density operator formalism provides a general description of the quantum state of a system through a positive semi-definite, Hermitian operator $\rho$. In this framework, the state is interpreted as a mixed state, that is, a classical statistical mixture of pure states $\{\ket{\mu}\}_\mu$,
\begin{align}
    \rho = \sum_{\mu} p_{\mu}\ket{\mu}\bra{\mu},
\end{align}
where $\sum_{\mu} p_{\mu}=1$ and the coefficients $p_{\mu}$ represent the probabilities associated with the pure states $\ket{\mu}$ in the ensemble.

The time evolution of the density operator is governed by the extension of the Schr\"odinger equation to mixed states, namely the Liouville--von Neumann equation (using units such that $\hbar=1$)
\begin{align}
    \frac{d}{dt}\rho(t) &= -i\left[H(t), \rho(t)\right] \overset{\text{def.}}{=} \mathcal{L}(t)\rho(t)\ ,\label{eq:Liouville}
\end{align}
where $H(t)$ is the Hamiltonian of the total, closed system, and $\mathcal{L}(t)$ is the so-called \emph{Liouvillian} superoperator.

This allows for an integral representation of the equation of motion, namely
\begin{equation}
    \rho(t)=U(t,0)\rho(0)U^\dagger(t,0)\equiv \mathcal{T_\leftarrow}\exp\left({\int_0^t \mathcal{L}(t') dt'}\right)\rho(0),
\end{equation}
with $U(t,0)=\mathcal{T_\leftarrow}\exp\left(-i\int_0^t H(t') dt'\right)$ the evolution operator and $\mathcal{T_\leftarrow}$ the chronological time ordering operator.

Given a bipartition of the degrees of freedom into a system $S$ and an environment $E$, the total Hamiltonian is typically decomposed as
\begin{align}
    H &= H_S + H_E + H_I\ ,
    \label{Eq:partion_hamiltonian}
\end{align}
where $H_S$ denotes the free system Hamiltonian, $H_E$ the free environment Hamiltonian, and $H_I$ the system--environment interaction Hamiltonian.

When only interested in the system dynamics, the system degrees of freedom can be isolated by introducing the reduced system density operator ${\rho_S = \mathrm{Tr}_E[\rho]}$ where $\mathrm{Tr}_E$ denotes a partial trace over the degrees of freedom of the environment.
The equation of motion of the system density operator is hence obtained via the partial trace of \eqref{eq:Liouville} over the environment
\begin{align}
    \frac{d}{dt}\rho_S(t) &= \Tr_E\left[\mathcal{L}(t)\rho(t)\right]\ , \label{eq:traced-Liouville}
\end{align}
where the initial state is typically assumed, as we shall do later, to be a product state of system and environment $\rho(0) = \rho_S(0)\rho_E(0)$ and the environment to be initially in a Gibbs state $\rho_E(0) \propto \exp\left(-\beta H_E\right)$, where $\beta = (k_BT)^{-1}$ is the inverse temperature. In the following, it is also assumed that the interaction Hamiltonian has vanishing expectation value on the initial Gibbs state, i.e., $\mathrm{Tr}_E\left[H_I\rho_E(0)\right] = 0$. This assumption can generally be satisfied by absorbing the trace of the interaction term into the system Hamiltonians.

\eqref{eq:Liouville} can equivalently be written as
\begin{equation}
\label{eq:dynmap}
    \rho_S(t)=\Lambda(t,0)\rho_S(0),
\end{equation}
where the quantum dynamical maps
\begin{align}
    \Lambda(t,0)\rho_S(0) &= \\
    \Tr_E&\left[\mathcal{T_\leftarrow}\exp\left(\int_0^t\mathcal{L}(t') dt' \right)\rho_S(0)\,\rho_E(0)\right]\nonumber,
\end{align}
are introduced.

Expressing the equation of motion in terms of dynamical maps, as in \eqref{eq:dynmap}, allows for a practical definition of Markovianity. In particular, the dynamics can be regarded as Markovian when the maps are completely positive, trace-preserving, and divisible, i.e., $\Lambda(t,0)=\Lambda(t,s)\Lambda(s,0)$ for $t>s$~\cite{rivas_open_2011, breuer_theory_2009}. Under these conditions, the equation of motion of the system reduces to the so-called Gorini--Kossakowski--Sudarshan--Lindblad (GKSL) master equation~\cite{lindblad_generators_1976, GoriniJMathPhys1976}
\begin{equation}
    \frac{d}{dt}{\rho}_S  = -i\left[H_\text{eff}, \rho_S\right] +\sum_{j=1}^{d^2-1}\gamma_j \left( L_j\rho_SL^\dagger_j- \frac{1}{2}\left\{L^\dagger_jL_j ,\rho_S\right\} \right)\ ,\label{eq:LindbladMasterEquation}
\end{equation}
where $\gamma_j\geq0$ are the decay rates associated with the system jump operators $L_j$, $H_\text{eff}$ is the effective system Hamiltonian, and $d$ is the system Hilbert space dimension.
In many cases, the system's equation of motion cannot be reduced to the GKSL master equation, and other approaches have to be used.
Importantly, the notion of Markovianity adopted throughout this review is of a practical nature, while formal definitions and quantitative measures can be found in more comprehensive references~\cite{Breuer2009,Rivas2010,Rivas_2014,pollock_operational_markov_2018,pollock_non-markovian_2018,Milz2019}.
In the following formalisms, our discussion of Markovianity will be heuristic and motivated by its practical and intuitive nature.
Although not equivalent, the different notions of Markovianity we outline below provide a physically consistent picture of how temporal correlations affect system dynamics.

Even though, unless stated otherwise, the following discussion will not refer to a specific system or environment, one case of interest that will often be considered in Sec.~\ref{sec:numerical-methods} is the bosonic harmonic oscillator environment with a coupling $H_I$ linear in the oscillator position, and an initial environment state $\rho_E$ quadratic in the position and momenta of the oscillators.
Such an environment is said to be \emph{Gaussian}. 
The typical Hamiltonian model is written as
\begin{align}
    H_E &= \int_{0}^\infty\!\!\! d\omega\, \omega a_\omega^\dagger a_\omega\ ,\label{eq:Hbath}\\
    H_{I} &= {O} \int_{0}^\infty \!\!\! d\omega\, \sqrt{J(\omega)} \left(a_\omega + a^\dagger_\omega\right)\ ,\label{eq:HSE} 
\end{align}
where we have introduced the annihilation (creation) operator $a^{(\dagger)}_\omega$ of a bosonic mode with frequency $\omega$, the spectral density $J(\omega)$, and a system coupling operator $O$. 
For now, we restrict ourselves to a single linear coupling, while more general forms where several system operators couple to independent or shared baths with distinct spectral densities will be discussed in Sec.~\ref{sec:numerical-methods}.

In the Hamiltonian above in \eqref{eq:HSE}, the spectral density $J(\omega)$, which fully characterizes the environment's influence on the system dynamics through its frequency-dependent coupling to the system, is considered given. 
Therefore, when addressing a real-world system, the spectral density must be characterized in advance.
This is typically done through a microscopic description that may go beyond that employed in the Hamiltonian model considered here. Such bosonic spectral densities can be estimated in various ways, both using first-principles methods~\cite{RengerJPCB2012,CokerJPCL2016,RheePCCP2018,KleinekathoferPR2023} and experimentally, for example, through spectroscopic techniques such as fluorescence line narrowing~\cite{RatsepJL2007} or resonance Raman~\cite{GustinPNAS2023}.
As a consequence, when simulating real-world systems, predictive power strongly depends on the accuracy of the spectral density being used.
We refer to an environment with a highly structured (structureless) spectral density as a highly structured (structureless) environment. This can include multiple peaks in the spectral density or a rapidly changing gradient.
In such cases, the practical notion of Markovianity that allows for the derivation of the GKSL master equation is the absence of environmental structure, such as a flat spectral density, associated with weak system--environment interactions~\cite{breuer_theory_2009}.

In the following, we comment on three routes for handling \eqref{eq:traced-Liouville} to obtain the system's dynamics without the assumption of Markovianity.

\subsubsection{Memory kernels}\label{sec:memory-kernel}

A first route relies on manipulating \eqref{eq:traced-Liouville} to obtain a differential equation for the system density operator $\rho_S$.
This can be achieved via the projection operator technique formulated by Nakajima and Zwanzig~\cite{10.1143/PTP.20.948, 10.1063/1.1731409}.
Introducing the projection superoperator onto the system degrees of freedom and the initial bath state, defined as $\mathcal{P}\bullet = \Tr_E[\bullet]\rho_E(0)$, together with its complement $\mathcal{Q}=1-\mathcal{P}$, leads to the formally exact Nakajima--Zwanzig equation~\cite{rivas_open_2011}
\begin{align}
    \frac{d}{dt}\mathcal{P}\tilde{\rho}(t) &= \int_0^t \mathcal{K}(t,t') \mathcal{P}\tilde{\rho}(t')\, dt', \label{eq:NZ}
\end{align}
where $\tilde{\rho}(t) = e^{i (H_S + H_E) t} \rho(t) e^{-i (H_S+ H_E) t}$ is the density operator in the interaction picture and the superoperator-valued kernel $\mathcal{K}(t,t')$ is explicitly given by
\begin{align}
    \mathcal{K}(t,t') = \mathcal{P}\mathcal{L}_I(t)\,\mathcal{T_\leftarrow}\exp\!\left[\int_{t'}^t d\tau\, \mathcal{Q}\mathcal{L}_I(\tau)\right]\mathcal{Q}\mathcal{L}_I(t')\mathcal{P},
\end{align}
is referred to as the Nakajima--Zwanzig \emph{memory kernel}. Here, $\mathcal{L}_I(t)$ denotes the Liouvillian in the interaction picture, $\mathcal{P}\tilde{\rho}(t)=\tilde{\rho}_S(t)\rho_E(0)$.
Within the assumptions discussed in Sec.~\ref{sec:OQS}, the memory kernel depends only on time separations, i.e., $\mathcal{K}(t,t')=\mathcal{K}(t-t')$. Within this formalism, a heuristic notion of Markovianity can be introduced, given by the regime in which the kernel has temporal support on a timescale much shorter than the characteristic timescale of the system. In such a case, the temporal behavior of the kernel may be approximated, on the timescale of the system dynamics, by a Dirac delta
\begin{align}
    \mathcal{K}(t-t') \propto\delta(t-t').
\end{align}

Assuming a large environment with weak system--environment interactions, a perturbative expansion of the memory kernel can be performed. 
Performing this expansion to second order leads to the Redfield master equation~\cite{breuer_theory_2009,REDFIELD19651}. 
Performing a further `secular' approximation, where rapidly oscillating terms in the master equation are neglected, we recover the GKSL master equation introduced in \eqref{eq:LindbladMasterEquation}.

Furthermore, discretization of the Nakajima-Zwanzig equation, \eqref{eq:NZ}, draws a formal connection to the transfer-tensor method (TTM)~\cite{cerrillo_non-markovian_2014} for the exact and efficient long-time extrapolation of OQS dynamics. Under the assumption of an initial product state of the system and environment, the system state at time $t_n$ can be written as
\begin{align}
    {\rho}(t_n) = \sum_{k=1}^K T_{k}{\rho}(t_{n-k}),
\end{align}
where $K$ is a cutoff determined by a finite memory time and $T_{k}$ is the $k$-th transfer tensor which can be obtained via 
\begin{align}
    T_k = \Lambda(t_k, 0) - \sum_{n=1}^{k-1}T_{k-n}\Lambda(t_n, 0),
\end{align}
where the dynamical map $\Lambda(t, 0)$ is obtained from a numerically exact method of choice, like HEOM~\cite{cerrillo_non-markovian_2014} or TEDOPA~\cite{rosenbach_efficient_2016}, or, in the interaction picture, via the Nakajima-Zwanzig memory kernel as
\begin{align}
    T_k = \mathcal{K}(t_k)\Delta t^2.
\end{align}
The TTM has since been the tool of choice in various OQS studies~\cite{kananenka_accurate_2016,chen_non-markovian_2020,y49y-7vhp, gherardini_transfer-tensor_2022, PhysRevA.102.052206, pollock_tomographically_2018} and can be adapted for capturing initial system--bath correlations~\cite{PhysRevA.96.062122}. It has inspired the development of other extrapolation approaches \cite{cygorek_time-nonlocal_2025, 10.1063/5.0228428, strachan2025acceleratedcalculationimpuritygreens}.

\subsubsection{Influence functionals}\label{sec:IF}

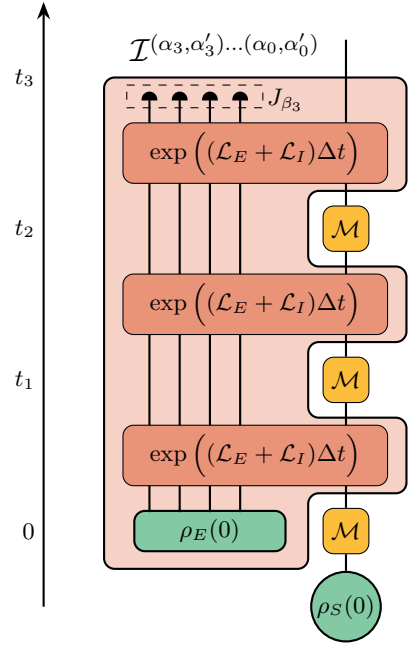
\begin{figure}
    \centering
    \begin{tikzpicture}[
        >=Stealth,
        wire/.style={thick},
        unitary/.style={
            draw=black,
            fill=paletteRed_Light1,
            rounded corners,
            minimum width=3.5cm,
            minimum height=0.8cm
        },
        op/.style={
            draw=black,
            fill=paletteYellow,
            rounded corners,
            minimum size=0.6cm
        },
        comb/.style={
            draw=black,
            fill=paletteRed_Light3,
            thick,
            rounded corners
        },
        state/.style={
            draw=black,
            fill=paletteGreen,
            thick
        },
        every node/.style={font=\small}
    ]

    \draw[comb]
        (-0.2,0.5)
        -- (2.5,0.5)
        -- (2.5,1.5)
        -- (3.8,1.5)
        -- (3.8,2.5)
        -- (2.5,2.5)
        -- (2.5,3.5)
        -- (3.8,3.5)
        -- (3.8,4.5)
        -- (2.5,4.5)
        -- (2.5,5.5)
        -- (3.8,5.5)
        -- (3.8,7)
        -- (-0.2,7)
        -- cycle;

    \draw[dashed, draw=black]
        (0.1,6.9)
        -- (1.9,6.9)
        -- (1.9,6.6)
        -- (0.1,6.6)
        -- cycle;

    \node at (2.2,6.7) {$J_{\beta_3}$};

    \node[circle, state, inner sep=0pt, minimum size=9.2mm] (S) at (3,0) {$\rho_S(0)$};

    \node[
        rectangle,
        rounded corners,
        state,
        minimum height=0.5cm,
        minimum width=2cm
    ] (E) at (1.2,1) {$\rho_E(0)$};

    \draw[->, thick] (-1,0) -- (-1,8);
    \node[left] at (-1,1) {$0$};
    \node[left] at (-1,3) {$t_1$};
    \node[left] at (-1,5) {$t_2$};
    \node[left] at (-1,7) {$t_3$};

    \node[unitary] (U1) at (1.8,2) {$\exp\Big((\mathcal{L}_E + \mathcal{L}_I)\Delta t\Big)$};
    \node[unitary] (U2) at (1.8,4) {$\exp\Big((\mathcal{L}_E + \mathcal{L}_I)\Delta t\Big)$};
    \node[unitary] (U3) at (1.8,6) {$\exp\Big((\mathcal{L}_E + \mathcal{L}_I)\Delta t\Big)$};

    \node[op] (A0) at (3,1) {$\mathcal{M}$};
    \node[op] (A1) at (3,3) {$\mathcal{M}$};
    \node[op] (A2) at (3,5) {$\mathcal{M}$};

    \foreach \x in {0.4,0.8,1.2,1.6} {
        \draw[wire] (\x,1.25) -- (\x,1.6);
        \draw[wire] (\x,2.4) -- (\x,3.6);
        \draw[wire] (\x,4.4) -- (\x,5.6);
        \draw[wire] (\x,6.4) -- (\x,6.7);
        \node[
            shape=semicircle,
            inner sep=0.04cm,
            draw=black,
            fill=black,
            thick
        ] at (\x,6.75) {};
    }

    \draw[thick] (S.90) -- (A0.south);
    \draw[thick] (A0.north) -- (3,1.6);
    \draw[thick] (3,2.4) -- (A1.south);
    \draw[thick] (A1.north) -- (3,3.6);
    \draw[thick] (3,4.4) -- (A2.south);
    \draw[thick] (A2.north) -- (3,5.6);
    \draw[thick] (3,6.4) -- (3,7.5);

    \node at (1.4,7.4) {\large$\mathcal{I}^{(\alpha_3,\alpha_3')\ldots(\alpha_0,\alpha_0')}$};

    \end{tikzpicture}
    \caption{Three time-step evolution of a system in the initial state $\rho_S(0)$ under the influence of an environment initially in the state $\rho_E(0)$. The free system evolution is given by the superoperator $\mathcal{M} = \exp(\mathcal{L}_S \Delta t)$ and the red-shaded area $\mathcal{I}^{(\alpha_3, \alpha_3')\ldots(\alpha_0, \alpha_0')}$ encompasses the free evolution of the environment as well as its interaction with the system. The superoperator $J_{\beta_3}$ is a partial trace over the final environmental degrees of freedom. The superoperator $\mathcal{I}$ is the so-called \emph{process tensor} of the environment; in the continuous time limit, it corresponds to the Feynman-Vernon influence functional.}
    \label{fig:PT-comb}
\end{figure}

Within the path integral formalism~\cite{feynman_theory_1963,weiss_quantum_2012,Caldeira_1981}, the time propagator that is the solution to the time-independent Schr\"odinger equation $U(t,0)$ can be expressed as an integral over trajectories in configuration space, namely
\begin{equation*}
    \langle x_f|U(t,0)|x_i\rangle
    =
    \int_{x(0)=x_i}^{x(t)=x_f}\mathcal{D}[x]\, e^{i\mathcal{S}[x]},
\end{equation*}
where $\mathcal{D}[x]$ denotes the integration measure over the space of paths $x$, and $\mathcal{S}[x]$ is the classical action functional associated with $H$, evaluated along the trajectory $x$.
Using the same approach, the solution of \eqref{eq:Liouville} can be written formally as
\begin{equation*}
\begin{aligned}
\rho(t,\chi,\chi') =
\int_{\substack{\Gamma(t)=\chi\\
                 \Gamma'(t)=\chi'}}
&\mathcal{D}[\Gamma]\mathcal{D}[\Gamma']\,
\rho\!\left(0,\Gamma(0),\Gamma'(0)\right)\\
&\times
\exp\!\left( i\left[\mathcal{S}[\Gamma]-\mathcal{S}[\Gamma']\right]\right).
\end{aligned}
\end{equation*}
Given that $\rho(t,\chi,\chi')=\langle\chi|\rho(t)|\chi'\rangle$, and $\chi$ and $\chi'$ label joint system--environment configurations.

Integrating away the paths associated with the environment leads to
\begin{equation*}
\begin{aligned}
\rho_S(t,\alpha,\alpha') =
\iint_{\substack{\gamma(t)=\alpha\\
                 \gamma'(t)=\alpha'}}
&\mathcal{D}[\gamma]\mathcal{D}[\gamma']\,\rho_S\!\left(0,\gamma(0),\gamma'(0)\right)\\
&\times \mathcal{I}[\gamma,\gamma']
\exp\!\left( i\left[\mathcal{S}_S[\gamma]-\mathcal{S}_S[\gamma']\right]\right),
\end{aligned}
\end{equation*}
where $\alpha$, $\alpha'$ label system configurations, $\mathcal{S}_S[\gamma]$ is the classical action functional induced by $H_S$ along the trajectory $\gamma$, and $\mathcal{I}[\gamma, \gamma']$ is the \emph{influence functional} defined as~\cite{feynman_theory_1963}
\begin{equation*}
\begin{aligned}
\mathcal{I}[\gamma,\gamma'] =
&\int d\sigma
\iint_{\substack{s(t)=\sigma\\
                 s'(t)=\sigma}}
\mathcal{D}[s]\mathcal{D}[s']\, \rho_E\!\left(0,s(0),s'(0)\right)\\
&\times\exp\!\left( i\left[\mathcal{S}_E[s]+\mathcal{S}_I[\gamma,s]
-\mathcal{S}_E[s']-\mathcal{S}_I[\gamma',s']\right]\right),
\end{aligned}
\end{equation*}
where $\mathcal{S}_E[s]$ ($\mathcal{S}_I[\gamma, s]$) is the classical action induced by $H_E$ ($H_{I}$).

In a superoperator formalism, where the system density matrix $[\rho_S(t_n)]_{i_nj_n}$ with Hilbert space indices $i_n, j_n$ is represented as a vector $\rho_{\alpha_n}$ with index $\alpha_n = (i_n, j_n)$, the reduced dynamics can be reformulated as a time-discretized version of the above
\begin{align}
[\rho_S(t)]_{\alpha_N} =&
\lim_{\begin{smallmatrix}
    &{\Delta t\to0}\\
    &t = N\Delta t
\end{smallmatrix}}
\sum_{\substack{
    \{\alpha_n\}, \{\alpha'_n\}
}}\nonumber\\
&\mathcal{I}^{(\alpha_N, \alpha'_N) \ldots (\alpha_0, \alpha'_0)}
\!\left(
    \prod_{l=1}^{N} \mathcal{M}^{\alpha'_l \alpha_{l-1}}
\right)
[\rho_S(0)]_{\alpha_0}.
\label{ReducedDensityMatrix}
\end{align}
The free dynamics of the system are given by
\begin{equation}
    \mathcal{M}^{\alpha^{\prime}_{n}\alpha_{n-1}} = (\alpha_{n}^{\prime}|e^{\mathcal{L}_{S}\Delta t}|\alpha_{n-1}),\label{eq:internal_dynamics}
\end{equation}
where the round brackets $|\alpha)$ indicate a vector state within Liouville space, i.e., the space of operators on the Hilbert space. 
The effects of the general non-Markovian bath are captured in the influence tensor $\mathcal{I}^{(\alpha_{n},\alpha_{n}^{\prime})\ldots(\alpha_{0},\alpha_{0}^{\prime})}$, which stores all non-Markovian information of the dynamics of the environment and its interaction with the system for all previous time steps as
\begin{align}
\mathcal{I}^{(\alpha_N,\alpha_N^\prime)\!\ldots\!(\alpha_0,\alpha_0^\prime)} &=\nonumber\\
\sum_{\beta_N\ldots\beta_0}
J_{\beta_N}\!
\prod_{n=1}^{N}\!&
(\alpha_n,\beta_n|e^{(\mathcal{L}_E +\mathcal{L}_I)\Delta t}|\alpha_n^\prime,\beta_{n-1})
[\rho_{E}(0)]_{\beta_0},
\label{eq:explicitPT}
\end{align}
where $\beta$ labels a basis of the environment, $J_{\beta_{N}}$ takes the partial trace over environmental degrees of freedom at the last time step, effectively ending the propagation of the environment.
When the continuous time limit is not taken, the influence tensor $\mathcal{I}^{(\alpha_{n},\alpha_{n}^{\prime})\ldots(\alpha_{0},\alpha_{0}^{\prime})}$ is called the \emph{process tensor} (PT) of the environment~\cite{pollock_non-markovian_2018}. 
The PT was introduced by \textcite{PhysRevLett.101.060401, PhysRevA.80.022339} and \textcite{pollock_non-markovian_2018} as a tool for characterizing and tomographically reconstructing \cite{White2020, PRXQuantum.3.020344} non-Markovian quantum processes.
The PT can be thought of as a map of an arbitrary set of control operations on a quantum system to the final state of the quantum system, enabling the calculation of multi-time correlations~\cite{PRXQuantum.3.020344,taranto2025higherorderquantumoperations}.
As a tool for open quantum system simulation, the PT approach has gained traction with the development of ACE and PT-TEMPO, which are presented in sections \ref{sec:ACE} and \ref{sec:TEMPO} of this review, respectively. Its main benefit lies in its ability to store the relevant information about the system--environment interaction in a compressed form that can be reused irrespective of the internal system dynamics. Figure~\ref{fig:PT-comb} shows a schematic of the discrete time-evolution of an open system with a PT. \textcite{keeling_process_2026} provides a recent review of PT methods for non-Markovian open quantum systems.

A simplification of path integral formulations can be achieved when in the presence of \emph{Gaussian} environments. Within these additional assumptions, the formal solution for the dynamics of the system density operator $\tilde{\rho}_S(t)$ in the interaction picture can be written in a closed form~\cite{feynman_theory_1963, aurell_operator_2020}
\begin{align}
    \tilde{\rho}_S(t) = \mathcal{T}_{\leftarrow}\exp\Bigg\{&
    -\int_0^t d s\int_0^s d s^\prime O^c(s)\Big[C_R(s-s^\prime)O^c(s^\prime) \nonumber\\
    &+ i C_I(s-s^\prime)O^a(s^\prime)\Big]
    \Bigg\}
    \rho_S(0)\ ,\label{eq:feynman-vernon-path-integral}
\end{align}
where we define the superoperators 
\begin{align}
    O^c(t)\bullet = [O(t), \bullet],~\textrm{and}~O^a(t)\bullet = \{O(t), \bullet\}.
\end{align}
In \eqref{eq:feynman-vernon-path-integral}, the influence functional is not expressed as a path integral but rather as a superoperator acting on the initial system state.
Thus, for bosonic Gaussian environments, their impact on the system in a path integral formulation is determined only by the environment correlation function, also known as the bath correlation function (BCF), $C(t) = C_R(t)+iC_I(t)$, defined as
\begin{align}
    C(t) = \int_{0}^{\infty} d\omega J(\omega)\left[\coth\left(\frac{\beta\omega}{2}\right)\cos(\omega t) -i\sin(\omega t) \right].\label{eq:Environment_Correlation_Function}
\end{align}
We note that an equivalent result to \eqref{eq:feynman-vernon-path-integral} holds for fermionic Gaussian environments~\cite{chen_theory_1987, cirio_canonical_2022, Thoenniss2023}.

Using the result in \eqref{eq:feynman-vernon-path-integral}, one can also introduce the heuristic notion of Markovianity, related to the timescale over which the environmental correlation function $C(t)$ in \eqref{eq:Environment_Correlation_Function} decays. 
When these correlations decay rapidly, excitations emitted by the system into the environment are rapidly dispersed and therefore do not affect the system at later times. 
The time-nonlocal kernels appearing in the Feynman-Vernon influence functional in \eqref{eq:feynman-vernon-path-integral} were later interpreted as memory kernels in close analogy with the convolution kernels $\mathcal{K}(t-t')$ arising in the projection operator formalism of Nakajima-Zwanzig~\cite{Ivander2024}.
Although these two approaches were originally developed independently, they were subsequently recognized as two mathematically equivalent ways of encoding the environment memory effects in the reduced system's dynamics. 

Different microscopic environments can lead to the same correlation function $C(t)$, and thus to equivalent system dynamics. 
For simulation methods, this opens the opportunity for optimization techniques based on an optimal, effective description of the environment; this notion of different representations of the environment generating identical system dynamics is central to most of the numerical techniques discussed in this review. 

\subsubsection{Extended systems}\label{sec:extended-sys}

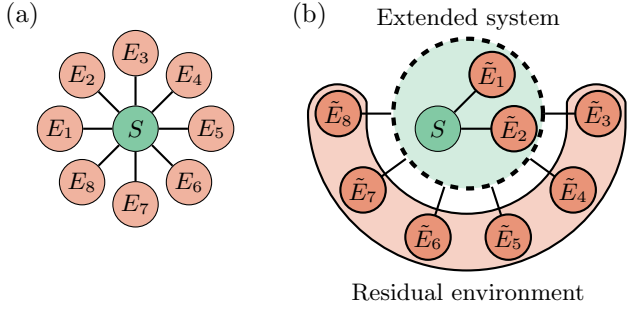
\begin{figure}
    \centering
    \begin{tikzpicture}[
        panel/.style={
            font=\normalsize
        },
        system/.style={
            circle,
            minimum size=0.6cm,
            draw=black,
            fill=paletteGreen,
            inner sep=0pt
        },
        bath/.style={
            circle,
            minimum size=0.6cm,
            draw=black,
            fill=paletteRed_Light2,
            inner sep=0pt
        },
        relevant bath/.style={
            bath,
            fill=paletteRed_Light1
        },
        extended area/.style={
            circle,
            fill=paletteGreen_Light3,
            inner sep=7.0mm
        },
        boundary/.style={
            circle,
            draw=black,
            ultra thick,
            dashed,
            inner sep=7.0mm
        },
        residual area/.style={
            draw=black,
            fill=paletteRed_Light3,
            thick,
            rounded corners
        },
        bond/.style={
            draw=black,
            thick
        }
    ]

    \node[panel] at (-1.5,1.5) {(a)};
    
    \node[system] (Sa) at (0,0) {$S$};
    
    \foreach \ang/\lab in {
        0/5,
        315/6,
        270/7,
        225/8,
        180/1,
        135/2,
        90/3,
        45/4
    } {
        \node[bath] (node\lab) at (\ang:1cm) {$E_{\lab}$};
        \draw[bond] (Sa) -- (node\lab);
    }

    \node[panel] at (2.3,1.5) {(b)};
    
    \node[extended area, label=Extended system] at (4.4,0.2) {};
    
    \node[system] (Sb) at (4,0) {$S$};
    
    \draw[bond] (Sb.45) -- ++(0.5,0.5)
        node[relevant bath] (Eb1) {$\tilde{E}_1$};
    
    \draw[bond] (Sb.east) -- ++(0.7,0)
        node[relevant bath] (Eb2) {$\tilde{E}_2$};
    
    \node[boundary] (box) at (4.4,0.2) {};
    
    \coordinate (awC) at ($(box.west)+(-0.7,0)$);
    
    \def\resShiftX{-0.4cm}
    \def\resOuterX{21.0mm}
    \def\resOuterY{20.75mm}
    \def\resInnerR{13.25mm}
    \def\resCapR{3.875mm}
        
    \draw[residual area]
        (awC) ++(\resShiftX,0)
        arc (-180:0:{\resOuterX} and {\resOuterY})
        arc (0:180:{\resCapR})
        arc (0:-180:{\resInnerR})
        arc (0:180:{\resCapR});
    
    \node at (4.4,-2.15) {Residual environment};
    
    \def\resleg{0.70cm} 
    
    \foreach[count=\k from 3] \ang in {0,-36,-72,-108,-144,-180} {
        \draw[bond]
            (box.\ang) -- ++(\ang:\resleg)
            node[relevant bath] {$\tilde{E}_{\k}$};
    }

    \end{tikzpicture}

    \caption{(a) An open system is interacting with---possibly infinite or continuous---environmental degrees of freedom. (b) The boundary between the `system' and the environment (dashed black line) is redefined to incorporate some \emph{`relevant'} environmental degrees of freedom and the system of interest $S$ within an \emph{extended system} which couples to the `residual' environmental degrees of freedom. Under some conditions, this residual environment can be traced out to obtain a GKSL master equation for the extended system. Such a situation corresponds to a \emph{Markovian embedding} of the extended system.}
    \label{fig:schematics-EE}
\end{figure}

A third route consists of extending the system under consideration---rather than tracing out all environmental degrees of freedom---to include part of the environment explicitly in the dynamical description. The resulting strategy is based on solving the dynamics of an \emph{extended system}. The purpose of these techniques is to reformulate the original problem of the joint \{system + environment\} evolution in a tractable way, by identifying a finite number of relevant environmental degrees of freedom that require explicit treatment.
Figure~\ref{fig:schematics-EE} shows a schematic of the extended system approach.

Two leading strategies have been developed for constructing such an extended system.
One of them identifies effective environments of reduced size by constructing effective environmental degrees of freedom that reproduce the influence of the physical environment on the system dynamics.
Several of these approaches fall under the umbrella of \emph{Markovian embeddings}, in which the environment is partitioned into a non-Markovian core, retaining the relevant memory effects, and a residual Markovian environment, which can---by construction---be traced out. This leads to a GKSL master equation for the extended system obtained by supplementing the original system with the non-Markovian degrees of freedom.
Archetypal examples are the \emph{reaction coordinate} methods~\cite{garg_effect_1985, strasberg_nonequilibrium_2016, martinazzo_communication_2011, nazir_reaction_2018,Iles-Smith2014}, the polaron and variational polaron transformations~\cite{McCutcheon_2010,PhysRevB.84.081305,nazir_modelling_2016,Burgess_2026,Silbey_Harris_1984,Bundgaard-Nielsen_2026}, and the pseudomode theory. In its standard form, pseudomode theory replaces a bosonic environment with an effective environment composed of a finite set of damped, mutually uncoupled effective modes, such that the extended system undergoes GKSL dynamics~\cite{imamoglu_stochastic_1994,garraway_cavity_1996,Garraway1997,TamascelliPRL2018,pleasance_generalized_2020}. The pseudomode model has been further extended to allow for a possible reduction in the number of pseudomode required for representing a target environment. This includes pseudomodes with correlated dissipators~\cite{zhou_systematic_2024}, quasi-Lindblad and non-Hermitian pseudomode formulations~\cite{park_quasi_2024,menczel_non-hermitian_2024}, or by allowing couplings between pseudomodes~\cite{MascherpaPhysRevA2020,huang_coupled_2026}, for which an upper bound on the number of pseudomodes required for reaching a target accuracy has been formalized~\cite{huang_coupled_2026}.
 In principle, fermionic environments can also be treated via a pseudomode representation \cite{h8g7-bmng, PhysRevB.89.165105,Alford_2026,Chen_2014,Brenes_2020, PhysRevB.107.035150}.

In practice, effective-environment methods require a proper characterization of the effective degrees of freedom. Although these approaches are often used in an approximate form, by retaining only a few non-Markovian degrees of freedom and therefore introducing a residual environment that is not guaranteed to be Markovian by construction, numerically exact methodologies can also be developed. In the latter case, the effective environment is constructed to accurately reproduce the influence of the physical environment on the system dynamics. As discussed in Sec.~\ref{sec:IF}, this can be reduced to constructing the effective environment by matching the corresponding bath correlation function. In this case, practical notions of Markovianity are then the same as defined for the path integral formulation as in Sec.~\ref{sec:IF}.

The second of the two strategies consists of extracting a finite number of degrees of freedom from the actual environment.
Several of these approaches rely on a \emph{discretization} of the environment, wherein an optimal discretization has been identified via the chain-mapping procedure~\cite{de_vega_how_2015}.
In chain-mapping approaches, a unitary transformation is applied to the environmental modes, mapping them onto a one-dimensional non-uniform chain with nearest-neighbor interactions~\cite{chin_exact_2010, tamascelli_efficient_2019}. In the presence of a linear coupling between system and environment as in \eqref{eq:HSE}, and for an environment spectrum of finite support (see Sec.~\ref{sec:TEDOPA} for more details), the chain mapping allows one to identify a causal light-cone dynamics within the resulting one-dimensional system. This makes it possible, for any finite evolution time, to identify a finite number of environmental degrees of freedom involved in the dynamics.
Considering a Gaussian environment as in Eqs.~(\ref{eq:Hbath}) and (\ref{eq:HSE}), a different but closely related approach is obtained by expressing the environmental modes in the time domain through the Fourier transform $a(t)=\int d\omega\, e^{-i\omega t} a_\omega$.
In this case, within the interaction picture, the interaction Hamiltonians can be written as an interaction between the system operator $O(t)$ and the time-modes $a(t')$ with a strength $g(t-t')$ given by the Fourier transform of the usual coupling strength in the frequency domain.
The formalism of collisional models~\cite{ciccarello_quantum_2022, lacroix_making_2025,Rau1963} provides a finite representation of the environment by introducing a coarse-graining timescale $\tau$, and defining the \emph{time-bin} modes as an average of the time-modes over this timescale.
Within the framework of collisional models, the notion of memory of the environment is directly related to the number of time-bin modes that interact with the system at each time step. When the dynamics can be described in terms of collisions with a single time-bin at each step, the resulting collisional model is Markovian and, in many instances, its continuous-time limit yields a GKSL master equation for the system~\cite{ciccarello_quantum_2022,Rau1963}.

Overcoming the exponential growth of the Hilbert space or of the histories is the central motivation for the development of the numerically exact methods discussed in this review. These methods explicitly make use of tensor networks, discussed in the following section, because they allow the efficient representation both of the path integral and of explicit environmental degrees of freedom. 

\subsection{Tensor network formalism}\label{sec:TN}

In this section, we give a brief introduction to the tensor network (TN) formalism. 
For a more in-depth discussion, we refer the interested reader to previous reviews and topical books~\cite{orus_practical_2014, bridgeman_hand-waving_2017, simone_montangero_introduction_2018, cirac_matrix_2021}.

While the tensor network formalism, which expresses high-rank tensors as contractions of multiple lower-rank ones, is a broadly applicable concept~\cite{kolda_tensor_2009}, it has developed particularly deep roots in the field of many-body physics. There, as briefly discussed below, the tensor network formalism has provided an efficient and physically motivated framework for representing key physical quantities, such as quantum states and operators, that are central to the analysis of both equilibrium and dynamical properties of complex quantum systems.

In this section, we focus on the representation of quantum states and operators as TNs.
However, as we will see in the subsequent sections of this review, the concepts and tools described here apply to other multidimensional objects (e.g., the process tensor).

Consider a multipartite quantum system composed of $N$ subsystems, which we refer to as \emph{sites}, each associated with a local Hilbert space $\mathcal{H}_k$, for $k=1,\dots,N$. For simplicity, we consider all local Hilbert spaces to have the same dimension $d$. The full Hilbert space of the many-body system is then $\mathcal{H} = \bigotimes_{k=1}^N\mathcal{H}_k$, and given $\{|\phi_{i_k}\rangle\}_{i_k=1,\ldots, d}$ a local basis for the Hilbert space of site $k$, a generic many-body quantum state $|\psi\rangle$ can be expressed as 
\begin{equation}
\label{eq:fullHilbertstate}
    \ket{\psi} = \sum_{\{i_k\}}c_{i_1\ldots i_N}\ket{\phi_{i_1}}\ldots\ket{\phi_{i_N}}\ .
\end{equation}
Here, the amplitudes $c_{i_1\ldots i_N}$ form a rank-$N$ tensor that encodes the full wavefunction. 

One can readily see that this tensor contains $d^N$ elements, reflecting a fundamental challenge arising in many-body physics: the exponential growth of the Hilbert space with the number of particles. As an example, storing the state of one of the simplest many-body systems, a spin chain of 50 sites, already requires around 18 Petabytes.
It is therefore unimaginable to practically perform any simulation on the full Hilbert space, even on such a simple system. While such scaling renders direct many-body system simulations intractable for large $N$, it often reflects a large amount of superfluous information associated with states that are physically irrelevant or inaccessible in finite time. 

A key role in the analysis of physically relevant states is played by the correlations between bipartitions of the systems, which can be quantified via the entanglement entropy. 
For states represented in the full Hilbert space as in \eqref{eq:fullHilbertstate}, when the states are distributed according to the Haar measure, the entanglement entropy scales with the size of the bipartition, giving rise to the so-called \emph{volume law}~\cite{HaydenCommunMathPhys2006}. This corresponds to a scenario where contiguous groups of sites can become arbitrarily highly correlated.

This behavior stems from the fact that the Hilbert space, in its full generality, imposes no constraints on the physicality of the described state. For instance, the locality/finite-range of physical interactions places strong restrictions on the evolution of correlations in realistic many-body systems~\cite{poulin_quantum_2011, EisertRevModPhys2010}. As a result, only a subset of the Hilbert space, referred to as the \emph{physically relevant} subspace, is actually necessary for the characterization of equilibrium or dynamically accessible states.

Of particular interest is the case in which the number of parameters required to describe the physically relevant subspace scales polynomially in the number of particles, thereby reopening the possibility of feasible exact simulations of many-body quantum systems. Among these, particular attention has been devoted to scenarios where the entanglement entropy of a bipartition---which, despite the name, quantifies entanglement only when the global many-body state is pure, while for mixed states it also includes classical correlations and local mixedness--- scales, rather than with the volume, with the area of the boundary of the bipartition, leading to the so-called \emph{area law}. This behavior characterizes situations in which correlations between two subsystems are primarily mediated by sites near the boundary, while contributions from distant sites are negligible.

Over the past decades, numerous analytical results have established the validity of area-law scaling in various settings, most notably for ground (in one-dimensional systems) and thermal states of gapped local Hamiltonians on lattice systems~\cite{BombelliPRD1986, AudenaertPRA2002, srednicki_entropy_1993, hastings_area_2007}.
For a comprehensive exposition, see the review by \textcite{EisertRevModPhys2010}.

Beyond equilibrium, it has been shown that generating a generic many-body quantum state through the evolution under a local Hamiltonian requires a time that grows exponentially with the number of sites, far beyond experimentally accessible timescales~\cite{poulin_quantum_2011}. As a consequence, the dynamics over polynomial timescales remain confined to an exponentially small subspace. Related to this, locality constrains the speed at which correlations can spread. Starting from an uncorrelated initial state, appreciable correlations can develop only between degrees of freedom lying within an effective light cone whose radius grows linearly with time. Outside this light cone, correlations are exponentially suppressed, reflecting the existence of a bounded speed of their propagation~\cite{LiebCommunMathPhys1972}. In a similar spirit, it has been shown that \textit{temporal} correlations within the time-discretized Feynman-Vernon influence functional grow slowly, underpinning the possibility of representing it as an MPO in time \cite{PhysRevB.104.035137, PhysRevLett.128.220401}.

The tensor network formalism has emerged as a tunable framework capable of parameterizing only the physically relevant subspace, thereby avoiding the exponential complexity faced by formalisms that parametrize the entire Hilbert space.
In the following sections, we will identify a class of states and a representation thereof that is particularly well suited to capturing states that follow an area law.

\subsubsection{Graphical representation}
\begin{figure}[h]
    \centering
    \begin{tikzpicture}
    [tensor/.style={circle,draw=black,fill=paletteBlue_Light3,
                 inner sep=0pt,minimum size=6mm}]
    \node (scalar) at (-0.25,0) [tensor] {$a$};
    \node (eq) at (-1.0, 0.0) {$a \to$};
    \node [below] at (-0.5,-0.5) {scalar};
    \node [above] at (-1.25,0.25) {(a)};
    
    \node (vector) at (2.5,0) [tensor] {$v$};
    \node (eq) at (1.0, 0.0) {$v_i \to$};
    \node  at (1.75,0.25) {$i$};
    \draw (1.5,0)--(vector.west);
    \node [below] at (1.75,-0.5) {vector};
    \node [above] at (0.75,0.25) {(b)};
    
    \node (matrix) at (5.5,0) [tensor] {$A$};
    \node (eq) at (4.0, 0.0) {$A_{ij} \to$};
    \node  at (4.75,0.25) {$i$};
    \node  at (6.25,0.25) {$j$};
    \draw (4.50,0)--(matrix.west);
    \draw (matrix.east)--(6.50,0);
    \node [below] at (5.0,-0.5) {matrix};
    \node [above] at (3.75,0.25) {(c)};
    
    \node (3-tensor) at (1,-2) [tensor] {$T$};
    \node at (-0.8, -2) {$T_{ijk} \to $};
    \node (i) at (-0.1, -2) {$i$};
    \node (j) at (2, -2) {$j$};
    \node (k) at (1, -3) {$k$};
    \draw (i.east)--(3-tensor.west);
    \draw (3-tensor.east)--(j.west);
    \draw (3-tensor.south)--(k.north);
    \node [below] at (-0.2,-2.4) {rank-3 tensor};
    \node [above] at (-1.25,-1.8) {(d)};

    \node (eq) at (3.8,-2.2) {$\mathrm{tr}[M] = $} ;
    \node (M) at (5,-2.2) [tensor] {$M$};
    \draw (M.east) edge [in=-210, out=30,looseness=6] (M.west) ;
    \node [above] at (3,-1.8) {(e)};

    \node (eqf) at (0.0,-4) {$v_i = \sum_j A_{ij}u_j \to$};
    
    \node (v) at (2.5,-4) [tensor] {$v$};
    \draw (1.5,-4) --  (v.west) ;
    \node (eq) at (3.2,-4) {=};
    \node[fill=paletteBlue!75] (matrix) at (4.5,-4) [tensor] {$A$} ;
    \node (u) at (5.7,-4) [tensor] {$u$};
    \draw (3.5,-4) -- (matrix.west) ;
    \draw (matrix.east) -- (u.west) ;
    \node [above] at (-1.25,-3.8) {(f)};
    
    \end{tikzpicture}
    \caption{(a) A scalar is a rank-0 tensor which is represented by a plain geometrical shape. (b) A vector is a rank-1 tensor; thus it is represented by a geometrical shape with one leg. (c) A matrix has two indices; thus its representation has two legs. (d) A rank-3 tensor is an object with 3 indices; its representation thus has three legs. (e) The trace of a  $n\times n$-matrix $M$ is depicted by contracting the two legs of the matrix together. (f) The matrix-vector multiplication $\vec{v} = A\vec{u}$, where $A$ is a $n\times m$-matrix and $\vec{u}$ is a $m$-dimensional vector and $\vec{v}$ is a $n$-dimensional vector.}
    \label{fig:diagrammatic}
\end{figure}
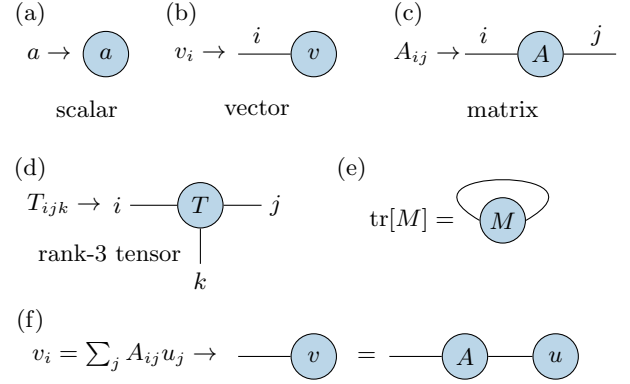

Before proceeding, it is useful to introduce a graphical representation of tensors and tensor operations, based on the Penrose graphical notation~\cite{penrose1971}, which is widely adopted in the tensor network community. In this diagrammatic notation, a rank-$k$ tensor is represented as a geometric shape, such as a circle or a box, with $k$ legs, each corresponding to an index of the tensor. For example, a scalar (rank-0 tensor), a vector (rank-1 tensor), and a matrix (rank-2 tensor) are depicted with $0$, $1$, and $2$ legs, respectively, as illustrated in Fig.~\ref{fig:diagrammatic}(a--c).

Operations on tensor networks typically involve contractions between different tensors, such as, for example, the action of a matrix $A$ on a vector $\mathbf{u}$, yielding a new vector $\mathbf{v}$ with components $v_i = \sum_k A_{ik} u_k$. In the graphical representation, this operation is expressed by connecting the legs corresponding to the summed index $k$ during the contraction.
These examples are shown in Fig.~\ref{fig:diagrammatic}.

Using this graphical representation, the tensor $c$ of probability amplitudes in \eqref{eq:fullHilbertstate} is represented diagrammatically as
\begin{equation}
    \begin{tikzpicture}
    \node (C) at (-2.5,0) {$c_{i_1\ldots i_N} = $};
    \node (c) at (0,0) [ellipse,
                        draw=black,
                        fill=paletteBlue_Light3,
                        minimum width = 2cm,
                        minimum height = 1cm] {\Large c};
    \draw (c.200) -- node[left=0.25] {$i_1$} +(200:0.5);
    \draw (c.220) -- node[below] {$i_2$} +(220:0.5);
    \node[below] at (c.south) {\Large $\ldots$};
    \draw (c.-20) -- node[right=0.25] {$i_N$} +(-20:0.5);
    \node at (2,0) {.};
    \end{tikzpicture}\label{eq:c}
\end{equation}

\subsubsection{Schmidt decomposition}
When considering a many-body system, the correlations between a bipartition $A/B$ of the system can be conveniently characterized by expressing the state of the system in terms of a Schmidt decomposition.

Considering a generic state of the form given in \eqref{eq:fullHilbertstate}, a Schmidt decomposition can be obtained by means of a singular value decomposition (SVD). The SVD is a factorization of a rectangular matrix $A$ of dimension $d_1 \times d_2$ into a product of three matrices: a $d_1 \times d_1$ unitary matrix $U$, a $d_1 \times d_2$ diagonal matrix $S$, and a $d_2 \times d_2$ unitary matrix $V^\dagger$
\begin{equation}
    \label{eq:svd}
    \begin{tikzpicture}[tensor/.style={circle,draw=black,fill=paletteBlue_Light3,
                     inner sep=0pt,minimum size=6mm}]
        \node (A) at (0,0) [tensor] {$A$};
        \draw (-0.7,0) --  (A.west) ;
        \draw (A.east) -- (0.7,0) ;
        \node (eq) at (1,0) {=};
        \node (U) at (2, 0) [tensor, fill=paletteBlue_Light2, draw=black] {$U$} ;
        \node[shape=diamond, draw=black, fill=paletteYellow_Light2, minimum width=4mm] (S) at (3,0) {$S$};
        \node (Vd) at (4, 0) [tensor, fill=paletteRed_Light2, draw=black] {$V^\dagger$} ;
        \draw (1.3,0) -- (U.west) ; 
        \draw (U.east) -- (S.west) ;
        \draw (S.east) -- (Vd.west) ;
        \draw (Vd.east) -- (4.7, 0) ;
    \end{tikzpicture}\ .
\end{equation}

The diagonal elements $\{s_i\}_{i=1,\ldots,r}$ of the matrix $S$, with $r \leq \mathrm{min}(d_1, d_2)$, are non-negative and are called the singular values of the matrix $A$, usually ordered in descending magnitude. Since only the first $r\times r$ block of $S$ is non-zero, the SVD can also be equivalently formulated in a reduced manner, with $U$ a $d_1\times r$ left-unitary matrix ($U^\dagger U=\mathbbm{1}$), $S$ a $r\times r$ diagonal matrix, and $V^\dagger$ a $r\times d_2$ right-unitary matrix ($V^\dagger V=\mathbbm{1}$). 

From a numerical point of view, this decomposition is made possible by the existence of optimized algorithms such as the Golub-Kahan one~\cite{golub_calculating_1965}, divide-and-conquer SVD~\cite{laug}, or randomized SVD~\cite{halko_finding_2011, tamascelli_improved_2015, kohn_probabilistic_2018}.

By grouping together the contributions of partitions $A$ and $B$, the state can be written as 
\begin{equation}
\label{eq:schmidt_2}
    \ket{\psi} = \sum_{ij} c_{ij} \ket{\psi_i^A}  \ket{\phi_j^B}\ ,
\end{equation}
where $i = (i_1, \dots, i_A)$ and $j = (i_{A+1}, \dots, i_N)$. 
Performing an SVD on the matrix of coefficients $c_{ij}$ such that 
\begin{align}
    c_{ij} &= \sum_{\alpha=1}^r U_{i\alpha} S_{\alpha\alpha} V^\dagger_{\alpha j}\ ,
\end{align}
the state can be rewritten in Schmidt form as 
\begin{align}
    \ket{\psi} &= \sum_{\alpha=1}^r S_{\alpha\alpha} \Big( \sum_{i} U_{i\alpha} \ket{\psi_i^A} \Big)  \Big( \sum_{j} V^\dagger_{\alpha j} \ket{\phi_j^B} \Big) \\
    &= \sum_{\alpha=1}^r s_\alpha \ket{\psi_\alpha^A}  \ket{\psi_\alpha^B}\ .
\end{align}
With $\{\ket{\psi_\alpha^A}\}_\alpha$ and $\{|\psi_\alpha^B\rangle\}_\alpha$ denoting orthonormal bases of the Hilbert spaces of partitions $A$ and $B$, respectively, and $s_\alpha$ are the Schmidt coefficients, which satisfy the normalization condition $\sum_{\alpha=1}^r s_\alpha^2 = 1$. The $r$ non-zero Schmidt coefficients are directly related to the bipartite entanglement entropy $S(\rho)$, where $\rho$ is the density operator of one of the two partitions, which for such a state takes the form
\begin{equation}
    S(\rho) = - \sum_{\alpha=1}^r s_\alpha^2 \log(s_\alpha^2)\ .
\end{equation}
For example, if the two partitions are uncorrelated, then $r = 1$, $s_1 = 1$, and hence $S(\rho) = 0$.

Importantly, although the number of Schmidt coefficients $r$ can in principle scale exponentially with the system size, ${r \le \mathrm{min}(\mathrm{dim}(\mathcal{H}_A), \mathrm{dim}(\mathcal{H}_B))}$, many physically relevant states exhibit only limited correlations. In such a case, a much smaller value of $r$ can still provide an excellent approximation to the physical state. For example, when the entanglement entropy between all contiguous bipartitions obeys an area law, or, more precisely, all Rényi entropies $S_\alpha$, with $\alpha \leq 1$ obey an area law, the required Schmidt rank $r$ can scale polynomially, rather than exponentially, in the number of subsystems, while still allowing an accurate approximation to the full quantum state~\cite{SchuchPRL2008}.

The Schmidt decomposition provides a natural way to parametrize a quantum state with a limited amount of correlations between two subsystems by imposing a cutoff value $\chi$ on the number of singular values. This cutoff can be tuned to accommodate higher levels of correlation as needed. However, such a truncation constrains correlations only across a specific bipartition, and particles within partitions $A$ or $B$ can still exhibit correlations that scale according to a volume law.

\subsubsection{Matrix product states and operators}

To obtain an ansatz that describes states satisfying an area law, a restriction on correlations must be imposed for every possible bipartition of the system into contiguous blocks. For one-dimensional many-body systems, as we will motivate in more detail below, this can be achieved by performing a sequence of SVDs, one for each bipartition, and imposing the cutoff $\chi$ at each step~\cite{evenbly_practical_2022}. Through this iterative procedure, a tensor network representation of the state naturally emerges, where the tensor $c$ of a quantum state $\ket{\psi}$ can be decomposed into a product of $N$ smaller rank-3 tensors $T_{k}$
\begin{equation}
   \ket{\psi} = \sum_{\{i_k\}}\sum_{\{\alpha\}} T^{\alpha_0\alpha_1}_{1;i_1}T^{\alpha_1\alpha_2}_{2;i_2}\ldots T^{\alpha_{N-1}\alpha_N}_{N;i_N} \ket{\phi_{i_1}}\ldots\ket{\phi_{i_N}}.\label{eq:mps}
\end{equation}
This decomposition of the tensor of the amplitudes of a quantum state into a product of smaller rank tensors is called a \emph{Matrix Product State} decomposition.
In this form, the local tensor $T_k$ contains the information on the quantum state on site $k$ and its relation (especially the entanglement) with the neighboring sites.
The contracted indices $\alpha_k$ between the tensors are called \emph{inner indices} and carry information about the correlations between bipartitions of the state at bond $k$.
The number of different values an inner index can take is called the \emph{bond dimension} and is denoted $D$.
The free indices $i_k$ associated with local quantum states are called \emph{physical indices}.

Diagrammatically, the decomposition of a quantum state as an MPS presented in Eq.~(\ref{eq:mps}) takes the form
\begin{align}
    \label{eq:mps_diagram}
    \begin{tikzpicture}
        [tensor/.style={circle,draw=black,fill=paletteBlue_Light3,
                     inner sep=0pt,minimum size=6mm}]
        \node (c) at (-0.25,0) {$\ket{\psi}$};
        \node (eq) at (0.25,0) {$=$} ;
        \node[tensor] (T1) at (1,0) {$T_1$} ;
        \node (i1) at (1, -1) {$i_1$} ;
        \node[tensor] (T2) at (2,0) {$T_2$} ;
        \node (i2) at (2, -1) {$i_2$} ;
        \node[tensor] (T3) at (3,0) {$T_3$} ;
        \node (i3) at (3, -1) {$i_3$} ;
        \node (dots) at (4,0) {$\ldots$} ;
        \node[tensor] (TN) at (5.4,0) {$T_N$} ;
        \node (iN) at (5.4, -1) {$i_N$} ;
        \draw (i1) -- (T1.south) ; \draw (T1.east) -- node[above] {$\alpha_1$} (T2.west) ; 
        \draw (i2) -- (T2.south) ; \draw (T2.east) -- node[above] {$\alpha_2$} (T3.west) ;
        \draw (i3) -- (T3.south) ; \draw (T3.east) -- node[above] {$\alpha_3$} (dots.west) ;
        \draw (iN) -- (TN.south) ; \draw (dots.east) -- node[above] {$\alpha_{N-1}$} (TN.west) ;
        \node at (6,0) {.} ;
    \end{tikzpicture}
\end{align}
Note that, although here one tensor is associated with each system site, the same construction can be generalized to other partitions of the degrees of freedom of the many-body system.

Importantly, if no singular values are truncated during the successive SVDs, the bond dimension generated by this procedure can grow exponentially with the system size, reaching $D \propto d^{N/2}$ across the central bipartition. This reflects the fact that the procedure merely reformulates the generic state in \eqref{eq:fullHilbertstate}, which belongs to an exponentially large Hilbert space~\cite{orus_practical_2014, barthel_closedness_2022}.
Hence, in practice, MPSs are used as ans\"atze for many-body quantum states by introducing a truncation $\chi$ in the bond dimension $D$, restricting the formalism to spanning a subspace of the total Hilbert space whose size can be controlled by tuning the value of the bond dimension.
In particular, for one-dimensional systems, an area law implies that the upper bound for the entanglement entropy is independent of the system size. As a consequence, choosing a maximum bond dimension $\chi$ that is independent of $N$ restricts the MPS to states whose entanglement entropy satisfies an area-law bound. The number of coefficients required to describe such an MPS scales as $\mathcal{O}(Nd\chi^2)$ and therefore grows only linearly--and no longer exponentially--with the system size.

Most importantly, a system-size-independent bond dimension cannot, in general, represent states satisfying an area law for more than one-dimensional systems. Indeed, the successive SVD procedure described above can only generate, by construction, a one-dimensional tensor network. Although it can also be applied to higher-dimensional systems, the resulting MPS always corresponds to a one-dimensional unraveling of their degrees of freedom. Thus, sites that are close in the original geometry may become widely separated along the MPS chain, thereby obscuring the locality underlying the area-law structure. We will return to higher-dimensional systems in the following sections.

To compute expectation values of observables or apply unitary transformations to a quantum state, we need a TN representation of operators.
In the same fashion as a one-dimensional quantum state can be represented as an MPS, operators acting on those states can be represented as \emph{Matrix Product Operators} (MPO).
For an operator $O$, such as a density operator $\rho$, an MPO can be defined as follows
\begin{equation}
    O = \sum_{i, i', w}W_{1;\,i_1\ i_1'}^{w_0w_1}\ldots  W_{N;\,i_N\ i_N'}^{w_{N-1}w_N} \ket{\phi_{i_1'}}\bra{\phi_{i_1}}  \ldots \ket{\phi_{i_N'}}\bra{\phi_{i_N}} ,
\end{equation}
where $i$, $i^{'}$ and $w$ are respectively shorthand notations for $\{i_k\}$, $\{i^{'}_k\}$ and $\{w_k\}$.
The terminology used to name the indices of the MPO is the same as for the MPS.
A diagrammatic formulation of an MPO representation of an operator is
\begin{equation}
    \begin{tikzpicture}
    [tensor/.style={rectangle,draw=black,fill=paletteBlue_Light3,
                     inner sep=0pt,minimum size=6mm}]
    \node at (-0.5, 0) {$O$} ;
    \node (eq) at (0,0) {$=$};
    \node (l) at (0,0) {} ;
    \node[tensor] (w1) at (1,0) {$W_1$} ;
    \node[tensor] (w2) at (2.5,0) {$W_2$} ;
    \node[tensor] (w3) at (4,0) {$W_3$} ;
    \node (dots) at (5.5,0) {$\ldots$} ;
    \node[tensor] (wN) at (7,0) {$W_N$} ;
    \node (r) at (8,0) {} ;
    \draw (w1.west) ; \draw (w1.east) -- node[above] {$w_1$} (w2.west) ; \draw (w2.east) -- node[above] {$w_2$} (w3.west) ; \draw (w3.east) -- node[above] {$w_3$} (dots.west) ; \draw (dots.east) -- node[above] {$w_{N-1}\ $} (wN.west) ; \draw (wN.east);
    \draw (w1.south) -- node[below=0.25cm] {$i_1^\prime$} (1,-1) ; \draw (w1.north) -- node[above=0.25cm] {$i_1$} (1,1) ;
    \draw (w2.south) -- node[below=0.25cm] {$i_2^\prime$} (2.5,-1) ; \draw (w2.north) -- node[above=0.25cm] {$i_2$} (2.5,1) ;
    \draw (w3.south) -- node[below=0.25cm] {$i_3^\prime$} (4,-1) ; \draw (w3.north) -- node[above=0.25cm] {$i_3$} (4,1) ;
    \draw (wN.south) -- node[below=0.25cm] {$i_N^\prime$} (7,-1) ; \draw (wN.north) -- node[above=0.25cm] {$i_N$} (7,1) ;
    \node at (7.75,0) {.};
\end{tikzpicture}
\end{equation}

The tensors defining MPSs (and MPOs) are not uniquely defined. It is straightforward to see that, by inserting identities of the form $\mathbbm{1} = U U^{-1}$, with $U$ an invertible matrix, between any pair of neighboring tensors, one obtains an equivalent MPS where $\tilde{T} = U^{-1} T U$
\begin{align}
    \label{eq:gauge-freedom}
    \begin{tikzpicture}
        [tensor/.style={circle,draw=black,fill=paletteBlue_Light3,
                     inner sep=0pt,minimum size=6mm},
        tensor2/.style={circle,draw=black,fill=paletteBlue_Light3,
                     inner sep=0pt,minimum size=6mm},
        matrix/.style={draw=black,fill=paletteGreen_Light1,inner sep=0pt,minimum size=7mm}]
        \node (dotsL) at (0,0) {$\ldots$} ;
        \node[tensor] (T1) at (1,0) {$T$} ;
        \node[tensor] (T2) at (2,0) {$T$} ;
        \node (dotsR) at (3,0) {$\ldots$} ;
        \draw (dotsL.east) -- (T1.west) ;
        \draw (T1.east) -- (T2.west) ;
        \draw (T2.east) -- (dotsR.west);
        \draw (T1.south) -- (1, -0.8) ; 
        \draw (T2.south) -- (2, -0.8) ; 
        \node (eq) at (1.5,-1.25) {$=$} ;
        \node (dotsL2) at (-2,-2) {$\ldots$} ;
        \node[matrix] (Ud0) at (-1,-2) {$U^{-1}$} ;
        \node[tensor] (T12) at (0,-2) {$T$} ;
        \node[matrix] (U1) at (1,-2) {$U$} ;
        \node[matrix] (Ud1) at (2,-2) {$U^{-1}$} ;
        \node[tensor] (T22) at (3,-2) {$T$} ;
        \node[matrix] (U2) at (4,-2) {$U$} ;
        \node (dotsR2) at (5, -2) {$\ldots$} ;
        \draw (dotsL2.east) -- (Ud0.west) ;
        \draw (Ud0.east) -- (T12.west) ;
        \draw (T12.east) -- (U1.west) ;
        \draw (U1.east) -- (Ud1.west) ;
        \draw (Ud1.east) -- (T22.west);
        \draw (T22.east) -- (U2.west) ;
        \draw (U2.east) -- (dotsR2.west);
        \draw (T12.south) -- (0, -2.8) ; 
        \draw (T22.south) -- (3, -2.8) ;
        \node at (1.5, -3) {$=$} ;
        \node (dotsL3) at (0,-3.75) {$\ldots$} ;
        \node[tensor2] (T13) at (1,-3.75) {$\tilde{T}$} ;
        \node[tensor2] (T23) at (2,-3.75) {$\tilde{T}$} ;
        \node (dotsR3) at (3,-3.75) {$\ldots$} ;
        \draw (dotsL3.east) -- (T13.west) ;
        \draw (T13.east) -- (T23.west) ;
        \draw (T23.east) -- (dotsR3.west);
        \draw (T13.south) -- (1, -4.55) ; 
        \draw (T23.south) -- (2, -4.55) ;  
        \node at (5,-4.55) {.} ;
    \end{tikzpicture}
\end{align}
This gauge freedom motivates the introduction of \emph{canonical forms} for MPSs. The most common canonical forms are the left- and right-canonical forms, obtained when the tensors $T_{i_k}^{\alpha_{k-1},\alpha_k}$ correspond, respectively, to left- or right-isometric matrices, i.e., $T_{(i_k\alpha_{k-1}),\alpha_k} = U_{(i_k\alpha_{k-1}),\alpha_k}$ or $T_{\alpha_{k-1},(i_k\alpha_k)} = V^\dagger_{\alpha_{k-1},(i_k\alpha_k)}$, see \eqref{eq:svd}, which can significantly simplify contractions between MPSs because the products of isometries and their conjugates result in an identity. These canonical structures naturally arise when constructing an MPS through successive singular value decompositions (see \eqref{eq:svd}). 
A particularly useful canonical form is the \emph{mixed-canonical form}, where, with respect to a chosen site $k$, the tensors $T_{i_q}$ with $q < k$ are left-canonical, while those with $q > k$ are right-canonical. In this form, the tensor $T_{i_k}$ does not have a defined canonicity and contains the singular value matrix $S$ corresponding to the bipartition $(1,\dots,k)$ and $(k+1,\dots,N)$.
For instance, the case $k = 3$ reads
\begin{align}
    \label{eq:mixed-gauge}
    \begin{tikzpicture}
        [tensor/.style={circle,draw=black,fill=paletteBlue_Light3, inner sep=0pt,minimum size=6mm},
        tensor1/.style={circle,draw=black,fill=paletteBlue_Light2, inner sep=0pt,minimum size=6mm},
        tensor2/.style={circle,draw=black,fill=paletteRed_Light2, inner sep=0pt,minimum size=6mm}]
        \node (c) at (-0.25,0) {$\ket{\psi}$};
        \node (eq) at (0.25,0) {$=$} ;
        \node[tensor1] (T1) at (1,0) {$U_1$} ;
        \node (i1) at (1, -1) {$i_1$} ;
        \node[tensor1] (T2) at (2,0) {$U_2$} ;
        \node (i2) at (2, -1) {$i_2$} ;
        \node[tensor] (T3) at (3,0) {$T_3$} ;
        \node (i3) at (3, -1) {$i_3$} ;
        \node[tensor2] (T4) at (4,0) {$V_4^\dagger$} ;
        \node (i4) at (4, -1) {$i_4$} ;
        \node (dots) at (5,0) {$\ldots$} ;
        \node[tensor2] (TN) at (6,0) {$V_N^\dagger$} ;
        \node (iN) at (6, -1) {$i_N$} ;
        \draw (i1) -- (T1.south) ; \draw (T1.east) -- (T2.west) ; 
        \draw (i2) -- (T2.south) ; \draw (T2.east) -- (T3.west) ;
        \draw (i3) -- (T3.south) ; \draw (T3.east) -- (T4.west) ;
        \draw (i4) -- (T4.south) ; \draw (T4.east) -- (dots.west) ;
        \draw (iN) -- (TN.south) ; \draw (dots.east) -- (TN.west) ;
        \node at (6.5,0) {.} ;
    \end{tikzpicture}
\end{align}
Another well-known canonical form is the \emph{Vidal canonical form}~\cite{VidalPRL2003}, where the singular value matrices $S_k$ associated with the bipartition between sites $(1,\dots,k)$ and $(k+1,\dots,N)$ are explicitly inserted between the tensors $T_{i_k}$ and $T_{i_{k+1}}$ for each $k$
\begin{align}
    \label{eq:vidal-gauge}
    \begin{tikzpicture}
        [tensor/.style={circle,draw=black,fill=paletteBlue_Light3,inner sep=0pt,minimum size=6mm},
        sv/.style={shape=rectangle, rotate=45, draw=black,fill=paletteYellow_Light2,inner sep=0pt,minimum size=3mm}]
        \node (c) at (-0.25,0) {$\ket{\psi}$};
        \node (eq) at (0.25,0) {$=$} ;
        \node[tensor] (T1) at (1,0) {$\Gamma_1$} ;
        \node (i1) at (1, -1) {$i_1$} ;
        \node[sv] (S1) at (1.75,0) {\rotatebox{-45}{\scriptsize$S_1$}};
        \node[tensor] (T2) at (2.5,0) {$\Gamma_2$} ;
        \node (i2) at (2.5, -1) {$i_2$} ;
        \node[sv] (S2) at (3.25,0) {\rotatebox{-45}{\scriptsize$S_2$}};
        \node[tensor] (T3) at (4,0) {$\Gamma_3$} ;
        \node (i3) at (4, -1) {$i_3$} ;
        \node[sv] (S3) at (4.75,0) {\rotatebox{-45}{\scriptsize$S_3$}};
        \node (dots) at (5.5,0) {$\ldots$} ;
        \node[tensor] (TN) at (6.3,0) {$\Gamma_N$} ;
        \node (iN) at (6.3, -1) {$i_N$} ;
        \draw (i1) -- (T1.south) ; \draw (T1.east) -- (S1.135); \draw (S1.-45) -- (T2.west) ; 
        \draw (i2) -- (T2.south) ; \draw (T2.east) -- (S2.135); \draw (S2.-45) -- (T3.west) ; 
        \draw (i3) -- (T3.south) ; \draw (T3.east) -- (S3.135); \draw (S3.-45) -- (dots.west) ; 
        \draw (iN) -- (TN.south) ; \draw (dots.east) -- (TN.west) ;
        \node at (6.8,0) {.} ;
    \end{tikzpicture}
\end{align}
Employing states in a canonical form can greatly reduce the computational cost associated with manipulating MPSs, since contractions can be carried out efficiently by exploiting the left- or right-unitary property of the tensors.

\subsubsection{Tree tensor network states and operators}
Generalizations of the SVD to tensors of rank larger than 2 can allow the construction of tensor networks of different geometry.
In Tucker decomposition, for example, a tensor is factorized into a core tensor $C$ contracted with a unitary matrix $U^i$ along each of its legs~\cite{kolda_tensor_2009}
\begin{equation}
    \label{eq:tucker}
    \begin{tikzpicture}[tensor/.style={circle,draw=black,fill=paletteBlue_Light3,
                     inner sep=0pt,minimum size=6mm}]
        \node (3-tensor) at (0,0) [tensor] {$T$};
        \node (eq) at (1,0) {=}; 
        \node (A) at (2,0) [tensor, fill=paletteBlue_Light2] {$U^1$};
        \node[shape=diamond, minimum size=2mm, draw=black, fill=paletteYellow_Light2] (G) at (3, 0) {$C$};
        \node (B) at (4,0) [tensor, fill=paletteBlue_Light2] {$U^2$};
        \node (C) at (3,-1) [tensor, fill=paletteBlue_Light2] {$U^3$};
        \draw (-0.7,0)--(3-tensor.west);
        \draw (3-tensor.east)--(0.7,0);
        \draw (3-tensor.south)--(0,-0.7);
        \draw (1.3,0) -- (A.west);
        \draw (A.east) -- (G.west);
        \draw (G.east) -- (B.west);
        \draw (B.east) -- (4.7,0);
        \draw (G.south) -- (C.north);
        \draw (C.south) -- (3,-1.7);
    \end{tikzpicture}\ .
\end{equation}

Via successions of such SVD, it is possible to create a hierarchical (i.e.~tree-like) decomposition of an initial high-rank tensor~\cite{grasedyck_hierarchical_2010, hackbusch_new_2009}, such as the (loop-free) tree tensor network (TTN) state~\cite{shi_classical_2006}. In a TTN, contrary to MPSs, a site can be connected to any arbitrary number of child sites, realizing the quasi-1D structure
\begin{equation}
\begin{tikzpicture}[
    tensor/.style={circle, draw=black, fill=paletteBlue_Light3, minimum size=4mm},
    line/.style={draw, thick},
    leg/.style={draw, thick}
    ]
\node (c) at (-0.25,3) {$\ket{\psi}$};
\node (eq) at (0.5,3) {$=$} ;
\node [tensor] (p1) at (0,0) {};
\node [tensor] (p2) at (1,0) {};
\node [tensor] (p3) at (2,0) {};
\node [tensor] (p4) at (3,0) {};
\node [tensor] (p5) at (4,0) {};
\node [tensor] (p6) at (5,0) {};
\node [tensor] (p7) at (6,0) {};
\node [tensor] (p8) at (7,0) {};

\node [tensor] (m1) at (0.5,1.2) {};
\node [tensor] (m2) at (2.5,1.2) {};
\node [tensor] (m3) at (4.5,1.2) {};
\node [tensor] (m4) at (6.5,1.2) {};

\node [tensor] (u1) at (1.5,2.4) {};
\node [tensor] (u2) at (5.5,2.4) {};

\node [tensor] (r) at (3.5,3.6) {};

\draw[line] (p1) -- (m1);
\draw[line] (p2) -- (m1);
\draw[line] (p3) -- (m2);
\draw[line] (p4) -- (m2);
\draw[line] (p5) -- (m3);
\draw[line] (p6) -- (m3);
\draw[line] (p7) -- (m4);
\draw[line] (p8) -- (m4);
\draw[line] (m1) -- (u1);
\draw[line] (m2) -- (u1);
\draw[line] (m3) -- (u2);
\draw[line] (m4) -- (u2);
\draw[line] (u1) -- (r);
\draw[line] (u2) -- (r);

\foreach \x in {p1,p2,p3,p4,p5,p6,p7,p8,m1,m2,m3,m4,u1,u2,r} {
    \draw[leg] (\x) -- ++(0.,-0.8);
}
\node at (7.5,3) {.};
\end{tikzpicture}
\label{eq:TTN}
\end{equation}
Equation~(\ref{eq:TTN}) represents the decomposition of a multipartite quantum state as a binary tree.
These TTNs can be constructed from the initial high-rank tensor $c_{i_1\ldots i_N}$ using a hierarchical Tucker decomposition~\cite{grasedyck_hierarchical_2010, hackbusch_new_2009}, or can be assumed \emph{a priori} when considering a product state.
For scaling reasons, TTN states where tensors have at most three legs (including both physical and inner ones) are computationally favorable~\cite{gunst_t3ns_2018}.
In TTNs, the entanglement entropy associated with a bipartition is bounded by the number of virtual bonds cut by the partition and by their dimensions. Accordingly, imposing a maximum bond dimension $\chi$ that is independent of the system size allows TTNs to approximately represent states satisfying an area law in one-dimensional and tree-like geometries~\cite{evenbly_tensor_2011}.
The operator version of the TTN state is defined accordingly.
Like MPOs, TTN operators can be constructed manually, but without guarantees of finding the exact representation minimizing the operator inner bond dimension, or with algorithmic approaches~\cite{milbradt_state_2024, li_optimal_2024, cakir_optimal_2025}.
TTNs have been extensively used in the context of quantum chemistry as they are the cornerstone of the Multi-Layer Multi-Configuration Time-Dependent Hartree method~\cite{larsson_tensor_2024} (see Sec.~\ref{sec:MLMCTDH}) and in condensed matter~\cite{murg_simulating_2010}.

\subsubsection{Higher-dimensionality tensor networks}
In principle, one can define a tensor network ansatz of arbitrary dimensionality. In this construction, a lattice with the desired dimensionality is identified and each lattice site is associated with a tensor connected to its nearest neighbors through inner bonds, whose bond dimension can be tuned to accommodate a target maximum level of correlations. This approach has been realized, for example, in the Projected Entangled Pair States (PEPS) formalism~\cite{VerstraetePRL2006}.

However, higher-dimensional tensor networks face significant challenges, in particular related to the computational costs associated with operations. 
In particular, since the computational advantage of tensor networks stems from the use of low-rank tensors, it is crucial to ensure that the operations being performed do not generate high-rank tensors. When dealing with two- or higher-dimensional tensor networks, the computation of network contractions, which involve successive pairwise contractions of tensors, can lead to intermediate tensors whose rank increases with the number of contractions performed. 
This growth in tensor rank substantially increases the computational cost of these operations and, as a result, can significantly diminish the efficiency of the tensor network formalism. 
Furthermore, due to the existence of loops in these higher-dimensional networks, there are no clear generalizations of the canonical forms that exist for (quasi-)one-dimensional networks~\cite{evenbly_gauge_2018}.

\begin{figure}[ht]
    \centering
    \resizebox{\linewidth}{!}{%
        \begin{tikzpicture}[
    scale=0.75,
    transform shape,
    tensorLower/.style={circle,draw=black,fill=paletteBlue_Light2,minimum size=4.0mm,inner sep=0pt},
    tensorUpper/.style={circle,draw=black,fill=paletteGreen_Light1,minimum size=4.0mm,inner sep=0pt},
    tensorG/.style={circle,draw=black,fill=paletteYellow,minimum size=4.0mm,inner sep=0pt},
    bond/.style={draw=black},
    phys/.style={draw=black},
    physred/.style={draw=paletteRed, very thick},
    arr/.style={->,thick}
]
\def\sx{0.8}
\def\dx{0.3}
\def\dy{0.42}
\def\sep{2.8}
\def\roff{2.2mm}

\begin{scope}
\foreach \i in {0,...,5} {
    \foreach \j in {0,...,4} {
        \pgfmathsetmacro{\X}{\i*\sx+\j*\dx}
        \pgfmathsetmacro{\Yl}{\j*\dy}
        \pgfmathsetmacro{\Yu}{\j*\dy+\sep}
        \coordinate (ACL\i\j) at (\X,\Yl);
        \coordinate (ACU\i\j) at (\X,\Yu);
    }
}
\foreach \i in {0,...,5} {
    \foreach \j in {0,...,4} {
        \node[tensorLower] (AL\i\j) at (ACL\i\j) {};
    }
}
\foreach \i in {0,...,4} {
    \pgfmathtruncatemacro{\ip}{\i+1}
    \foreach \j in {0,...,4} {
        \draw[bond] (AL\i\j) -- (AL\ip\j);
    }
}
\foreach \i in {0,...,5} {
    \foreach \j in {0,...,3} {
        \pgfmathtruncatemacro{\jp}{\j+1}
        \draw[bond] (AL\i\j) -- (AL\i\jp);
    }
}
\foreach \i in {0,...,5} {
    \foreach \j in {0,...,4} {
        \draw[phys] ($(ACL\i\j)+(0,\roff)$) -- ($(ACU\i\j)+(0,-\roff)$);
    }
}
\draw[physred] ($(ACL54)+(0,\roff)$) -- ($(ACU54)+(0,-\roff)$);
\foreach \i in {0,...,4} {
    \pgfmathtruncatemacro{\ip}{\i+1}
    \foreach \j in {0,...,4} {
        \draw[bond] (ACU\i\j) -- (ACU\ip\j);
    }
}
\foreach \i in {0,...,5} {
    \foreach \j in {0,...,3} {
        \pgfmathtruncatemacro{\jp}{\j+1}
        \draw[bond] (ACU\i\j) -- (ACU\i\jp);
    }
}
\foreach \i in {0,...,5} {
    \foreach \j in {0,...,4} {
        \node[tensorUpper] (AU\i\j) at (ACU\i\j) {};
    }
}
\end{scope}

\draw[arr] (5.5,2.2) -- (6.6,2.2);

\begin{scope}[shift={(7.0,0)}]
\foreach \i in {0,...,5} {
    \foreach \j in {0,...,4} {
        \pgfmathsetmacro{\X}{\i*\sx+\j*\dx}
        \pgfmathsetmacro{\Yl}{\j*\dy}
        \pgfmathsetmacro{\Yu}{\j*\dy+\sep}
        \coordinate (BCL\i\j) at (\X,\Yl);
        \coordinate (BCU\i\j) at (\X,\Yu);
    }
}
\coordinate (BG) at ($(BCL54)!0.5!(BCU54)$);
\draw[physred] (BCU44) -- (BG);
\foreach \i in {0,...,5} {
    \foreach \j in {0,...,4} {
        \ifnum\i=5
            \ifnum\j=4\else\node[tensorLower] (BL\i\j) at (BCL\i\j) {};\fi
        \else
            \node[tensorLower] (BL\i\j) at (BCL\i\j) {};
        \fi
    }
}
\foreach \i in {0,...,4} {
    \pgfmathtruncatemacro{\ip}{\i+1}
    \foreach \j in {0,...,4} {
        \ifnum\i=4
            \ifnum\j=4\else\draw[bond] (BL\i\j) -- (BL\ip\j);\fi
        \else
            \draw[bond] (BL\i\j) -- (BL\ip\j);
        \fi
    }
}
\foreach \i in {0,...,5} {
    \foreach \j in {0,...,3} {
        \pgfmathtruncatemacro{\jp}{\j+1}
        \ifnum\i=5
            \ifnum\j=3\else\draw[bond] (BL\i\j) -- (BL\i\jp);\fi
        \else
            \draw[bond] (BL\i\j) -- (BL\i\jp);
        \fi
    }
}
\foreach \i in {0,...,5} {
    \foreach \j in {0,...,4} {
        \ifnum\i=5
            \ifnum\j=4\else\draw[phys] ($(BCL\i\j)+(0,\roff)$) -- ($(BCU\i\j)+(0,-\roff)$);\fi
        \else
            \draw[phys] ($(BCL\i\j)+(0,\roff)$) -- ($(BCU\i\j)+(0,-\roff)$);
        \fi
    }
}
\foreach \i in {0,...,4} {
    \pgfmathtruncatemacro{\ip}{\i+1}
    \foreach \j in {0,...,4} {
        \ifnum\i=4
            \ifnum\j=4\else\draw[bond] (BCU\i\j) -- (BCU\ip\j);\fi
        \else
            \draw[bond] (BCU\i\j) -- (BCU\ip\j);
        \fi
    }
}
\foreach \i in {0,...,5} {
    \foreach \j in {0,...,3} {
        \pgfmathtruncatemacro{\jp}{\j+1}
        \ifnum\i=5
            \ifnum\j=3\else\draw[bond] (BCU\i\j) -- (BCU\i\jp);\fi
        \else
            \draw[bond] (BCU\i\j) -- (BCU\i\jp);
        \fi
    }
}
\foreach \i in {0,...,5} {
    \foreach \j in {0,...,4} {
        \ifnum\i=5
            \ifnum\j=4\else\node[tensorUpper] (BU\i\j) at (BCU\i\j) {};\fi
        \else
            \node[tensorUpper] (BU\i\j) at (BCU\i\j) {};
        \fi
    }
}
\node[tensorG] (G) at (BG) {};
\draw[bond] (BU53) -- (G);
\draw[bond] (BL44) -- (G);
\draw[bond] (BL53) -- (G);
\end{scope}

\draw[arr] (9.0,-0.5) -- (8.0,-1.6);

\begin{scope}[shift={(3.5,-6.5)}]
\foreach \i in {0,...,5} {
    \foreach \j in {0,...,4} {
        \pgfmathsetmacro{\X}{\i*\sx+\j*\dx}
        \pgfmathsetmacro{\Yl}{\j*\dy}
        \pgfmathsetmacro{\Yu}{\j*\dy+\sep}
        \coordinate (CCL\i\j) at (\X,\Yl);
        \coordinate (CCU\i\j) at (\X,\Yu);
    }
}
\coordinate (CG0) at ($(CCL54)!0.5!(CCU44)$);
\coordinate (CG) at ($(CG0)+(0.45,-0.65)$);

\draw[bond] (CCU34) -- (CG);
\draw[bond] (CCU43) -- (CG);
\draw[bond] (CCU53) -- (CG);
\draw[bond] (CCL44) -- (CG);
\draw[bond] (CCL53) -- (CG);

\foreach \i in {0,...,5} {
    \foreach \j in {0,...,4} {
        \ifnum\i=5
            \ifnum\j=4\else\node[tensorLower] (CL\i\j) at (CCL\i\j) {};\fi
        \else
            \node[tensorLower] (CL\i\j) at (CCL\i\j) {};
        \fi
    }
}
\foreach \i in {0,...,4} {
    \pgfmathtruncatemacro{\ip}{\i+1}
    \foreach \j in {0,...,4} {
        \ifnum\i=4
            \ifnum\j=4\else\draw[bond] (CL\i\j) -- (CL\ip\j);\fi
        \else
            \draw[bond] (CL\i\j) -- (CL\ip\j);
        \fi
    }
}
\foreach \i in {0,...,5} {
    \foreach \j in {0,...,3} {
        \pgfmathtruncatemacro{\jp}{\j+1}
        \ifnum\i=5
            \ifnum\j=3\else\draw[bond] (CL\i\j) -- (CL\i\jp);\fi
        \else
            \draw[bond] (CL\i\j) -- (CL\i\jp);
        \fi
    }
}
\foreach \i in {0,...,5} {
    \foreach \j in {0,...,4} {
        \ifnum\i=5
            \ifnum\j=4\else
                \ifnum\i=4
                    \ifnum\j=4\else\draw[phys] ($(CCL\i\j)+(0,\roff)$) -- ($(CCU\i\j)+(0,-\roff)$);\fi
                \else
                    \draw[phys] ($(CCL\i\j)+(0,\roff)$) -- ($(CCU\i\j)+(0,-\roff)$);
                \fi
            \fi
        \else
            \ifnum\i=4
                \ifnum\j=4\else\draw[phys] ($(CCL\i\j)+(0,\roff)$) -- ($(CCU\i\j)+(0,-\roff)$);\fi
            \else
                \draw[phys] ($(CCL\i\j)+(0,\roff)$) -- ($(CCU\i\j)+(0,-\roff)$);
            \fi
        \fi
    }
}
\foreach \i in {0,...,4} {
    \pgfmathtruncatemacro{\ip}{\i+1}
    \foreach \j in {0,...,4} {
        \ifnum\j=4
            \ifnum\i=3\else
                \ifnum\i=4\else\draw[bond] (CCU\i\j) -- (CCU\ip\j);\fi
            \fi
        \else
            \draw[bond] (CCU\i\j) -- (CCU\ip\j);
        \fi
    }
}
\foreach \i in {0,...,5} {
    \foreach \j in {0,...,3} {
        \pgfmathtruncatemacro{\jp}{\j+1}
        \ifnum\i=4
            \ifnum\j=3\else\draw[bond] (CCU\i\j) -- (CCU\i\jp);\fi
        \else
            \ifnum\i=5
                \ifnum\j=3\else\draw[bond] (CCU\i\j) -- (CCU\i\jp);\fi
            \else
                \draw[bond] (CCU\i\j) -- (CCU\i\jp);
            \fi
        \fi
    }
}
\foreach \i in {0,...,5} {
    \foreach \j in {0,...,4} {
        \ifnum\i=5
            \ifnum\j=4\else\node[tensorUpper] (CU\i\j) at (CCU\i\j) {};\fi
        \else
            \ifnum\i=4
                \ifnum\j=4\else\node[tensorUpper] (CU\i\j) at (CCU\i\j) {};\fi
            \else
                \node[tensorUpper] (CU\i\j) at (CCU\i\j) {};
            \fi
        \fi
    }
}
\node[tensorG] (H) at (CG) {};
\end{scope}
\end{tikzpicture}
    }
    \caption{First two steps of the contraction of a pair of two-dimensional tensor networks. The red bond represents the bond under contraction. The number of indices of the resulting yellow tensor grows rapidly after each contraction, resulting in a computational bottleneck of two-dimensional TN.}
    \label{Fig_2DTN}
\end{figure}
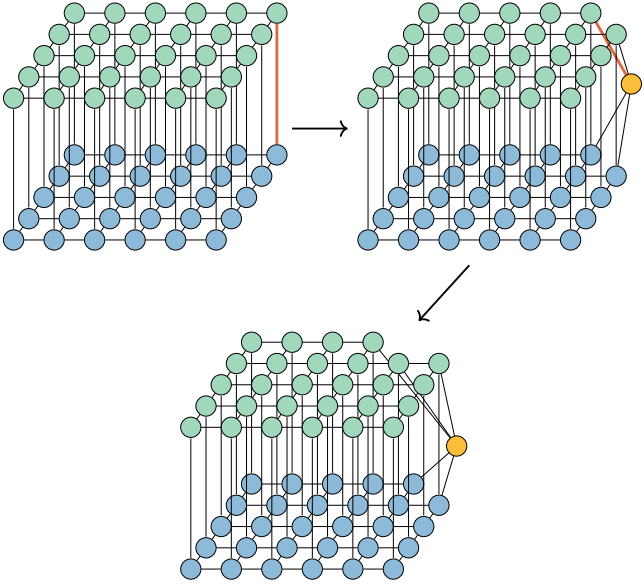

As a simple illustrative example, Fig.~\ref{Fig_2DTN} shows how, during the contraction of two two-dimensional tensor networks, higher-rank tensors can be rapidly generated, rendering such operations computationally unfeasible. 
Numerous advances have been reported in recent years aimed at reducing the computational cost of higher-dimensional tensor networks~\cite{ran2020tensor, gray_hyper-optimized_2021, gray_hyperoptimized_2024, schindler_algorithms_2020}.

\subsubsection{Time-evolution methods}
\label{sec:TDVP}
To study the dynamics of a quantum state $\ket{\psi(t)}$ starting from a given initial condition $\ket{\psi(0)}$, one has to be able to perform time-evolution on a tensor network state (TNS).
The general solution of the Schr\"odinger equation for a time-independent Hamiltonian is
\begin{align}
    \ket{\psi(t)} &= e^{-i\h t}\ket{\psi(0)}\ .
\end{align}
The straightforward solution would thus be to write down the evolution operator $U(t) = \exp(-i\h t)$ as an MPO or a TTN operator.
However, this is impractical for long-time simulations as the size of the bond dimension would become too large.
A solution is to split up $U(t)$ into a product of $N$ one-time-step evolution operators $U(\Delta t)$, with $\Delta t = t/N$, and truncate the MPS/TTN bond dimension after each application
\begin{align}
    U(t) = \e^{-\i\h N\Delta t} &= \left( \e^{-\i\h\Delta t}\right)^{N} = \big(U(\Delta t)\big)^N \ .
\end{align}
The exponentiation of the full Hamiltonian might, however, still be computationally inefficient; that is why one usually decomposes the single time-step evolution operator further.
For example, let us say that the Hamiltonian $\h$ can be decomposed into two terms $\h = \h_1 + \h_2$ where the terms comprising each $\h_i$ commute with one another, but $\h_1$ and $\h_2$ don't necessarily commute.
Then, the single time step evolution operator can be split up using a Trotter decomposition~\cite{TrotterPAMS1959} 
\begin{align}
\label{eq:trotter}
    U(\Delta t) &= \e^{-\i\h_1 \Delta t} \e^{-\i\h_2 \Delta t} + \mathcal{O}(\Delta t^2)\ .
\end{align}
The single time step error induced by the Trotterization of the exponential can be decreased by considering higher-order decompositions~\cite{suzuki_generalized_1976}, such as to $\mathcal{O}(\Delta t^3)$ when considering the Suzuki-Trotter decomposition~\cite{Suzuki1985}.
\begin{align}
\label{eq:symmtrotter}
    U(\Delta t) &= e^{-iH_1 \Delta t/2} e^{-iH_2 \Delta t}e^{-iH_1 \Delta t/2} + \mathcal{O}(\Delta t^3)\ ,
\end{align}
here referred to as symmetric Trotterization. These errors are associated with a single time step. After $N=t/\Delta t$ time steps, the accumulated error scales one order lower in $\Delta t$.
The \emph{time-dependent DMRG}~\cite{white_real-time_2004, cazalilla_time-dependent_2002} and the \emph{Time-Evolving Block Decimation} (TEBD)~\cite{vidal_efficient_2004, shi_classical_2006} methods are based on this approach, where each exponential is written in the form of an MPO.
TEBD is adapted for Hamiltonians with short-range interactions that can be split into internally commuting parts. 
The impact of the time step error on the unitarity of the time evolution can be controlled by choosing a sufficiently small $\Delta t$.
The other source of error originates from the truncation of the bond dimension performed after each time step.
This truncation affects the unitarity of the time evolution, but convergence can be easily controlled by varying the number of singular values kept.\\

Another class of time evolution methods takes a different perspective.
Instead of finding an efficient MPO representation $U(\Delta t)$, they aim at approximating the \emph{action} of the time evolution operator on a quantum state.
This means that they offer a way to evolve $\ket{\psi(t)}$ into $\ket{\psi(t+\Delta t)}$ without constructing an explicit representation of $U(\Delta t)$.
The two main methods representative of this approach are the global Krylov method \cite{saad_iterative_2003, paeckel_time-evolution_2019} and the time-dependent variational principle (TDVP)~\cite{dirac_note_1930, frenkel_wave_1934, raab_diracfrenkelmclachlan_2000}.
The central idea of TDVP, in the modern tensor networks formulation, is that instead of solving the Schr\"odinger equation and then truncating the TNS representation of the quantum state, one can solve the equations of motion projected into a space of restricted bond dimension \cite{haegeman_time-dependent_2011, haegeman_unifying_2016, lubich_dynamical_2013, paeckel_time-evolution_2019, bauernfeind_time_2020}.
Within this approach, one looks for a solution $\ket{\varphi} \in \mathcal{M}$ of the Schr\"odinger equation where $\mathcal{M} \subset \mathcal{H}$ is a manifold of the total Hilbert space $\mathcal{H}$ in which one believes that the relevant physical states exist.
The manifold $\mathcal{M}$ corresponds to the space of MPS or TTN of a given bond dimension $D$.
The major advantage of this method is that it naturally preserves the unitarity of the time evolution and conserves the energy.
The downside is that the uniformity of the bond dimension across the entire MPS/TTN may not be an efficient representation of entanglement structure in the quantum state (i.e. some bonds have a too large bond dimension), slowing down the computations.
To address this downside, two-site variants have been devised where the contraction of two neighboring sites is evolved and then split using an SVD, thus adapting the bond dimension during the evolution, but at the cost of losing the exact unitarity and conservation of the energy.
Recently, adaptive one-site algorithms have combined the best of both worlds, allowing for locally increasing bond dimension on the fly while preserving the unitarity of the evolution~\cite{yang_time-dependent_2020, dunnett_efficient_2021, li_time-dependent_2024}.

\subsubsection{Software packages}

Several software packages efficiently implement tensor operations (e.g. contractions and SVD) and can be used to easily construct and manipulate tensor networks.
We will not try to be exhaustive, but will guide the reader toward those we consider the most prominent in different programming languages.

\paragraph{NumPy:} \texttt{NumPy} is the fundamental package for scientific computing in \texttt{Python}. Although it is not directly tailored for tensor networks, this library provides multidimensional array objects, a large library of functions that operate efficiently on these data structures, basic linear algebra, and supports the use of the Einstein summation convention.

\paragraph{TeNPy:} Tensor Network Python is a \texttt{Python} library providing MPS data structures and linear algebra for their manipulation. Additionally, it provides an implementation of DMRG and several time-evolution algorithms, including TEBD~\cite{tenpy2024}.

\paragraph{ITensor:} \texttt{ITensors.jl} and \texttt{ITensorMPS.jl} form an ecosystem of packages in the \texttt{Julia} programming language that includes a full set of algorithms involving MPS and MPO, such as the density matrix renormalization group (DMRG) and TEBD, and algorithms for summing, multiplying, and optimizing MPS and MPOs~\cite{fishman_itensor_2022}.

\paragraph{QuantumKitHub: } A \texttt{Julia} ecosystem for tensor networks and quantum many-body physics, including several packages such as \texttt{TensorOperations.jl} for fast tensor operations using a convenient Einstein index notation~\cite{TensorOperations.jl}, \texttt{MPSKit.jl}~\cite{devos_MPSKit_2026} and \texttt{PEPSKit.jl}~\cite{Brehmer_PEPSKit_2026} for MPS and PEPS simulations respectively.

\paragraph{Tensor Toolbox:} The Tensor Toolbox for \texttt{MATLAB} provides a suite of tools for working with multidimensional arrays. This toolbox provides many standard methods for decomposing tensors~\cite{bader_tensor_2025}.

\paragraph{cuTensorNet:} NVIDIA cuTensorNet is a library for tensor network computations on GPUs. It provides a set of APIs for creating and contracting tensor networks and performing tensor decomposition using QR or SVD~\cite{cuQuantum_2023}.\\

Libraries for the specific OQS methods that will be discussed in Sec.~\ref{sec:numerical-methods} will be presented in the relevant subsections and are summarized in Table~\ref{tab:Summary_Methods} in Sec.~\ref{sec:outlook}.

\section{NUMERICAL METHODS}\label{sec:numerical-methods}

In this section, we present the different tensor-network numerically exact methods introduced in Sec.~\ref{sec:Intro}.
These methods are presented in alphabetical order.

\subsection{ACE}\label{sec:ACE}

\subsubsection{Overview}

The \textit{Automated Compression of Environments} (ACE) algorithm offers a numerically exact framework for modeling open quantum systems \cite{mortiz_2022_ACE} utilizing process tensors (PTs). Therein, ACE constructs a PT from, in principle, an arbitrarily large set of discrete, non-interacting environmental subsystems. The PT, which contains the influence of the entire environment, is constructed through an iterative combination of each environmental subsystem utilizing TN compression techniques to construct an efficient representation of the PT in matrix product operator (MPO) form. We also note that, in tandem with the construction of the ACE algorithm, \textcite{Ye2021TensorNetworkIF} also proposed a numerically exact tensor network framework based on environmental discretization~\cite{park_tensor_2024}.

\subsubsection{Derivation} 
\begin{figure*}[t]
    \centering
    \resizebox{\textwidth}{!}{%
        \input{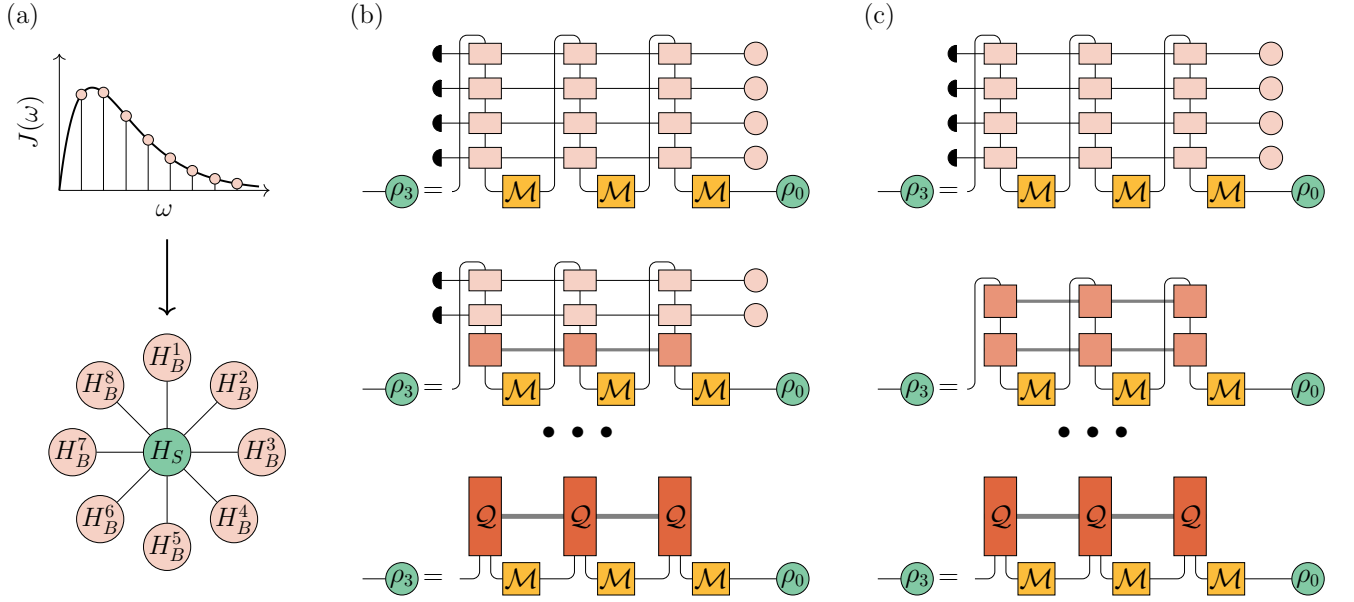}
    }
    \caption{Geometry of the system interacting with an arbitrary environment: (a) ACE decomposes an arbitrary environment into $N_{E}$ distinct non-interacting sub-units of environmental degrees of freedom referred to as modes. This structure, where the system couples to all the modes and the modes only couple to the system, is called the star geometry. For a system coupled to a continuum of modes, ACE requires a discretization of the continuous bath into $N_{E}$ discrete modes that faithfully reproduce the bath correlation function. ACE algorithm: (b-c) represents the two-dimensional tensor network for the ACE algorithm, where propagation in time goes from right to left, with red boxes representing the PT-MPO for each mode at every time step, red circles representing the initial state of the environment, and black caps representing tracing out the environmental degrees of freedom to end propagation. The yellow boxes represent the system propagator, with the green circles representing the states of the system. The evolution of the reduced dynamics is then calculated through tensor contraction of the full 2D tensor network. (b) Demonstrates the sequential combination of PT-MPOs through combining the environment Liouville space's (inner bonds) using \eqref{eq:ACEModeCombination} one by one, where SVD compression techniques are used after each combination to maintain a tractable inner bond until the full PT-MPO has been constructed. (c) Instead of using a sequential combination, a tree-like contraction method can be used where neighboring PT-MPOs are combined and compressed, leading to the combination and compression of smaller tensors compared to the sequential combination, allowing for a numerically more efficient construction of the full PT-MPO.}
    \label{fig:ACE_1}
\end{figure*}

Consider a generic open quantum system Hamiltonian that constitutes a system and its environment with which it interacts, given by \eqref{Eq:partion_hamiltonian}. For generality, we combine all environmental contributions and their coupling to the system into the Hamiltonian $H_{B} = H_E + H_I$, which is assumed to be decomposable into distinct non-interacting sub-units of environmental degrees of freedom, henceforth referred to as modes. That is,
\begin{equation}
    {H_{B} = \sum_{k =1}^{N_{E}}H_{B}^{{k}}}\label{eq:ACE_modes_hamiltonian},
\end{equation}
where the index $k$ indexes the mode.
This can be visualized as a star-like geometry where each mode couples solely to the system, as shown in Fig.~\ref{fig:ACE_1}(a).

Assuming that the initial state of the system is uncorrelated with the environment $\rho(0)=\rho_S(0)\rho_E$, the time-evolved density operator of the system is
\begin{equation}
    \rho_{S}{(t)} = \Tr_{E}[U(t) (\rho_{S}{(0)}  \rho_{E})U^{\dagger}(t)].
\end{equation}
A Trotterization of the propagator is then performed, which yields
\begin{equation}
    U(\Delta t) = e^{\mathcal{L}_{B}\Delta t} e^{\mathcal{L}_{S}\Delta t} + \mathcal{O}(\Delta t^{2}),
\end{equation}
where $\mathcal{L}_{S} = - i H_{S}^{c}$ is the Liouvillian superoperator for the system and $\mathcal{L}_{B} = -i H_{B}^{c}$ is the Liouvillian superoperator of the environment that includes the interaction Hamiltonian, and $H_i^c$ denotes the commutator with the respective Hamiltonian. Note that this formulation can easily be extended to include general time-local Liouville operators, for example, Lindblad operators. By formulating this using symmetric Trotterization, a lower error can be achieved (see \eqref{eq:symmtrotter}). 

By virtue of the assumption of uncoupled modes, see \eqref{eq:ACE_modes_hamiltonian}, the total Liouvillian can be written as a summation of the $N_{E}$ individual modes in Liouville space $\mathcal{L}_{B}  = \sum_{k=1}^{N_{E}}\mathcal{L}^{k}_{B}$.
Therefore, this allows for a further Trotterization of the environment propagator, allowing for distinct non-interacting time propagators for each mode
\begin{equation}
    e^{\mathcal{L}_{B} \Delta t} = e^{\mathcal{L}_{B}^{1}\Delta t}e^{\mathcal{L}_{B}^{2}\Delta t}\ldots e^{\mathcal{L}_{B}^{N_{E}}\Delta t} + \mathcal{O}(\Delta t^{2}).\label{eq:liouvillian_modes_trotter}
\end{equation}

 Inserting a complete basis for both the system and total environment, the reduced dynamics can be reformulated in the path integral form of \eqref{ReducedDensityMatrix}.\\

Thus, it follows directly from \eqref{eq:liouvillian_modes_trotter} that the PT given in \eqref{eq:explicitPT} naturally has a representation in the matrix product operator form 
\begin{align}
    \mathcal{I}^{(\alpha_{N}\alpha_{N}^{\prime})\ldots(\alpha_{1}\alpha_{1}^{\prime})} &=\sum_{d_{N},\ldots,d_{0}}\mathcal{Q}^{\alpha_{N}\alpha_{N}^{\prime}}_{d_{N}d_{N-1}} \mathcal{Q}^{\alpha_{N-1}\alpha_{N-1}^{\prime}}_{d_{N-1}d_{N-2}} \ldots\mathcal{Q}^{\alpha_{1}\alpha_{1}^{\prime}}_{d_{1}d_{0}},
\end{align}
where the system degrees of freedom indices $(\alpha_{n},\alpha_{n}^{\prime})$ are reinterpreted as the physical legs of the TN, whereas environmental degrees of freedom $d_{n}$ (with dimension of the environment Liouville space) correspond to the inner bonds of the TN relevant for describing the effects of the environment on the system's time evolution as shown in Fig.~\ref{fig:ACE_1}(b-c). Practically, this is achieved by identifying the following MPOs
\begin{equation} \label{eq:Q_def}
\mathcal{Q}^{(\alpha_n\alpha_n^{\prime})}_{d_nd_{n-1}} =
\begin{cases}
(e^{\mathcal{L}_B \Delta t}|\rho_{e}(0))_{(\alpha_1, d_1), (\alpha_1^{\prime}, d_{0})}, & n = 1, \\[0.5em]
(e^{\mathcal{L}_B \Delta t})_{(\alpha_n, d_n), (\alpha_n^{\prime}, d_{n-1})}, & 1 < n < N, \\[0.5em]
(\mathrm{Tr}_B \, e^{\mathcal{L}_B \Delta t})_{(\alpha_N, d_N), (\alpha_N^{\prime}, d_{N-1})}, & n = N,
\end{cases}
\end{equation}
with fixed indices at the boundaries ${d_{N} = d_{0} = 1}$. Through this recasting, we compute the dynamics via a series of tensor contractions, which correspond to propagating the system and environment degrees of freedom. However, the inner bonds of the environment can become computationally intractable, since it corresponds to the Liouville dimension space of the total environment, precluding their direct application to calculating system dynamics unless the Liouville space of the environment is small. However, as discussed in Sec.~\ref{sec:TN}, MPOs enable the use of compression techniques. Here, compression physically corresponds to a reduction to the degrees of freedom in the environment that are relevant to the reduced system's dynamics, which can be an exponentially small fraction of the total, allowing for efficient simulation of the system evolution. Thus, the key challenge is to be able to construct these truncated PT-MPOs without first calculating the full uncompressed PT.\\

The intractability of creating the full environment MPO even for a single time step necessitates an approach of consecutive construction, through the combination of modes and compression. This is made possible due to the decomposition of the environment into $N_{E}$ modes, see~\eqref{eq:ACE_modes_hamiltonian},
as shown in Fig.~\ref{fig:ACE_1}(a). This allows the construction of individual PT-MPOs for each environment mode using \eqref{eq:Q_def}, where in this case the uncompressed dimension of the inner bonds corresponds to the Liouville space of the $k$\textsuperscript{th} mode. This can be visualized as the two-dimensional tensor network depicted in Fig.~\ref{fig:ACE_1}(b-c). The entire PT-MPO is then constructed by sequentially combining the $N_E$ modes, typically one at a time as demonstrated in Fig.~\ref{fig:ACE_1}(b). Thus, the total PT-MPO for the modes $1$ to $M$ can be obtained by first combining the PT-MPO for modes $1$ and $2$. Then, the PT-MPO for mode $3$ is combined with the composite PT-MPO for modes $1$ and $2$. This process is iterated until all modes are combined, such that the final step combines the PT-MPO for the modes $1$ to $M-1$ with the PT-MPO for mode $M$. Compression of the PT-MPO is performed at each step. Consider, for example, the first two modes within Fig.~\ref{fig:ACE_1}~(c). The PT-MPO of the first (second) mode is defined with the tensors $\mathcal{Q}^{(\alpha_{n}, \alpha^{\prime \prime}_{n})}_{(e_{n}, e_{n-1})}$ ($\mathcal{P}^{(\alpha_{n}^{\prime \prime}, \alpha_{n}^{\prime})}_{(f_{n}, f_{n-1})}$). Their contraction reads 
\begin{widetext}
\begin{equation}
\mathcal{\tilde{Q}}^{(\alpha_{n} \alpha^{\prime}_{n})}_{(d_{n} d_{n-1})}
= \sum_{\substack{
    \alpha_n^{\prime\prime} \\
    (e_{n} e_{n-1}) \\
    (f_{n} f_{n-1})
}}
\delta_{d_{n} (e_{n}, f_{n})}
\mathcal{Q}^{\alpha_{n} \alpha_{n}^{\prime \prime}}_{e_{n} e_{n-1}}
\mathcal{P}^{\alpha_{n}^{\prime \prime} \alpha_{n}^{\prime}}_{f_{n} f_{n-1}}
\delta_{d_{n-1} (e_{n-1}, f_{n-1})}.
\label{eq:ACEModeCombination}
\end{equation}
\end{widetext}

The Kronecker $\delta$s allow for the definition of the dimension of the inner bonds ($d_{n} = (e_{n}, f_{n})$) of the contracted PT-MPO, which is given by the product of the inner bond dimension of the individual PT-MPOs. To maintain a tractable inner bond dimension after each combination, they are compressed using SVD. For this, a so-called forward sweep is performed, moving from time step $n = 1$ to $n = N$. At each step, the PT-MPO $Q^{\alpha_{n} \alpha^{\prime}_{n}}_{d_{n} d_{n-1}}$ is reshaped into a matrix $A_{d_{n} ((\alpha_{n}, \alpha_{n}^{\prime}), d_{n-1})}$, and a SVD is performed, resulting in
\begin{equation}
    Q^{\alpha_{n} \alpha^{\prime}_{n}}_{d_{n} d_{n-1}} \simeq A_{d_{n} ((\alpha_{n}, \alpha_{n}^{\prime}), d_{n-1})} = \sum_{j_{n}} U_{d_{n} j_{n}} \sigma_{j_{n}}V^{
    \dagger {\alpha_{n} \alpha_{n}^{\prime}}}_{j_{n} d_{n-1}},
\end{equation}
where the index $j_{n}$ is truncated such that only the singular values, satisfying $\sigma_{j} \ge \epsilon \sigma_{0}$, are kept, where $\epsilon$ is the specified truncation threshold, and $\sigma_{0}$ is the largest singular value. Therefore, $U_{d_{n}, j_{n}}$ and $V_{j_{n}, (\alpha_{n}, \alpha_{n}^{\prime}), d_{n-1}}$ are the reduced columns of the left and right unitary matrices. After truncation, we update the PT-MPO as follows
\begin{equation}
    \mathcal{\tilde{Q}}^{\alpha_{n} \alpha_{n}^{\prime}}_{j_{n} d_{n-1}} = V^{\dagger\alpha_{n} \alpha_{n}^{\prime}}_{j_{n} d_{n-1}}.
\end{equation}
The unitary matrix $U$ and the diagonal matrix $\sigma$ are propagated forward and incorporated into the next PT-MPO at timestep $n + 1$ as
\begin{equation}
    Q^{\alpha_{n+1} \alpha^{\prime}_{n+1}}_{d_{n+1} j_{n}} =  \sum_{d_{n}} Q^{\alpha_{n+1} \alpha^{\prime}_{n+1}}_{d_{n+1} d_{n}} U_{d_{n} j_{n}} \sigma_{j_{n}}.
\end{equation}
This forward sweep achieves compression while preserving the relevant contributions of the environment's interactions with the system over time. Subsequently, a backward sweep is performed, moving from $n = N$ to $n  = 2$. This has been found to optimize the compression of the combination of PT-MPO objects, as information about the final time further reduces the relevant information that needs to be stored in the environment.\\

\begin{figure}[htbp]
    \centering
    \includegraphics[width=\linewidth]{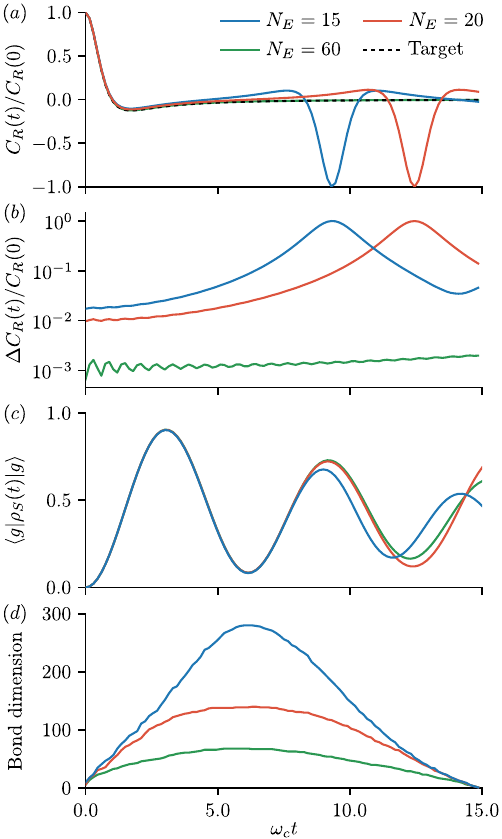}
    \caption{ (a) Convergence of the real part of the discretized BCF $C_R(t,N_E)$, (b) the relative error in the discretized BCF $\Delta C_R(t) = |C_R(t)-C_R(t,N_E)|$ with the continuous BCF and (c) the ground state population dynamics $\bra{g}\rho_S(t)\ket{g}$ of a two-level system coupled to a bosonic environment with an Ohmic spectral density and exponential cutoff, for different numbers of discretized modes, $N_{E}=\{15,20,60\}$ using equidistant sampling, at fixed simulation time $\omega_{c}t=15$. Strong deviations occur when the number of modes is below the heuristic criterion described below \eqref{Eq:ACE_Heuristic}, leading to discretization errors. (d) Growth of the inner bond dimension of the final PT-MPO throughout the simulation. When too few modes are used, the inner bond dimension becomes significantly larger to account for the discretization error in the bath correlation function.} 
    \label{fig:ACE_Convergence}
\end{figure}

The sequential mode combination in the standard ACE approach is often suboptimal in practice. This is because the sequential addition of modes results in PT-MPOs with large bond dimensions (which capture degrees of freedom of multiple modes) being combined with PT-MPOs whose bond dimensions are the size of the Liouville space for the $k$\textsuperscript{th} environment. This is problematic because rank-reducing algorithms, such as SVD, scale with the third power of the bond dimension. By contrast, if the PT-MPOs are organized such that neighboring modes have similar influences on the system, a tree-like contraction scheme can be employed \cite{mortiz_tree_like_2024}. Here, neighboring modes are combined and compressed, and the resulting PT-MPOs are then combined with adjacent modes of comparable influence and size, as illustrated in Fig.~\ref{fig:ACE_1}(c). Through this approach, the compression step can maintain relatively small bond dimensions at each stage, significantly improving efficiency.\\ 
\subsubsection{Discretization of a continuous environment}
Because ACE naturally requires a system interacting with a discrete set of modes, simulating a system coupled to a continuum of modes requires some form of discretization. As depicted in Fig.~\ref{fig:ACE_1}~(a), discretization amounts to replacing the continuous distribution of bath modes with a finite set of modes. For a bosonic environment, this can be done by sampling the environment's spectral density. A standard approach would be an equidistant sampling via
\begin{align}
    g_q = \sqrt{J(\omega_q)\Delta\omega},
    \label{Eq:ACE_Heuristic}
\end{align}
with the frequency discretization $\Delta\omega = \omega_\textrm{max}/N_E$. The necessary number of modes, $N_E$, typically depends on the simulation time and has to be chosen such that it satisfies a heuristic estimate given by ${N_E \geq 0.4 (\omega_\text{max} - \omega_\text{min}) T}$, where $T$ is the simulation time and $\omega_\text{max(min)}$ are the maximum (minimum) frequencies that are sampled in the spectral density, to avoid aliasing effects~\cite{moritz_ACE_coding}. It should be noted that equidistant sampling has an error that scales as $\mathcal{O}(1/N_{E}^{2})$ whereas other discretisation methods can lead to more optimal representations with fewer modes ~\cite{de_vega_how_2015}. This issue of discretizing the environment is also present in the ML-MCTDH method (see Sec.~\ref{sec:MLMCTDH}).\\ 

We illustrate this by showing the convergence of the bath correlation function and of the system's dynamics in a spin--boson model as the number of discretized modes changes. Thus, we consider a two-level system that interacts with a bosonic environment with the following Hamiltonian \\
\begin{align}
    H = &\frac{\omega_0}{2}\sigma^{z} + \frac{\Delta}{2} \sigma^{x} +\int_0^\infty d\omega \omega \awd\aw\nonumber\\
    &+ \sigma^{+}\sigma^{-}\int_0^\infty d\omega \sqrt{J(\omega)}(\aw + \awd)\ ,
\end{align}
with $\omega_{0} = 0$, $\Delta = \omega_{c}$. The spectral density is considered as an Ohmic bath with an exponential cutoff $J(\omega) = 2 \alpha \omega \exp(-\omega /\omega_{c})$ with $\omega_{c}$ the bath cutoff frequency and $\alpha = 0.1$, and we initialize the two-level system with an excitation in the excited state. Using the heuristic estimate, the minimum number of modes required to accurately discretize the bath correlation function for a propagation time of $\omega_{c}t = 15$ is $N_{E} = 60$, which produces the correct ground state system population dynamics as shown in Fig.~\ref{fig:ACE_Convergence}~(a-c). Instead, using $N_{E} = 15$ modes inadequately represents the environment at later propagation times and leads to discretization errors in the bath correlation function, as highlighted in Fig.~\ref{fig:ACE_Convergence}~(a-b). These discretization errors introduce additional features in the bath correlation function that deviate from the analytical solution, resulting in additional spurious non-Markovian environmental effects on the system's dynamics. 

Maybe counterintuitively at first, undersampling also results in the PT-MPO inner bonds increasing in size to account for the unphysical non-Markovianity of the environment on the system's dynamics, as seen in Fig.\ref{fig:ACE_Convergence}~(d). Thus, undersampling not only produces inaccurate reduced system dynamics, but the full PT-MPO also becomes computationally more expensive to construct due to combining and compressing tensors with large inner bond dimensions.

\subsubsection{Applications and implementations} 
Owing to its generality, the ACE approach can be applied to a wide range of open quantum system problems. For example, the original article~\cite{mortiz_2022_ACE} proved the algorithm's efficiency for simulating a central spin coupled to an environment of a thousand spins by effectively combining and compressing modes. Furthermore, ACE has been shown to handle anharmonic environments, which are notoriously challenging to solve numerically~\cite{anharmonic}, and has been applied to calculating the dynamics of fermionic environments~\cite{mortiz_tree_like_2024}.  Despite its broad applicability, the efficiency of ACE depends strongly on the structure of the environment under consideration. ACE is particularly advantageous for environments that can be represented by a moderate number of discrete modes, such as spin or fermionic environments. For continuous environments, its performance depends critically on the chosen discretization scheme, as inadequate sampling may cause the generated PT to miss important features of the spectral density, leading to inaccurate reduced-system dynamics. For harmonic oscillator environments, the local Hilbert-space truncation of each mode introduces an additional convergence parameter. Achieving convergence can become particularly challenging at higher temperatures or under strong system--environment coupling, where higher Fock states acquire significant occupation.

A practical advantage of ACE, and PT methods more generally is that once the PT has been constructed, it can be reused for different system Hamiltonians and initial conditions without recomputing the environmental dynamics. This substantially reduces the computational cost of parameter sweeps, optimization tasks, and the evaluation of multi-time correlation functions. The method is most efficient when environmental correlations are sufficiently compressible, yielding PT-MPOs with moderate bond dimensions. In contrast, highly structured environments often exhibit long memory times, resulting in larger bond dimensions and increased computational cost.

In this context, the PT-MPO bond dimension provides a quantitative measure of environmental complexity. Relatively structureless environments can be represented with small bond dimensions, whereas highly structured environments require larger bond dimensions to accurately capture their influence on the system. Furthermore, ACE has revealed a physical interpretation of these bond dimensions: they correspond to the effective subspace of the environment's Liouville space containing those environmental excitations that can influence the system dynamics. This connection can be made through a lossy transformation whose pseudoinverses enable the extraction of environmental observables ~\cite{cygorek_inner_bonds}.  As ACE utilizes the PT paradigm and efficiently combines and compresses PTs for generic environmental modes, it is straightforward to use it to calculate the influence of multiple environments interacting with a system. For example, ACE has been utilized for calculating the dynamics of several quantum dots coupled to both a common photon and individual phonon environments \cite{wiercinski_role_2024}.

The ACE algorithm has the following convergence parameters:
\begin{itemize}
  \item the timestep $\Delta t$,
  \item the singular value cutoff threshold $\epsilon$,
  \item the number of modes $N_{E}$ when considering continuous environments,
  \item the local Hilbert space of the modes when considering bosonic environments.
\end{itemize}

Note that these convergence parameters are generally interdependent. For example, the necessary singular value cutoff threshold for convergence depends non-trivially on the timestep, potentially leading to degraded convergence for a given timestep. This is because the information stored in the $\mathcal{L}_{B}^{N_{E}}$ for each mode is scaled by the timestep $e^{\mathcal{L}_{B}^{N_{E}}\Delta t} \approx \mathbbm{1} + \mathcal{L}_{B}^{N_{E}}\Delta t + ...$ which for small timesteps can lead to values much smaller than the unity values of the identity matrix. Therefore, smaller compression thresholds can be required to accurately account for the effects of each mode of the environment for a given timestep. \\

The ACE method is implemented in the open-source C++ ACE package~\cite{moritz_ACE_coding}, which provides the framework to build and propagate the PT-MPOs for any given system--environment coupling. The package also allows for the creation of PT-MPOs using different TEMPO-based methods (see Sec.~\ref{sec:TEMPO}). PTs for different environments produced from these methods can then be efficiently combined and compressed using ACE.

\subsection{DAMPF}\label{sec:DAMPF}
\subsubsection{Overview}

The \emph{dissipation-assisted matrix product factorization} (DAMPF) method~\cite{SomozaPRL2019} is designed to simulate a finite set of system degrees of freedom coupled to a discrete set of harmonic oscillators undergoing Lindblad relaxation, known as \emph{pseudomodes}.  Owing to a rigorous mathematical equivalence~\cite{TamascelliPRL2018, SmirneOSID2022} and analytical error bounds~\cite{MascherpaPRL2017, trivedi_convergence_2021}, environments consisting of a finite set of pseudomodes can be used effectively to reproduce the influence of continuous bosonic baths on system dynamics with arbitrarily high accuracy, particularly effective in tackling finite-time simulations of environments with a high degree of structure. In general, DAMPF supports the presence of groups of system degrees of freedom, each interacting with distinct local environments, and can even model non-linear system--environment couplings~\cite{ZhangPRB2025}.

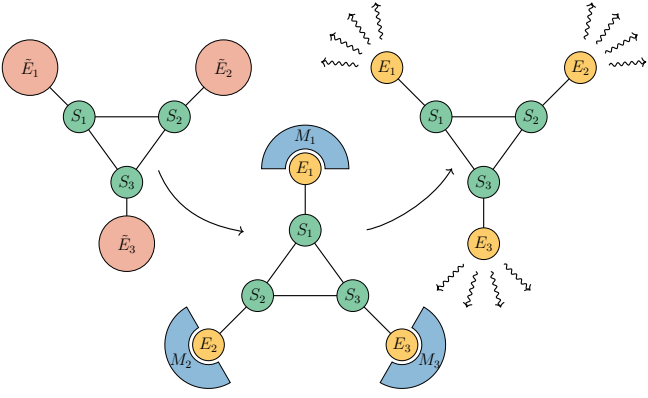
\begin{figure}[t]
    \centering
    \resizebox{\linewidth}{!}{%
        \begin{tikzpicture}[
    every node/.style={font=\large},
    sys/.style={circle,draw=black,fill=paletteGreen,minimum size=8mm,inner sep=0pt},
    env/.style={circle,draw=black,fill=paletteYellow_Light1,minimum size=8mm,inner sep=0pt},
    tenv/.style={circle,draw=black,fill=paletteRed_Light2,minimum size=14mm,inner sep=0pt},
    bond/.style={draw=black,thick},
    memfill/.style={draw=black,fill=paletteBlue_Light2},
    flow/.style={draw=black,->,thick},
    wig/.style={
        draw=black,
        ->,
        thick,
        decorate,
        decoration={snake,amplitude=1.2pt,segment length=5pt}
    }
]

\def\triW{2.4}
\def\triY{0.9}
\def\triB{-0.75}
\def\rSE{1.55}
\pgfmathsetmacro{\dxSE}{0.8*\rSE}
\pgfmathsetmacro{\dySE}{0.8*\rSE}
\pgfmathsetmacro{\dySElow}{1.0*\rSE}

\def\Rin{0.5}
\def\Rout{1.1}
\pgfmathsetmacro{\Rmid}{0.5*(\Rin+\Rout)}

\newcommand{\HalfAnnulusText}[3]{%
    \begin{scope}[shift={(#1)},rotate=#2]
        \path[memfill,even odd rule]
            (-\Rout,0)
            arc[start angle=180,end angle=0,radius=\Rout]
            --
            (\Rin,0)
            arc[start angle=0,end angle=180,radius=\Rin]
            -- cycle;
        \node at (0,\Rmid) {#3};
    \end{scope}
}

\begin{scope}[shift={(0,0)}]
    \coordinate (S1L) at (0,\triY);
    \coordinate (S2L) at (\triW,\triY);
    \coordinate (S3L) at (0.5*\triW,\triB);

    \coordinate (E1L) at ($(S1L)+(-\dxSE,\dySE)$);
    \coordinate (E2L) at ($(S2L)+(\dxSE,\dySE)$);
    \coordinate (E3L) at ($(S3L)+(0,-\dySElow)$);

    \draw[bond] (S1L) -- (S2L);
    \draw[bond] (S1L) -- (S3L);
    \draw[bond] (S2L) -- (S3L);

    \draw[bond] (S1L) -- (E1L);
    \draw[bond] (S2L) -- (E2L);
    \draw[bond] (S3L) -- (E3L);

    \node[sys] at (S1L) {$S_1$};
    \node[sys] at (S2L) {$S_2$};
    \node[sys] at (S3L) {$S_3$};

    \node[tenv] at (E1L) {$\tilde{E}_1$};
    \node[tenv] at (E2L) {$\tilde{E}_2$};
    \node[tenv] at (E3L) {$\tilde{E}_3$};
\end{scope}

\draw[flow] 
    (2.0,-0.45) 
    .. controls (2.5,-1.7) and (3.65,-1.9) .. 
    (4.15,-2.00);

\begin{scope}[shift={(4.5,-2.55)},rotate around={180:(1.2,-0.08)}]
    \coordinate (S1M) at (0,\triY);
    \coordinate (S2M) at (\triW,\triY);
    \coordinate (S3M) at (0.5*\triW,\triB);

    \coordinate (E1M) at ($(S1M)+(-\dxSE,\dySE)$);
    \coordinate (E2M) at ($(S2M)+(\dxSE,\dySE)$);
    \coordinate (E3M) at ($(S3M)+(0,-\rSE)$);

    \draw[bond] (S1M) -- (S2M);
    \draw[bond] (S1M) -- (S3M);
    \draw[bond] (S2M) -- (S3M);

    \draw[bond] (S1M) -- (E1M);
    \draw[bond] (S2M) -- (E2M);
    \draw[bond] (S3M) -- (E3M);

    \HalfAnnulusText{E1M}{60}{$M_3$}
    \HalfAnnulusText{E2M}{-60}{$M_2$}
    \HalfAnnulusText{E3M}{180}{$M_1$}

    \node[sys] at (S1M) {$S_3$};
    \node[sys] at (S2M) {$S_2$};
    \node[sys] at (S3M) {$S_1$};

    \node[env] at (E1M) {$E_3$};
    \node[env] at (E2M) {$E_2$};
    \node[env] at (E3M) {$E_1$};
\end{scope}

\draw[flow]
    (7.25,-1.95)
    .. controls (7.8,-1.85) and (8.90,-1.25) ..
    (9.40,-0.40);

\begin{scope}[shift={(9.0,0)}]
    \coordinate (S1R) at (0,\triY);
    \coordinate (S2R) at (\triW,\triY);
    \coordinate (S3R) at (0.5*\triW,\triB);

    \coordinate (E1R) at ($(S1R)+(-\dxSE,\dySE)$);
    \coordinate (E2R) at ($(S2R)+(\dxSE,\dySE)$);
    \coordinate (E3R) at ($(S3R)+(0,-\rSE)$);

    \draw[bond] (S1R) -- (S2R);
    \draw[bond] (S1R) -- (S3R);
    \draw[bond] (S2R) -- (S3R);

    \draw[bond] (S1R) -- (E1R);
    \draw[bond] (S2R) -- (E2R);
    \draw[bond] (S3R) -- (E3R);

    \node[sys] at (S1R) {$S_1$};
    \node[sys] at (S2R) {$S_2$};
    \node[sys] at (S3R) {$S_3$};

    \node[env] at (E1R) {$E_1$};
    \node[env] at (E2R) {$E_2$};
    \node[env] at (E3R) {$E_3$};

    \draw[wig] ($(E1R)+(150:0.72)$) -- ++(150:0.95);
    \draw[wig] ($(E1R)+(125:0.72)$) -- ++(125:0.95);
    \draw[wig] ($(E1R)+(100:0.72)$) -- ++(100:0.95);
    \draw[wig] ($(E1R)+(175:0.72)$) -- ++(175:0.95);

    \draw[wig] ($(E2R)+(30:0.72)$)  -- ++(30:0.95);
    \draw[wig] ($(E2R)+(55:0.72)$)  -- ++(55:0.95);
    \draw[wig] ($(E2R)+(80:0.72)$)  -- ++(80:0.95);
    \draw[wig] ($(E2R)+(5:0.72)$)   -- ++(5:0.95);

    \draw[wig] ($(E3R)+(225:0.72)$) -- ++(225:0.95);
    \draw[wig] ($(E3R)+(255:0.72)$) -- ++(255:0.95);
    \draw[wig] ($(E3R)+(285:0.72)$) -- ++(285:0.95);
    \draw[wig] ($(E3R)+(315:0.72)$) -- ++(315:0.95);
\end{scope}

\end{tikzpicture}
    }
    \caption{Conceptual representation of the pseudomode environment approach, in which the actual environment is partitioned into a non-Markovian core, $E$, and a Markovian remainder, $M$, the latter being successively traced out exactly.}
    \label{Fig_DAMPF_1}
\end{figure}

In practice, as shown in Fig.~\ref{Fig_DAMPF_1}, the pseudomode representation partitions each environment into two components: a non-Markovian core that retains memory effects, and a Markovian reservoir responsible for its relaxation. By extending the system to include the non-Markovian core, the residual Markovian reservoir can be traced out exactly, yielding a Lindblad description for the pseudomode dynamics. This strategy can significantly reduce the dimensionality of the relevant environmental degrees of freedom, while still fully capturing non-Markovian effects and hence allowing for numerically exact simulations. A one-to-one correspondence between the pseudomode approach and HEOM has recently been established~\cite{muller2026onetoonecorrespondencehierarchicalequations}.

The equivalence between pseudomode models and continuous bosonic environments has been formally established both for environments represented by uncoupled pseudomodes~\cite{TamascelliPRL2018}, and in the presence of linear couplings between pseudomodes~\cite{MascherpaPhysRevA2020}. The specific choice of pseudomode model depends on the degree of structure in the spectral density (see Sec.~\ref{sec:pseudomodes}).

At the Hamiltonian level, the uncoupled pseudomode model adopts the same system--environment coupling structure as that in Eqs.~(\ref{eq:Hbath})-(\ref{eq:HSE}), with the continuous bath replaced by a finite number of discrete environmental modes:
\begin{equation}
\label{eq_1_DAMPF}
H = H_S + \sum_{n=1}^N\sum_{q=1}^{Q(n)} \omega_{nq} a_{nq}^\dagger a_{nq} + \sum_{n=1}^N O^S_{n} \sum_{q=1}^{Q(n)} g_{nq}(a_{nq}^\dagger + a_{nq}).
\end{equation}
Here, the index $n$ labels different system sites, $nq$ indexes the $q$th pseudomode coupled to site $n$, and $Q(n)$ is the number of pseudomodes coupled to site $n$. To account for environmental dissipation, each pseudomode is subject to a Lindblad dissipator of the form~\eqref{eq:LindbladMasterEquation}:
\begin{equation}
\begin{aligned}
\label{eq_dampf_diss}
\mathcal{D}[\rho] = \sum_{n=1}^N\sum_{q=1}^{Q(n)} &\gamma_{nq}(1 + \bar{n}_{nq}) \left( a_{nq} \rho a_{nq}^\dagger - \frac{1}{2} \{ a_{nq}^\dagger a_{nq}, \rho \} \right) \\
&+ \gamma_{nq} \bar{n}_{nq} \left( a_{nq}^\dagger \rho a_{nq} - \frac{1}{2} \{ a_{nq} a_{nq}^\dagger, \rho \} \right),
\end{aligned}
\end{equation}
with $\bar{n}_{nq}\equiv n_{nq}(\omega_{nq})=(\exp(\beta_{nq}\omega_{nq})-1)^{-1}$ being the thermal population of the pseudomode $nq$ (see Sec.~\ref{sec:pseudomodes}). When combined with tensor network techniques, and in particular the matrix product state representation, DAMPF enables the efficient simulation of quantum dynamics involving tens or even hundreds of pseudomodes, as the Lindbladian relaxation actively counters the growth of correlations and hence the bond dimension (see Sec.~\ref{sec:dampf_mps}).

Importantly, when tackling continuous bosonic environments, the pseudomode environment yields an effective representation of the underlying environmental degrees of freedom; consequently, DAMPF can provide direct access only to the system degrees of freedom.

\subsubsection{Constructing effective pseudomode environments}
\label{sec:pseudomodes}

The construction of effective pseudomode environments relies on the fact that, when a system couples to a harmonic bath or to a pseudomode environment, initially prepared in a Gaussian state, such as a thermal state, the influence of the environment on the system dynamics is fully determined by the bath correlation functions $C_n(t)$ (see Section \ref{sec:OQS}). Consequently, an effective environment can induce the same system dynamics as the actual environment, provided that their BCFs match~\cite{TamascelliPRL2018}. 

For the sake of clarity, in the following, we restrict our attention to the case of a single environment coupled to a single system transition, so that only one bath correlation function, $C(t)$, is required. For a continuous bosonic environment at inverse temperature $\beta=(k_BT)^{-1}$, characterized by a spectral density $J(\omega)$, the bath correlation function is given by Eq.~(\ref{eq:Environment_Correlation_Function}). 

In contrast, an environment described in terms of pseudomodes gives rise to a bath correlation function $C_{\rm ps}(t)$ of the form~\cite{Lemmer_2018}
\begin{equation}
C_{\rm ps}(t) = \sum_{q=1}^Q\sum_{\pm}\frac{g_q^2}{2}\Big[\Big(\coth\Big(\frac{\beta_q\omega_q}{2}\Big) \pm 1\Big)e^{\mp i\omega_qt-\frac{\gamma_q}{2}|t|}\Big].
\label{pseudomode_eq1}
\end{equation}
Its frequency-domain expression is
\begin{equation}
C_{\rm ps}(\omega) = \sum_{q=1}^Q\sum_{\pm}\frac{g_q^2}{4\pi}\frac{\big[\coth\big(\beta_q\omega_q/2\big) \pm 1\big]\gamma_q}{(\gamma_q/2)^2+(\omega\mp\omega_q)^2},
\label{pseudomode_eq2}
\end{equation}
from which it is clear that, at finite temperature, each pseudomode contributes two thermally weighted Lorentzian peaks, centered at frequencies $\pm\omega_q$ and with full width at half maximum $\gamma_q$.
On a practical level, the equivalence between an effective pseudomode environment and the target continuous bosonic bath is established by fitting the pseudomode parameters, namely frequency $\omega_q$, coupling strength $g_q$, and dissipation rate $\gamma_q$, to achieve a matching between the BCFs, i.e., $C_{\rm ps}(t)\simeq C(t)$. For the moment, we consider the pseudomode temperature as fixed and equal to that of the continuous environment. The accuracy of this mapping can be arbitrarily improved by increasing the number $Q$ of pseudomodes employed, and is discussed in more detail below.

Although the number of fitting parameters may be large, the pseudomode bath correlation function consists of decaying complex exponentials (see~\eqref{pseudomode_eq1}), which allows for an efficient parameter estimation using numerical techniques based on the Prony algorithm~\cite{Potts01032011}. In particular, the Prony method is capable of fitting non-increasing oscillatory functions as a sum of complex exponentials with complex amplitudes, even when a large number of terms is involved, by reformulating the fitting problem into a significantly less computationally demanding eigenvalue problem. 

This Prony expansion closely resembles the structure of the pseudomode bath correlation function, with the main difference that for pseudomodes the amplitudes are real, and pairs of frequencies $\pm \omega_q$ with relative amplitudes $(1 + n(\omega))/n(\omega)$ must be identified. By imposing such constraints on the Prony method, this can directly yield valid pseudomode parameters or provide initial guesses amenable to subsequent optimization procedures, with the latter approach typically limiting the practical number of pseudomodes to several tens. In this way, DAMPF simulations were recently performed with environments comprising tens of pseudomodes~\cite{lorenzoni_full_2025}.

\subsubsection{Reducing computational costs}

Techniques that are capable of generating pseudomode environments at finite temperature, both of smaller size and with lower associated computational costs, have been developed.

The computational costs can be significantly affected by changes in the environmental temperature. In particular, when thermally populated pseudomodes are present, an increase in temperature can lead to a rapid growth in the number of relevant excited states to be included, thus requiring a higher cutoff in their local dimension. Conversely, when the temperature is decreased, this cutoff can be reduced, but correlations between pseudomodes may become stronger due to the reduced Lindblad relaxation rates. This also hinders DAMPF as it relies on a tensor network implementation of the pseudomode formalism, discussed below; hence, the computational costs increase due to such correlations.

As a result, the most computationally favorable regime for DAMPF simulations is often found at intermediate temperatures. In the T-TEDOPA approach~\cite{tamascelli_efficient_2019}, the temperature dependence is absorbed into the spectral density $J(\omega)$, yielding a so-called \emph{thermalized spectral density} $J_\beta(\omega)\equiv C(\omega)$ associated with a zero-temperature environment (see Sec.~\ref{sec:T-TEDOPA}). On a practical level, by writing
\begin{equation}
J_\beta(\omega)=[1+n_\beta(\omega)]J(\omega)H(\omega)+n_\beta(|\omega|)J(|\omega|)H(-\omega),
\end{equation}
where $H$ is the Heaviside step function, one can readily see that using the thermalized spectral density rather than the physical spectral density has two main consequences.

First, a negative frequency sector appears, determined by the physical spectral density $J(|\omega|)$ rescaled by the thermal occupation $n_\beta(|\omega|)$. Second, the positive frequency sector is rescaled by the thermal factor $[1+n_\beta(\omega)]$. Both sectors have an amplitude increasing with respect to the temperature $T$ of the physical environment.

Inspired by this construction, DAMPF implements an intermediate approach in which the temperatures $\{{T_q}\}_{q=1,\dots,Q}$ of the thermally populated pseudomodes can be independently reduced to intermediate values $0\leq T_q\leq T$. In practice, lowering the pseudomode temperatures requires compensating for the difference between the reduced temperatures $\{{T_q}\}_{q=1,\dots,Q}$ and the physical temperature $T$. This is achieved by increasing the corresponding couplings $\{{g_q}\}_{q=1,\dots,Q}$, to preserve the positive frequency sector $[1+n_\beta(\omega)]J(\omega)$, and by introducing zero-temperature pseudomodes at negative frequencies, to accurately reproduce the negative frequency sector $n_\beta(|\omega|)J(|\omega|)$.

This approach was employed by~\textcite{lorenzoni_full_2025}, where excitation energy transfer (EET) in the Fenna--Matthews--Olson (FMO) photosynthetic complex from green sulfur bacteria, a prototypical model system for studying quantum effects in biology, was simulated at $77\,\mathrm{K}$ and $300\,\mathrm{K}$. The $77\,\mathrm{K}$ spectral density was represented using 31 pseudomodes with temperatures $T_q \lesssim 100\,\mathrm{K}$, while the $300\,\mathrm{K}$ case was modeled by reducing the temperature of the three lowest-frequency $\omega_{q}$, and thus most thermally populated, pseudomodes to $T_q < 100\,\mathrm{K}$, similar to the $77\,\mathrm{K}$ case, with only a mild increase in their coupling strength. This adjustment required the inclusion of two additional negative-frequency, zero-temperature pseudomodes, for a total of 33 pseudomodes. As a result, simulations at $300\,\mathrm{K}$ could be performed with only marginally higher computational costs than those at $77\,\mathrm{K}$.

Concerning the generation of pseudomode environments of smaller size, this can be achieved when simulating a timescale $\tau$ shorter than the memory timescale of the environment, which is encoded in the decay timescale of the bath correlation function $C(t)$. This situation commonly occurs in the presence of highly structured environments, such as those describing the vibrational dynamics of molecular systems or the electromagnetic response of structured photonic configurations~\cite{LorenzoniPRL2024,MedinaPRL2021}. In such cases, the effective pseudomode environment only needs to reproduce the target bath correlation function within the simulated time window, i.e., $C(t \leq \tau)$. 

\begin{figure}[ht]
    \centering
    \includegraphics[width=\linewidth]{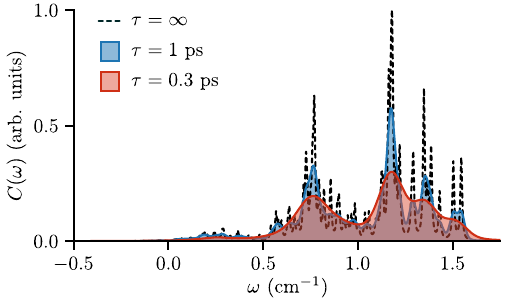}
    \caption{Systematic coarse-graining of the FMO experimentally estimated vibrational environment when simulation time is reduced from $\tau=\infty$ (black) to $\tau=1\, {\rm ps}$ (blue) and $\tau=300\, {\rm fs}$ (red).}
    \label{Fig_DAMPF_3}
\end{figure}

This approach allows one to discard any environmental information associated with longer-time dynamics, effectively introducing a systematic broadening in the spectral domain corresponding to the finite simulation time. Such broadening can significantly reduce the number of pseudomodes required~\cite{LorenzoniPRL2024}. 
As demonstrated by~\textcite{LorenzoniPRL2024,lorenzoni_full_2025}, Fig.~\ref{Fig_DAMPF_3} shows how this coarse-graining procedure simplifies the Fourier transform of the bath correlation function for the highly structured, experimentally estimated spectral density of the Fenna--Matthews--Olson (FMO) complex at 77~K~\cite{RatsepJL2007}. The two cases shown correspond to $\tau=1\,{\rm ps}$ and $\tau=300\,{\rm fs}$, which are typical timescales for simulating, respectively, EET and linear spectra. This spectral broadening reduces the number of distinct environmental features, thereby enabling an accurate fit with a smaller number of pseudomodes.

\subsubsection{Accuracy of pseudomode representation}
When pseudomode environments are used to effectively model a continuous environment, a fitting of the pseudomode bath correlation function to a target one is required. Here, we comment on the impact of the fitting error on system dynamics, and thus on the reliability of the pseudomode representation in the context of numerically exact methods.

On the one hand, the deviation between the pseudomode and target spectral densities, defined as $\delta J(\omega) \equiv |J(\omega) - J_{\rm ps}(\omega)|$, can be used to establish an upper bound on the deviation of the system dynamics generated by the effective environment relative to the target one~\cite{MascherpaPRL2017}. On the other hand, in practice, and particularly in the presence of oscillating bath correlation functions typically generated by highly structured environments, the deviation of the bath correlation function, $\epsilon(t) \equiv C_{\rm ps}(t) - C(t)$, can oscillate rapidly around zero. As a result, the actual error is often significantly smaller than the analytical bound~\cite{MascherpaPRL2017}, which allows the usage of far fewer pseudomodes but requires convergence to be assessed numerically.

Naively, such convergence could be tested by repeating simulations using pseudomode environments of increasing accuracy; this approach can reduce the efficiency of the pseudomode approach. Ideally, one would instead identify a set of pseudomode parameters that are expected to yield high accuracy from the outset, thereby reducing the computational overhead associated with accuracy checks. 

This can be done through preliminary tests based either on easily accessible figures of merit or on simple benchmarks against other numerically exact methods. An example of the former was considered by \textcite{lorenzoni_full_2025}, where the coherence dynamics of an independent boson model, in which only one level interacts with the harmonic environment, were analyzed. With respect to the Hamiltonian in Eq.~(\ref{eq_1_DAMPF}), this corresponds to $H_S=\sum_{n=1}^2 \epsilon_n |n\rangle\langle n|$, with $O_1^S=0$ and $O_2^S=|2\rangle\langle 2|$, so that only the level $|2\rangle$ is coupled to the environment. A key advantage of this setting is that the coherence $\langle 2|\rho_s(t)|1\rangle$ can be calculated analytically for an arbitrary spectral density~\cite{ReinaPRA2002,RengerJPC2002, breuer_theory_2009}. Since its dynamics are determined entirely by the system--environment interaction, in the absence of internal system dynamics, this quantity is particularly sensitive to deviations in the environmental spectral density. Moreover, the same coherence can be computed efficiently with DAMPF even in the presence of hundreds of pseudomodes because, as discussed in the next section, the method can be restricted to the propagation of coherences only. It therefore provides an efficient and reliable first test of the quality of the pseudomode fit.

\begin{figure}[ht]
    \centering
    \includegraphics[width=\linewidth]{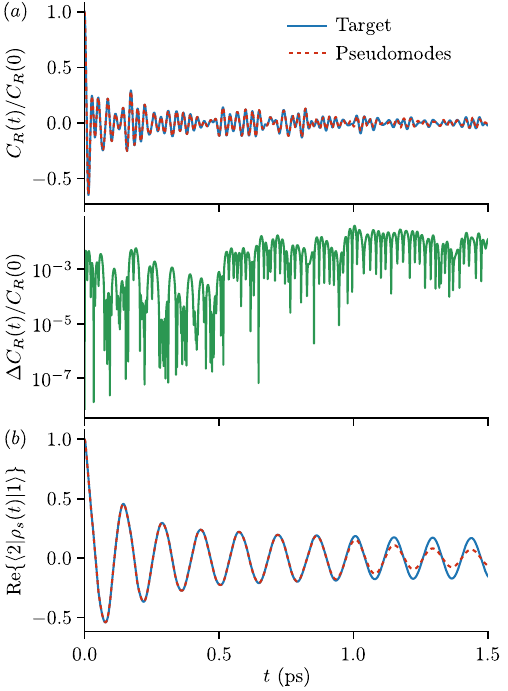}
    \caption{(a) Comparison between the bath correlation function derived from the experimentally estimated spectral density of the FMO complex at $77\,K$ and that obtained from an effective pseudomode environment constructed to match it up to $\tau = 1\,\mathrm{ps}$. 
    (b) Comparison of the real part of the coherence $\langle 2|\rho_S(t)|1\rangle$ of a 2-level system where only the second level is coupled to the environment ($O_1^S=0$, $O_2^S=|2\rangle\langle 2|$) using the experimentally estimated spectral density of the FMO complex and the corresponding effective pseudomode environment.}
    \label{Fig_DAMPF_4}
\end{figure}

As an illustrative example from~\textcite{lorenzoni_full_2025}, which considers the spectral density of the FMO complex at $77\,\mathrm{K}$, Fig.~\ref{Fig_DAMPF_4}(a) compares the target bath correlation function $C(t)$ over a simulation timescale $\tau=1\,\mathrm{ps}$ with its pseudomode counterpart $C_{\rm ps}(t)$ obtained using 31 pseudomodes. Figure~\ref{Fig_DAMPF_4}(b) then compares the corresponding coherences of the independent spin--boson model, showing good agreement over the same timescale. The supplementary material of Ref.~\cite{lorenzoni_full_2025} further shows that a comparable level of accuracy is retained also for dimeric systems, as confirmed by comparison with T-TEDOPA simulations. Nevertheless, the validity of these benchmarks must still be assessed for different system-parameter regimes.

\subsubsection{Structured and broad environments}
Since the Fourier transform of the bath correlation function associated with a single pseudomode is a Lorentzian, effective environments composed of uncoupled pseudomodes are particularly well suited for highly structured spectral densities, since accurate fits can be obtained using only a limited number of Lorentzian contributions. This is the case, for instance, in molecular systems, where highly structured spectral densities can often be captured with only a modest number of pseudomodes, depending on the target figure of merit~\cite{LorenzoniPRL2024, lorenzoni_full_2025}. 

By contrast, in the case of broader, less structured spectral densities, a larger number of Lorentzian peaks may be required for an accurate decomposition. To address this, alternative ans{\"a}tze employing coupled pseudomodes~\cite{MascherpaPhysRevA2020} have been investigated within DAMPF, showing that they can reduce the number of pseudomodes required. In current implementations~\cite{SomozaCommunPhys2023}, however, the use of coupled pseudomodes poses substantial computational challenges. These arise from the increased costs of time-dependent simulations associated with tensor-network operations in the presence of interaction terms between pseudomodes, and the non-trivial inversion problem associated with the characterization of pseudomode parameters through the fitting to a target bath correlation function ~\cite{MascherpaPRL2017}. However, in light of recent advances in the characterization of coupled-pseudomode ans{\"a}tze~\cite{huang_coupled_2026,park_quasi_2024,MedinaPRL2021}, together with alternative pseudomode models (see Sec.~\ref{sec:extended-sys}), further algorithmic developments may lead to a computational advantage with respect to the pseudomode ansatz considered here.

\subsubsection{Tensor network representation}
\label{sec:dampf_mps}

Since the number of pseudomodes required in DAMPF simulations typically ranges from tens to hundreds, when considering multiple environments, the pseudomode approach alone does not allow for feasible non-perturbative simulations of the extended system due to the exponential growth of the Hilbert space. To overcome this limitation, the pseudomode framework is combined with a matrix product state formalism, which makes many-pseudomode scenarios tractable. Due to the presence of Lindblad relaxation, the extended system consisting of the system and pseudomodes has to be treated within the density operator formalism. 

For a system of $N$ sites, within the single system excitation regime, the density matrix of the extended system (system + pseudomodes) is first expanded as
\begin{equation}
\label{eq_state_DAMPF}
    \rho(t)=\sum_{n,m=1}^N|n\rangle\langle m|\rho^{(n,m)}(t),
\end{equation}
where the $N^2$ density matrix constituents $\{\rho^{(n,m)}(t)\}_{n,m}$ describe the pseudomode density matrices conditional on every possible system configuration.

\begin{figure}[t]
    \centering
    \resizebox{\linewidth}{!}{%
        \begin{tikzpicture}[
    scale=0.65,
    every node/.style={font=\large},
    sys/.style={
        circle,
        draw=black,
        fill=paletteGreen,
        minimum size=6mm,
        inner sep=0pt
    },
    tensor/.style={
        circle,
        draw=black,
        fill=paletteBlue_Light2,
        minimum size=3.5mm,
        inner sep=0pt
    },
    bond/.style={draw=black,thick},
    tleg/.style={draw=black,thick},
    arr/.style={draw=black,->,thick},
    wave/.style={
        draw=black,
        ->,
        thick,
        decorate,
        decoration={snake,amplitude=1.2pt,segment length=5pt}
    }
]


\def\triW{1.65}
\def\triY{0.45}
\def\triB{-0.8}

\def\rOsc{1.95}

\def\rightX{6.3}
\def\legUp{0.90}

\def\yOne{2.5}
\def\yTwo{-0.5}
\def\yThree{-3.5}


\newcommand{\Oscillator}[1]{%
    \begin{scope}[shift={(#1)}]
        \path[draw=black,fill=paletteYellow_Light1,thick]
            (-0.36,0.26)
            --
            (0.36,0.26)
            --
            (0.26,-0.22)
            .. controls (0.14,-0.55) and (-0.14,-0.55) ..
            (-0.26,-0.22)
            -- cycle;

        \begin{scope}
            \clip
                (-0.36,0.26)
                --
                (0.36,0.26)
                --
                (0.26,-0.22)
                .. controls (0.14,-0.55) and (-0.14,-0.55) ..
                (-0.26,-0.22)
                -- cycle;

            \draw[thin] (-0.28, 0.05) -- (0.28, 0.05);
            \draw[thin] (-0.22,-0.13) -- (0.22,-0.13);
            \draw[thin] (-0.14,-0.31) -- (0.14,-0.31);
        \end{scope}
    \end{scope}
}

\newcommand{\Wave}[2]{%
    \draw[wave]
        ($(#1)+(#2:0.58)$)
        --
        ++(#2:1.04);
}

\newcommand{\LastLink}[1]{%
    \draw[bond]
        ({\rightX+\xC+0.12},#1)
        --
        ({\rightX+\xC+0.55},#1);

    \draw[dashed,thick]
        ({\rightX+\xC+0.55},#1)
        --
        ({\rightX+\xD-0.6},#1);

    \draw[bond]
        ({\rightX+\xD-0.6},#1)
        --
        ({\rightX+\xD-0.12},#1);
}


\coordinate (S1) at (0,\triY);
\coordinate (S2) at (\triW,\triY);
\coordinate (S3) at (0.5*\triW,\triB);

\coordinate (R11) at ($(S1)+(125:\rOsc)$);
\coordinate (R12) at ($(S1)+(165:\rOsc)$);
\coordinate (R13) at ($(S1)+(205:\rOsc)$);

\coordinate (R21) at ($(S2)+(55:\rOsc)$);
\coordinate (R22) at ($(S2)+(15:\rOsc)$);
\coordinate (R23) at ($(S2)+(-25:\rOsc)$);

\coordinate (R31) at ($(S3)+(215:\rOsc)$);
\coordinate (R32) at ($(S3)+(270:\rOsc)$);
\coordinate (R33) at ($(S3)+(325:\rOsc)$);

\draw[bond] (S1) -- (S2);
\draw[bond] (S1) -- (S3);
\draw[bond] (S2) -- (S3);

\draw[bond] (S1) -- (R11);
\draw[bond] (S1) -- (R12);
\draw[bond] (S1) -- (R13);

\draw[bond] (S2) -- (R21);
\draw[bond] (S2) -- (R22);
\draw[bond] (S2) -- (R23);

\draw[bond] (S3) -- (R31);
\draw[bond] (S3) -- (R32);
\draw[bond] (S3) -- (R33);

\Oscillator{R11}
\Oscillator{R12}
\Oscillator{R13}

\Oscillator{R21}
\Oscillator{R22}
\Oscillator{R23}

\Oscillator{R31}
\Oscillator{R32}
\Oscillator{R33}

\node[sys] at (S1) {$S_1$};
\node[sys] at (S2) {$S_2$};
\node[sys] at (S3) {$S_3$};

\Wave{R11}{120}
\Wave{R11}{145}

\Wave{R12}{165}
\Wave{R12}{190}

\Wave{R13}{210}
\Wave{R13}{235}

\Wave{R21}{60}
\Wave{R21}{35}

\Wave{R22}{15}
\Wave{R22}{-10}

\Wave{R23}{-30}
\Wave{R23}{-55}

\Wave{R31}{210}
\Wave{R31}{240}

\Wave{R32}{255}
\Wave{R32}{285}

\Wave{R33}{300}
\Wave{R33}{330}

\pgfmathsetmacro{\yBraceMid}{0.5*(\yOne+\yThree)}
\draw[arr]
    (3.9,0.05)
    to[out=5,in=180]
    ({\rightX-1.0},\yBraceMid);


\def\xA{0.0}
\def\xB{0.95}
\def\xC{1.90}
\def\xD{4.30}

\pgfmathsetmacro{\xDots}{0.5*(\xA+\xD)}

\draw[
    decorate,
    decoration={brace,amplitude=7pt,mirror},
    line width=0.9pt
]
    (\rightX-0.40,\yOne+0.60) -- (\rightX-0.40,\yThree-0.60);

\foreach \x in {\xA,\xB,\xC,\xD}{
    \draw[tleg]
        ({\rightX+\x},{\yOne+0.10})
        --
        ({\rightX+\x},{\yOne+\legUp});

    \draw[tleg]
        ({\rightX+\x},{\yTwo+0.10})
        --
        ({\rightX+\x},{\yTwo+\legUp});

    \draw[tleg]
        ({\rightX+\x},{\yThree+0.10})
        --
        ({\rightX+\x},{\yThree+\legUp});
}

\draw[bond] ({\rightX+\xA},\yOne) -- ({\rightX+\xB},\yOne);
\draw[bond] ({\rightX+\xB},\yOne) -- ({\rightX+\xC},\yOne);
\LastLink{\yOne}

\node at ({\rightX+\xDots},1.9) {$\cdot$};
\node at ({\rightX+\xDots},1.6) {$\cdot$};
\node at ({\rightX+\xDots},1.3) {$\cdot$};

\draw[bond] ({\rightX+\xA},\yTwo) -- ({\rightX+\xB},\yTwo);
\draw[bond] ({\rightX+\xB},\yTwo) -- ({\rightX+\xC},\yTwo);
\LastLink{\yTwo}

\node at ({\rightX+\xDots},-1.1) {$\cdot$};
\node at ({\rightX+\xDots},-1.4) {$\cdot$};
\node at ({\rightX+\xDots},-1.7) {$\cdot$};

\draw[bond] ({\rightX+\xA},\yThree) -- ({\rightX+\xB},\yThree);
\draw[bond] ({\rightX+\xB},\yThree) -- ({\rightX+\xC},\yThree);
\LastLink{\yThree}

\foreach \x in {\xA,\xB,\xC,\xD}{
    \node[tensor] at ({\rightX+\x},\yOne) {};
    \node[tensor] at ({\rightX+\x},\yTwo) {};
    \node[tensor] at ({\rightX+\x},\yThree) {};
}

\node at ({\rightX+\xA},{\yOne+\legUp+0.35}) {$\sigma_{i_1}$};
\node at ({\rightX+\xB},{\yOne+\legUp+0.35}) {$\sigma_{i_2}$};
\node at ({\rightX+\xC},{\yOne+\legUp+0.35}) {$\sigma_{i_3}$};
\node at ({\rightX+\xD},{\yOne+\legUp+0.35}) {$\sigma_{i_{NQ}}$};

\node at ({\rightX+\xA},{\yTwo+\legUp+0.35}) {$\sigma_{i_1}$};
\node at ({\rightX+\xB},{\yTwo+\legUp+0.35}) {$\sigma_{i_2}$};
\node at ({\rightX+\xC},{\yTwo+\legUp+0.35}) {$\sigma_{i_3}$};
\node at ({\rightX+\xD},{\yTwo+\legUp+0.35}) {$\sigma_{i_{NQ}}$};

\node at ({\rightX+\xA},{\yThree+\legUp+0.35}) {$\sigma_{i_1}$};
\node at ({\rightX+\xB},{\yThree+\legUp+0.35}) {$\sigma_{i_2}$};
\node at ({\rightX+\xC},{\yThree+\legUp+0.35}) {$\sigma_{i_3}$};
\node at ({\rightX+\xD},{\yThree+\legUp+0.35}) {$\sigma_{i_{NQ}}$};

\node[right] at ({\rightX+\xD+0.5},\yOne) {$(1,1)$};
\node[right] at ({\rightX+\xD+0.5},\yTwo) {$(n,m)$};
\node[right] at ({\rightX+\xD+0.5},\yThree) {$(N,N)$};

\end{tikzpicture}
    }
    \caption{Conceptual representation of the DAMPF method applied to continuous bosonic environments. {Each local environment is mapped to pseudomodes (damped harmonic oscillators), and their dynamics conditioned on a given system density matrix element $\ket{n}\bra{m}$ is represented as a matrix product state.}}
    \label{Fig_DAMPF_2}
\end{figure}
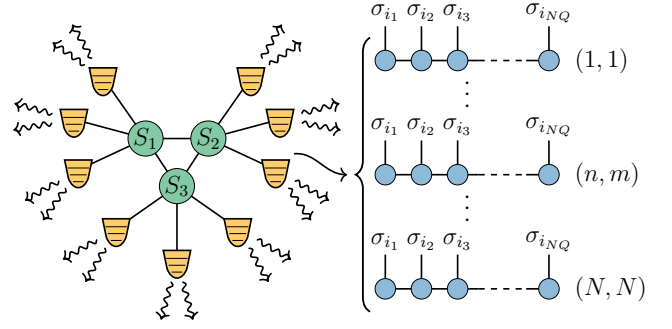

In DAMPF, each density matrix $\rho^{(n,m)}(t)$ is expanded as an MPS as
\begin{equation}
\label{eq_DAMPF}
    \rho^{(n,m)}(t)=\sum_{i_1\dots i_{NQ}}\hspace{-0.1cm}A^{(n,m)}_{1;i_1}\cdots A^{(n,m)}_{NQ;i_{NQ}}{\sigma_{i_1}}\cdots\sigma_{i_{NQ}},
\end{equation}
with $A^{(n,m)}_{q;i_q}$ matrices, where $\{\sigma_{i_q}\}_{i_q}$ denotes a basis of the Liouville space associated with pseudomode $q$, with $q = 1, \dots, NQ$. Hence, rather than representing the state in a single MPS, DAMPF represents it as a collection of $N^2$ MPSs, as shown in Fig.~\ref{Fig_DAMPF_2}.

The evolution of the density matrix, $\rho(t)=e^{\mathcal{L}t}\rho(0)$, is implemented via a Trotterization of the time propagator ${\rm exp}({\mathcal{L}t})$, separating the propagators associated with the Hamiltonian terms $H_S$, $H_E$, $H_{I}$ in~\eqref{eq_1_DAMPF} and with the dissipator $\mathcal{D}$ in~\eqref{eq_dampf_diss}. In particular, the bath Hamiltonian $H_E$ and the dissipator $\mathcal{D}$, which act only on the pseudomode degrees of freedom, generate time step propagators that act in the same way on all density matrix constituents $\rho^{(m,n)}(t)$, that is, 

\begin{equation}
    \rho^{(m,n)}(t)\to e^{\mathcal{L}_x\Delta t}\rho^{(m,n)}(t)\ ,
\end{equation}
with $x\in \{E,\mathcal{D}\}$ and $\mathcal{L}_\mathcal{D} = \mathcal{D}$.

The system Hamiltonian, on the other hand, acts only on the system degrees of freedom, and generates a time step propagator that induces a linear combination of the matrix constituents, that is,

\begin{equation}
    \rho^{(m,n)}(t)\to \sum_{n',m'}\Big(\langle m|e^{\mathcal{L}_S\Delta t}\big[|m'\rangle\langle n'|\big]|n\rangle\Big)\rho^{(m',n')}(t)\ .
\end{equation}

Lastly, considering the interaction Hamiltonian ${H_I=\sum_{n=1}^NO_n^SB_n}$, typical DAMPF simulations consider diagonal system coupling operator $O_n=|n\rangle\langle n|$. In this case, the associated time step propagator factorizes as a product of propagators acting on each density matrix constituent, that is

\begin{equation}
\label{DAMPF_evolv_inter}
    \rho^{(m,n)}(t)\to e^{-iB_m\Delta t}\rho^{(m,n)}(t)e^{iB_n\Delta t}.
\end{equation}

In the case of non-diagonal system coupling operators, a linear combination of the density matrix constituents from~\eqref{DAMPF_evolv_inter} also appears.

In the representation of the state in~\eqref{eq_DAMPF}, the MPS sites correspond exclusively to pseudomodes, which yields significant computational advantages.

As a first advantage, the bond dimension scales only with the correlations between pseudomodes, rather than with those between the system and the pseudomodes. In principle, this allows encoding large system--environment correlations with only a moderate increment of the computational cost. Moreover, when considering uncoupled pseudomodes, since their interaction is mediated by the system, the correlations developed between them are typically much smaller than the system–environment correlations, hence reducing the bond dimension employed in this formalism. Importantly, this formalism can be extended to encode more than a single excitation in the system, provided that the system Hilbert space remains polynomial with respect to the system size.

A second advantage, still related to the bond dimension, is that in this MPS formalism, the local Lindblad relaxation of the pseudomodes manifests as a local dissipator acting on the tensors. This uncorrelated source of dephasing further limits the build-up of correlations in the MPS formalism.

Finally, in the presence of some form of decoupling between system degrees of freedom $|n\rangle\langle m|$, e.g., a separation between coherence and population dynamics, DAMPF allows the separate simulation of each of these manifolds by solely considering the corresponding MPSs in the expansion as in \eqref{eq_DAMPF}.

Hence, as a consequence of the employed formalism, DAMPF simulations can typically be performed with relatively small bond dimensions. For example, absorption spectrum simulations~\cite{LorenzoniPRL2024} and excitation energy transfer dynamics~\cite{lorenzoni_full_2025} for the FMO complex have reached convergence with bond dimensions $\chi \lesssim 5$ and $\chi \lesssim 20$, respectively.

We note, however, that the formalism in \eqref{eq_DAMPF} is practical only when the number $N$ of relevant system degrees of freedom is limited, e.g., not exponentially large. This condition is, for instance, satisfied when an underlying symmetry, such as excitation-number conservation, allows the dynamics to be restricted to a reduced manifold of the Hilbert space.

Additionally, the mapping of the pseudomodes onto the one-dimensional MPS employed in DAMPF is not unique, since the pseudomodes can be arranged in an arbitrary order when constructing the chain. As a consequence, different orderings may lead to different computational costs, and the ordering that minimises the growth of the bond dimension is not always evident.

The DAMPF approach has the following convergence parameters
\begin{itemize}
    \item the number of pseudomodes $Q$ required for the finite expansion of the bath correlation function,
    \item the local Hilbert space dimension $d$ of each pseudomode,
    \item the bond dimension $\chi$ of the MPS representation,
    \item the time-step $\Delta t$ of the time-evolution.
\end{itemize}

\subsubsection{Applications and implementations}
The ability of DAMPF to treat several system sites together with up to hundreds of pseudomodes makes it suitable for simulations of open quantum systems in the presence of multiple environments with structured spectral densities. In molecular systems, this has enabled applications to charge-transfer~\cite{SomozaCommunPhys2023} and excitation energy-transfer~\cite{LorenzoniPRL2024,lorenzoni_full_2025} processes based on microscopic models, that fully incorporate experimentally measured or first-principle-method estimations of spectral densities. In such multi-site settings, DAMPF can provide a computationally favorable alternative to other non-perturbative approaches discussed in this review, whose cost may become substantial as the number of sites increases. An implementation of DAMPF on a quantum simulator has also been formalized~\cite{Lemmer_2018}, and extensions of the pseudomode approach to fermionic environments have been developed~\cite{CirioPRR2023}.

An open-source Python implementation of DAMPF is available in the DAMPyF package~\cite{LorenzoniArxiv2026}.

\subsection{HEOM}\label{sec:HEOM}

\subsubsection{Overview}
The hierarchical equations of motion (HEOM) provide a framework for modeling system dynamics that accounts for non-perturbative interactions between a system and its environment at finite temperature~\cite{Tanimura2020}. In this approach, the influence of the environment is accounted for by introducing auxiliary density operators of the same dimension as the system density operator.
The HEOM creates a hierarchy of coupled differential equations between these auxiliary density operators and the system density operator, enabling the time evolution of the system to be numerically exactly calculated. In the tensor network formulation of HEOM, the auxiliary density operators are represented as a tensor network. 

As a numerical method, HEOM requires a finite decomposition of the bath correlation functions~\cite{Ikeda2020}, and a closed hierarchy of the coupled differential equations derived from repeated time derivatives of the Feynman-Vernon influence functional. 

The HEOM formalism was originally introduced to describe quantum dissipation in systems coupled to a Drude--Lorentz spectral density, for which the BCF admits an analytical expansion in terms of exponential functions. In this case, Matsubara expansions provide closed-form expressions for the exponential series, typically requiring only a few terms at moderate temperatures, and reducing to a single exponential in the high-temperature limit ($T \rightarrow \infty$)~\cite{Tanimura1989}.
Nowadays, non-exponential decompositions and fitting methods enable the use of HEOM for general spectral densities~\cite{Qiang_Reduced_Quantum_Dynamics_2014,Ikeda2020,takahashi_high_2024}.

For many relevant systems, the correlation function does not exhibit simple decompositions. As such, the number of coupled differential equations may be large. The potentially large number of coupled differential equations for the auxiliary density operators arising in the HEOM formalism, along with the nearest-neighbor connectivity between these auxiliary density operators, has motivated the use of tensor-network techniques to enhance computational efficiency, enabling efficient simulations of larger and more complex systems.

\subsubsection{Derivation}

To derive the HEOM, we start by assuming the Hamiltonian for a system linearly coupled to a bosonic environment as in \eqref{eq:HSE}.
As outlined in Section~\ref{sec:OQS}, the associated time evolution of such a system is then determined by the Feynman-Vernon influence functional. In the interaction picture, this generates a system density operator propagator of the form in \eqref{eq:feynman-vernon-path-integral}.

\begin{figure}[t]
\centering
\begin{tikzpicture}[
    scale=0.75,
    every node/.style={font=\normalsize},
    rho/.style={
        circle,
        draw=black,
        fill=paletteGreen,
        minimum size=7.2mm,
        inner sep=0pt
    },
    heomarrow/.style={
        draw=black,
        -{Stealth[length=2.2mm,width=1.8mm]},
        thick,
        shorten >=2pt,
        shorten <=2pt
    },
    arrowlabel/.style={
        midway,
        font=\normalsize,
        inner sep=0.5pt
    }
]

\def\xmid{2.55}
\def\xbot{5.10}
\def\ysep{2.75}
\def\labelsep{6pt}
\def\arrowshift{30pt}

\node[rho] (rho00) at (0,0) {$\rho_{00}$};

\node[rho] (rho10) at (-\xmid,-\ysep) {$\rho_{10}$};
\node[rho] (rho01) at ( \xmid,-\ysep) {$\rho_{01}$};

\node[rho] (rho20) at (-\xbot,-2*\ysep) {$\rho_{20}$};
\node[rho] (rho11) at (0,-2*\ysep) {$\rho_{11}$};
\node[rho] (rho02) at (\xbot,-2*\ysep) {$\rho_{02}$};

\newcommand{\ParallelArrow}[5]{%
    \draw[heomarrow]
        ([xshift=#4]#1)
        --
        ([xshift=#4]#2)
        node[arrowlabel,#5=\labelsep] {#3};
}

\ParallelArrow{rho00}{rho10}{$O^c$}{-\arrowshift}{left}
\ParallelArrow{rho00}{rho01}{$O^c$}{ \arrowshift}{right}

\ParallelArrow{rho10}{rho20}{$O^c$}{-\arrowshift}{left}
\ParallelArrow{rho10}{rho11}{$O^c$}{-\arrowshift}{left}

\ParallelArrow{rho01}{rho02}{$O^c$}{ \arrowshift}{right}
\ParallelArrow{rho01}{rho11}{$O^c$}{ \arrowshift}{left}

\ParallelArrow{rho10}{rho00}{$O^c$}{ \arrowshift}{right}
\ParallelArrow{rho01}{rho00}{$O^a$}{-\arrowshift}{left}

\ParallelArrow{rho20}{rho10}{$2O^c$}{ \arrowshift}{right}
\ParallelArrow{rho11}{rho10}{$O^a$}{ \arrowshift}{right}

\ParallelArrow{rho02}{rho01}{$2O^a$}{-\arrowshift}{left}
\ParallelArrow{rho11}{rho01}{$O^c$}{-\arrowshift}{right}

\end{tikzpicture}

\caption{Connectivity diagram of the auxiliary density operators in the hierarchical equation of motion as given in \eqref{eqn:HEOM} for $N_R=N_I=1$ to a depth of two.}
\label{fig:HEOM}
\end{figure}
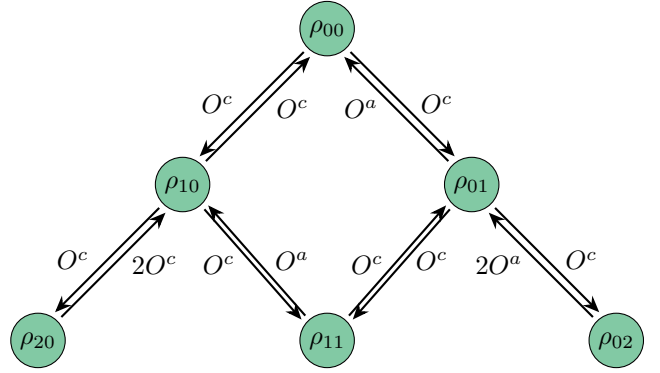

From this, the induced system dynamics for a Gaussian harmonic environment are uniquely determined by the BCF given by \eqref{eq:Environment_Correlation_Function}.
The original formulation for HEOM considered an exponential decomposition of the environment correlation function, which we shall present here. Other decompositions, e.g., using orthogonal polynomials, generate equivalent hierarchies~\cite{Rahman_2019}. 
We start by considering a decomposition of the correlation function into real and imaginary components
\begin{equation}
    C(t) = C_R(t)+iC_I(t),
\end{equation}
which are then each expanded over exponentials~\cite{Fruchtman2016}
\begin{align}
    C_R(t) &= \sum^{N_R}_{k=1}c_k^R e^{-\gamma_k^Rt},\\
    C_I(t) &= \sum^{N_I}_{k=1}c_k^I e^{-\gamma_k^It},
\end{align}
where the coefficients $c_k^{R,I}$ and frequencies $\gamma^{R,I}_k$ can, in principle, be complex-valued, provided that the whole function is real.
We chose this framework due to the simplicity of the generated HEOM.
To generate the HEOM, we apply consecutive time derivatives to the time-evolved system density operator described in \eqref{eq:feynman-vernon-path-integral}. The first derivative yields 
\begin{align}
    \frac{d}{dt}{\rho}_S(t) = -O^c(t)\mathcal{T}_\leftarrow&\bigg[\int ^t_0ds\Big(O^c(s)\sum_k^{N_R}c_k^Re^{-\gamma_k^R(t-s)}
    \nonumber\\
  & +iO^a(s)\sum_k^{N_I}c_k^Ie^{-\gamma_k^I(t-s)} \Big)\mathcal{F}(t)\bigg],\label{eqn:rhoDot}
\end{align}
where we have introduced the operator-valued function
\begin{align}
    \mathcal{F}(t) = \exp\Bigg\{&
    -\int_0^t d s\int_0^s d s^\prime O^c(s)\Big[C_R(s-s^\prime)O^c(s^\prime) \nonumber\\
    &+ i C_I(s-s^\prime)O^a(s^\prime)\Big]
    \Bigg\}
    \rho_S(0).
\end{align}
Transforming back to the Schr\"odinger frame, we have 
\begin{align}
    \frac{d}{dt}{\rho}_S(t) = &-iH_S^c \rho_S(t) \nonumber\\&-O^cU(t)\mathcal{T}_\leftarrow\bigg[\int ^t_0ds\big(O^c(s)\sum_k^{N_R}c_k^Re^{-\gamma_k^R(t-s)}
    \nonumber\\
   &+iO^a(s)\sum_k^{N_I}c_k^Ie^{-\gamma_k^I(t-s)} \big)\mathcal{F}(t)\bigg]U^\dag(t). 
\end{align}
We now introduce the first-order auxiliary density operators (ADOs) 
\begin{align}
    \rho^{1_{Rk}}(t) &= -i U(t)\mathcal{T}_\leftarrow\bigg[\int ^t_0dsO^c(s)c_k^Re^{-\gamma_k^R(t-s)}\mathcal{F}(t)\bigg]U^\dag(t),
    \nonumber\\
   \rho^{1_{Ik}}(t)&=U(t)\mathcal{T}_\leftarrow\bigg[\int ^t_0dsO^a(s)c_k^Ie^{-\gamma_k^I(t-s)}\mathcal{F}(t)\bigg]U^\dag(t).
\end{align}
This enables the differential equation for the reduced density matrix \eqref{eqn:rhoDot} to be reduced to the coupled equation between ADOs
\begin{equation}
    \frac{d}{dt}{\rho}_S(t) = -iH_S^c \rho_S(t) -iO^c\sum_{j=R,I}\sum_{k=1}^{{N_j}}\rho^{1_{jk}}(t).
\end{equation}
The system density operator evolution is now influenced by the ADOs $\rho^{1_{kj}}$. We can then iteratively take derivatives of the ADOs to generate a hierarchy of coupled differential equations of the form
\begin{align}
    \frac{d}{d t}{\rho}^\mathbf{n}(t) &= \left( -iH^c_S-\sum_{j=R,I}\sum_{k=1}^{N_j}n_{jk}\gamma^j_k \right)\rho^\mathbf{n}(t) \nonumber \\ &-i\sum^{N_R}_{k=1}c_k^R n_{Rk}O^c \rho^{\mathbf{n}^-_{Rk}}(t) + \sum^{N_I}_{k=1}c_k^In_{Ik}O^a\rho^{\mathbf{n}^-_{Ik}}(t) \nonumber \\
    &-i\sum_{j=R,I}\sum_{k=1}^{N_j}O^c \rho^{\mathbf{n}^+_{jk}}(t).\label{eqn:HEOM}
\end{align}
Each of the $\rho^{\mathbf{n}}$ represents an auxiliary density operator with dimension equal to that of the system density operator. We have also introduced the multi-index ${\mathbf{n} = \{(n_{Rk},n_{Ik})\}_k}$, with ${n_{jk}\in \{0,...,D_{jk}\}}$ and $j \in \{R, I\}$. The terms carrying an index $n^\pm_{jk}$ denote index pairs in which the $jk$th index is raised or lowered by one. The only `physical' density operator is the $(0,...,0)$ indexed operator, which identifies the system density operator. All other $\rho^\textbf{n}$ encapsulate environmental effects, accounting for $\mathbf{n}$th order derivatives of the influence functional. The superoperators $O^{c,a}$ couple different ADOs; a connectivity diagram of these ADOs is shown in Fig.~\ref{fig:HEOM}. While these ADOs do not directly correspond to degrees of freedom in the environment, they do encapsulate entanglement between the system and the environment.
The first term on the right-hand side of \eqref{eqn:HEOM}, proportional to $\rho^\mathbf{n}$, describes the unitary evolution and non-unitary decay of the density operators. The remaining terms can be interpreted as the creation and annihilation of fictitious quasiparticle excitations as in the bexciton formulation~\cite{Chen2024}. This interpretation is made explicit in the Dissipaton Equations of Motion (DEOM) formalism, which represents the environment directly in terms of quasiparticles known as \textit{dissipatons}, which are physical collective bath degrees of freedom, and enables environmental observables to be computed from their expectation values~\cite{Yan2016_DEOM}. Furthermore, a one-to-one correspondence between HEOM and the pseudomode approach, such as that employed in DAMPF, has recently been established~\cite{muller2026onetoonecorrespondencehierarchicalequations}.
Different choices for expanding the correlation function lead to different but formally equivalent formulations of the HEOM. A discussion of the relations between some of these formulations is given in ~\cite{Ikeda2020}. A relevant consideration is whether one expansion needs fewer ADOs than another, which can make the numerical solution more efficient. The inclusion of multiple baths with different couplings between them can be achieved by expanding the indexing of $\mathbf{n}$ to include the number of additional baths. While this extension is simple, it dramatically increases the number of coupled differential equations.
Formally, the depth of the hierarchy ($D_{jk}$) should be infinite; however, by making $D_{jk}$ finite, these become convergence parameters of the method. This leads to a number of coupled equations equal to 
\begin{equation}
    \# \text{ODEs} = \prod_{j=R,I}\prod_{k=1}^{N_j}(
    D_{jk}+1).
\end{equation}
The complexity of the HEOM clearly grows rapidly when accounting for complex decompositions of the correlation functions, multiple environments, or large system Hilbert spaces. In the following sections, we discuss methods for truncating these hierarchies.

\subsubsection{Finite expansion of the correlation function}
We will first present an analytical decomposition of the correlation function as a sum of exponentials, as introduced in the previous section, for the case of an underdamped spectral density
\begin{equation}
    J(\omega) = \frac{2\alpha\Gamma\omega}{\pi((\omega_c-\omega)^2+\Gamma^2)((\omega_c+\omega)^2+\Gamma^2)}\label{eqn:specUD}\ .
\end{equation}
This example illustrates why only a finite (but not necessarily small) number of terms need to be considered in the exponential expansion of the environment correlation function.
Furthermore, this example also motivates the introduction of alternative decompositions relying either on other mathematical expansions or fitting procedures.

Considering the spectral density in \eqref{eqn:specUD}, and using the Matsubara decomposition of the $\coth$ function
\begin{equation}
     \coth\left(\frac{\omega}{2T}\right) = 2T\left(\frac{1}{\omega}+ \sum_{k=1}^\infty \frac{2\omega}{\omega^2 +(2\pi k T)^2} \right),
\end{equation}
we can, utilizing the Cauchy residue theorem, calculate the correlation function $C(t)$ directly. 
From \eqref{eq:Environment_Correlation_Function}, this then yields
\begin{equation}
    C(t) = \sum_{k \in \mathbb{N}\cup\{\pm\}} c_k^R e^{-\gamma_k^R t}  +i\sum_{k=\pm} c^I_k e^{-\gamma_k^It},
\end{equation}
where the amplitudes of the real part of the correlation function are given by
\begin{equation}
    c_k^R = \begin{cases}
        \alpha \coth(\beta(\omega_c+\i\Gamma)/2)/4\omega_c,\hspace{20pt} &k= +\\
        \alpha \coth(\beta(\omega_c-\i\Gamma)/2)/4\omega_c,\hspace{20pt} &k= -\\
        \frac{-4\alpha\Gamma\gamma_k^R}{\beta((\i\Gamma-\omega_c)^2+\gamma_k^{R2})((\i\Gamma+\omega_c)^2+\gamma_k^{R2}))}, \hspace{19pt} &k\in \mathbb{N}
        
    \end{cases},
\end{equation}
and the associated rates are
\begin{equation}
    \gamma_k^R = \begin{cases}
        -i\omega_c + \Gamma, \hspace{10pt} &k = +\\
        i\omega_c + \Gamma, \hspace{17pt} &k = - \\
        2\pi k/\beta, \hspace{29pt} &k\in \mathbb{N}
    \end{cases}.
\end{equation}
While for the imaginary part of the correlation function, the coefficients are
\begin{equation}
    c_k^I = \begin{cases}
        -i\alpha /4\omega_c, \hspace{18pt} &k =+\\
        i\alpha /4\omega_c ,\hspace{10pt} &k =-
    \end{cases},
\end{equation}
with associated rates
\begin{equation}
    \gamma_k^I = \begin{cases}
        i\omega_c +\Gamma, \hspace{18pt} k =+\\
        -i\omega_c +\Gamma, \hspace{10pt} k =-
    \end{cases}.
\end{equation}
 
\begin{figure}[t]
    \centering
    \includegraphics[width=\linewidth]{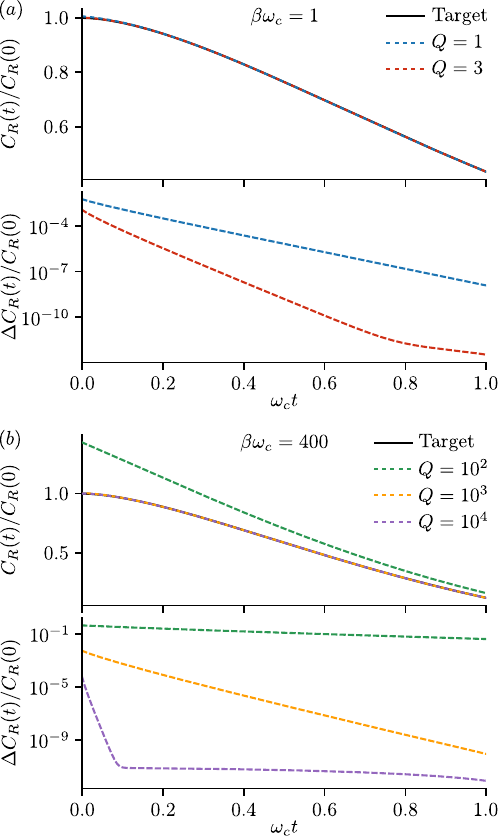}
    \caption{Convergence of the real part of the correlation function with the number $Q$ of kept Matsubara frequencies for an underdamped spectral density as given in \eqref{eqn:specUD} with parameters  $\Gamma=\alpha=\omega_c$ at high temperature $\beta\omega_c = 1$ (a) and low temperature $\beta\omega_c = 400$ (b).}
    \label{fig:Matsubara}
\end{figure}

In the high-temperature regime $k_B T \gtrsim \hbar\omega_c$, the Matsubara frequency terms $\gamma_k^R$ for $k\geq1$ can be ignored, and the series is truncated with just two real and imaginary terms. Conversely, at low temperatures, the number of Matsubara terms required for convergence grows rapidly.
The growth of the number of Matsubara frequencies required for convergence is shown in Fig.~\ref{fig:Matsubara}. We can see that in the high-temperature regime, few Matsubara frequencies are required to reproduce the correlation function. Conversely, in the low-temperature regime, $10^4$ Matsubara frequencies are required to get a commensurate accuracy. 
Such a large number of terms renders this expansion impractical for any numerical scheme, as the number of ADOs in the HEOM is strongly dependent on the number of terms in the expansion. 

This can be partly addressed via Markovian terminators as outlined below: For terms with large decay rates \( \Re\{\gamma^{R,I}_j\} \gg 1/\tau_S \) (where \( \tau_S \) is a characteristic system timescale), the corresponding contributions to the correlation function decay rapidly and can be treated as effectively Markovian. Ordering the rates such that \( |\Re\{\gamma_j^{R,I}\}| \leq |\Re\{\gamma_{j+1}^{R,I}\}| \), we can define a cutoff \( N_c^{R,I} \) beyond which this Markovian condition holds. For \( j > N_c^{R,I} \), we approximate
\begin{equation}
e^{-\gamma_j^{R,I} t} \approx \frac{\delta(t)}{\gamma_j^{R,I}} \, ,
\end{equation}
leading to the decomposition
\begin{align}
C_{R,I}(t) \approx\;& \sum_{j=1}^{N_c^{R,I}} c_j^{R,I} e^{-\gamma_j^{R,I} t} + \sum_{j=N_c^{R,I}+1}^{N_{R,I}} \frac{c_j^{R,I}}{\gamma_j^{R,I}} \delta(t) \\
=\;& C^{\text{NM}}_{R,I}(t) + C^{\text{M}}_{R,I}(t),
\end{align}
where \( C^{\text{NM}} \) and \( C^{\text{M}} \) denote the non-Markovian and Markovian contributions, respectively~\cite{Ishizaki2005_colored_noise}.

The total Markovian contribution can be incorporated via a Lindblad dissipator~\cite{breuer_theory_2009}:
\begin{equation}
-i H_S^c \rho \;\rightarrow\; -i \Tilde{H}_S^c \rho
+ \Gamma_M \left(2 O \rho O^\dagger - \{O^\dagger O, \rho\}\right),
\end{equation}
where we have the Markovian Lamb-shifted Hamiltonian $\tilde{H}_S = H_S+H_L$ where the shift is 
\begin{equation}
    H_L = \sum_{j=N_c^I+1}^{N_I} \frac{c_j^I}{\gamma_j^I}O^\dagger O,
\end{equation}
and the Markovian decay rate
\begin{equation}
\Gamma_M =\sum_{j=N_c^R+1}^{N_R} \frac{c_j^R}{\gamma_j^R}    .
\end{equation}
This allows fast-decaying terms to have a Markovian treatment, reducing the number of Matsubara components requiring explicit ADOs.

This analytical framework can be extended to structured spectral densities expressed as a sum of antisymmetrized Lorentzians as in \eqref{eqn:specUD}~\cite{Meier1999}. When combined with the Matsubara decomposition, such forms yield a systematic representation of the BCF: for $K$ antisymmetrized Lorentzian peaks, the imaginary part of the BCF requires $2K$ exponential terms, while the real part requires $2K + Q$ exponentials, where $Q$ denotes the number of Matsubara frequencies needed for convergence.
However, the number of Matsubara terms increases rapidly as the temperature decreases, rendering simulations computationally demanding in the low-temperature regime even with the Markovian cutoff. 
A low-temperature correction scheme for the termination of the hierarchy, for instance, using Zwanzig projection~\cite{Fay2022}, which provides greater consistency with weak coupling theory, can be used to further reduce the number of ADOs.

To reduce the number of terms in the expansion of the BCF, different expansion strategies have been devised.
Early work showed that the Matsubara frequencies can be truncated by using a Pad\'e approximation for the Bose-Einstein distribution~\cite{Hu2011}. 
Recently, a barycentric rational-function representation of the spectral density has been proposed~\cite{Xu2022}, where the complex poles of the spectral density are represented explicitly, thereby circumventing the Matsubara expansion and avoiding its exponential growth at low temperatures.
Generally, so-called `extended' or `generalized' HEOM approaches utilize a non-exponential decomposition of the BCF~\cite{Ikeda2020}. This enhances the ability to study directly environments with BCFs that do not readily admit an exponential decomposition, such as those of a general Ohmic type.

When analytical decompositions are not available, fitting techniques are employed to approximate the spectral density~\cite{takahashi_high_2024}. 
This generalizes the antisymmetrized Lorentzian approach: When the spectral density is fitted with $K$ (non antisymmetrized) Lorentzian peaks, the corresponding BCF contains $2K + Q$ exponential components, accounting for contributions at $\pm \omega_k$ together with $Q$ Matsubara terms. 
Fitting-based HEOM formulations often make use of algorithms originally developed for pseudomode constructions, such as the Prony method, which efficiently extracts exponential components directly from correlation data~\cite{Chen2022}. 
In addition, recent coarse-graining schemes \cite{LorenzoniPRL2024} have been proposed to systematically reduce the number of exponentials required, while retaining accurate dynamical behavior. 
Recently, benchmarking studies between T-TEDOPA and HEOM ~\cite{le_de_managing_2024} demonstrated that different expansion strategies---Lorentzian fitting, Matsubara decomposition, barycentric interpolation, and discretizing the spectral density---lead to variations in numerical performance across the methods, with merits and drawbacks to each in different parameter regimes. 
A broader perspective is provided in the recent review~\cite{bai_hierarchical_2024}, which discusses additional techniques including Padé decompositions and time-domain Prony fitting. 
Collectively, these developments have greatly improved the scalability and versatility of the HEOM formalism, extending its applicability to strongly coupled, structured, and low-temperature quantum environments.

\subsubsection{Truncation of the hierarchy depth}

\begin{figure}[t]
    \centering
    \includegraphics[width=\linewidth]{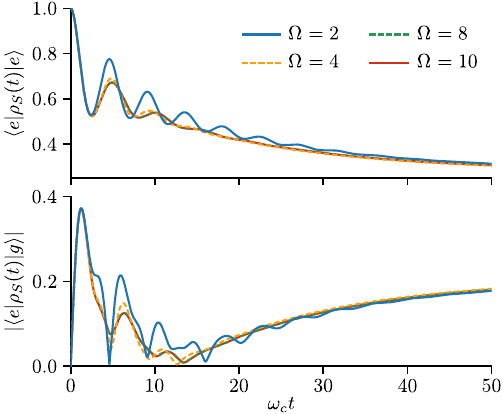}
    \caption{Convergence of the excited state population and coherence of a spin--boson model as a function of the hierarchy depth $\Omega$. The system Hamiltonian is ${H_S = \frac{\omega_0}{2}\left(\sigma^z+ \sigma^x\right)}$, the system coupling operator is $\sigma^z$ and we consider a Drude--Lorentz spectral density ${J(\omega)=2\omega\lambda\gamma/(\pi(\gamma^2+\omega^2))}$ with parameters $\gamma=0.2\omega_0$, $\lambda = 0.1\omega_0$ and $T=\omega_0$. The number of Matsubara terms required for convergence was $Q=3$.   }
    \label{fig:hierarchy}
\end{figure}

To close the depth of the hierarchy, previous works have truncated the hierarchy to a given depth $L$ such that $\sum_kn_k\leq L$, i.e., all $\rho^\mathbf{n}=0$ that satisfy $\sum_k n_k>L$
or, utilizing schemes to filter out the ADOs that contribute weakly to the system dynamics. Another scheme is to truncate each hierarchy to the same depth $D_{jk}= \Omega$, such that all $\rho^\mathbf{n}=0$ when any index $n_k>\Omega$. A more rigorous truncation of the hierarchy involves the Laplace transformation of the HEOM and utilizing the asymptotic behavior of terms deep in the hierarchy to create a closure~\cite{Tanimura2020}. In Fig.~\ref{fig:hierarchy}, we show the convergence of simulations of the spin--boson model dynamics with a Drude--Lorentz spectral density as we increase the symmetric hierarchy depth $\Omega$. We can see that when the hierarchy depth is insufficiently large, we see artificial non-Markovianity due to the reflection of information from within the effective mode hierarchy.

For symmetric truncation $D_{jk}=\Omega$, the resulting HEOM includes a total of  $(\Omega+1)^{N_I+N_R}$ ADOs. Clearly, when either the hierarchy depth $\Omega$ or the expansions $N_{R,I}$ are large, the number of ADOs required quickly becomes unmanageable. Noting also that the HEOM couples $\rho^\mathbf{n}$ to ADOs with at most a single unit index change, creating a nearest-neighbor topology, a tensor network approach becomes attractive.

\subsubsection{Tensor network representation}
Given the unfavorable scaling of HEOM with both the number of auxiliary modes and the number of excitations retained in each mode, considerable effort has been devoted to tensor-network representations that compress the hierarchy while preserving its accuracy. These approaches represent the auxiliary density operators (ADOs) in low-rank tensor forms.

\textcite{Shi2018} introduced the first MPS representation of HEOM, mapping the hierarchy onto a matrix product state and employing the time-dependent variational principle (TDVP) for propagation. In this framework, each of the elements of the ADOs is represented as an MPS
\begin{equation}
    \rho^\mathbf{n}_{ij} = \sum_{i_0,i_1,...,i_{\Omega}} A_{1n_1}^{i_0i_1}A_{2n_2}^{i_1i_2}\cdots A_{\Omega n_{\Omega}}^{i_{\Omega-1}i_{\Omega}}, \label{eq:MPS-HEOM}
\end{equation}
with $i_k \in\{1, \ldots, r_k^{ij}\}$, where $r_k^{ij}$ represents the bond dimension of the MPS and $r_0=r_{\Omega}=1$. We do not label the individual tensors $A_k$ with the $ij$ superscript for clarity.
In diagrammatic representation, the ADOs would have the form 
\begin{equation}
    \begin{tikzpicture}
        [tensor/.style={circle,draw=black,fill=paletteBlue_Light3,
                     inner sep=0pt,minimum size=5mm},
        line/.style={draw, thick},
        leg/.style={draw, thick}]
        \node (a) at (0, 0.5) {};
        \node (c) at (0.75,0) {$\rho^\mathbf{n}_{ij}$};
        \node (eq) at (1.25,0) {$=$} ;
        \node[tensor] (T1) at (2,0) {$A_1$} ;
        \node (i1) at (2, -1) {$n_1$} ;
        \node[tensor] (T2) at (3,0) {$A_2$} ;
        \node (i2) at (3, -1) {$n_2$} ;
        \node[tensor] (T3) at (4,0) {$A_3$} ;
        \node (i3) at (4, -1) {$n_3$} ;
        \node (dots) at (5,0) {$\ldots$} ;
        \node[tensor] (TN) at (6.4,0) {$A_{\Omega}$} ;
        \node (iN) at (6.4, -1) {$n_{\Omega}$} ;
        \draw (i1) -- (T1.south) ; \draw (T1.east) -- node[above] {$i_1$} (T2.west) ; 
        \draw (i2) -- (T2.south) ; \draw (T2.east) -- node[above] {$i_2$} (T3.west) ;
        \draw (i3) -- (T3.south) ; \draw (T3.east) -- node[above] {$i_3$} (dots.west) ;
        \draw (iN) -- (TN.south) ; \draw (dots.east) -- node[above] {$i_{\Omega-1}$} (TN.west) ;
        \node at (7,0) {.};
        \end{tikzpicture}
\end{equation}

Note that the partitioning performed here is between the pseudomodes $n_k$ and not the system Hilbert space. This formulation requires handling $N^2$ MPSs.
The representation of ADOs as an MPS in \eqref{eq:MPS-HEOM} enables a similar representation for the HEOM as 
\begin{equation}
    \frac{d}{dt}\rho^\mathbf{n}_{ij} = \sum_{kl}M^{ij,\mathbf{n}}_{kl,\mathbf{m}} \rho^\mathbf{m}_{kl},
\end{equation}
where the superoperator $M^{ij,\mathbf{n}}_{kl,\mathbf{m}}$ transforms between the MPS representations of the ADOs (though $\mathbf{m}$ is restricted to $\mathbf{n},\mathbf{n}_k^+,\mathbf{n}_k^-$). 
The superoperator $M^{ij,\mathbf{n}}_{kl,\mathbf{m}}$ is represented as an MPO, and the equations of motion can be solved using standard tensor network methods.
This demonstrated that tensor-network compression can efficiently capture the structure of the hierarchy even for systems with a large number of bath modes. Shortly thereafter, \textcite{Borelli2019} reformulated HEOM within a thermo-field dynamics framework, allowing the hierarchy to be represented as a wavefunction MPS. This formulation enabled adaptive low-rank tensor methods while preserving physical properties such as trace conservation. Subsequent developments improved the efficiency and robustness of this approach through adaptive low-rank Galerkin reduction~\cite{Borelli2021} and extended it to hybrid fermionic and bosonic environments by reformulating HEOM as a time-dependent Schrödinger-like equation~\cite{Ke2022}. Alternatively to the formulation presented above, one could also have written the vectorized ADOs, including the system, as an MPS~\cite{Shi2018, mangaud_survey_2023} or the purification ~\cite{Borelli2019, Borelli2021} as
\begin{equation}
    \begin{tikzpicture}
    [tensor/.style={circle,draw=black,fill=paletteBlue_Light3,
                     inner sep=0pt,minimum size=5mm},
        line/.style={draw, thick},
        leg/.style={draw, thick}]
        \node (c2) at (0.75,0) {$\left|\rho^\mathbf{n}\right)$};
        \node (eq2) at (1.25,0) {$=$} ;
        \node[tensor] (T12) at (2,0) {$A_0$} ;
        \node (i12) at (2, -1) {$\alpha$} ;
        \node[tensor] (T22) at (3,0) {$A_1$} ;
        \node (i22) at (3, -1) {$n_1$} ;
        \node[tensor] (T32) at (4,0) {$A_2$} ;
        \node (i32) at (4, -1) {$n_2$} ;
        \node (dots2) at (5,0) {$\ldots$} ;
        \node[tensor] (TN2) at (6.4,0) {$A_{\Omega}$} ;
        \node (iN2) at (6.4, -1) {$n_{\Omega}$} ;
        \draw (i12) -- (T12.south) ; \draw (T12.east) -- node[above] {$i_0$} (T22.west) ; 
        \draw (i22) -- (T22.south) ; \draw (T22.east) -- node[above] {$i_1$} (T32.west) ;
        \draw (i32) -- (T32.south) ; \draw (T32.east) -- node[above] {$i_2$} (dots2.west) ;
        \draw (iN2) -- (TN2.south) ; \draw (dots2.east) -- node[above] {$i_{\Omega-1}$} (TN2.west) ;
        \node at (7,0) {.};
    \end{tikzpicture}
\end{equation}
 A comprehensive overview of tensor-network approaches to HEOM is provided by ~\textcite{mangaud_survey_2023}.

While MPS approaches arrange auxiliary modes along a one-dimensional chain, the hierarchical structure of HEOM naturally motivates higher-dimensional tensor network topologies. \textcite{Yan2021} introduced Tucker and hierarchical Tucker decompositions for HEOM, demonstrating that hierarchical tensor representations can substantially reduce the dimensionality of the problem and outperform one-dimensional tensor decompositions for large numbers of effective bath modes.

In \textcite{Ke2023}, a tree tensor network state (TTNS) formulation of HEOM was developed. By recognizing that the exponential decomposition of the bath correlation function maps the environment onto a set of effective discrete bath modes, the HEOM hierarchy can be recast as a Schrödinger-like equation for an extended wavefunction governed by a non-Hermitian super-Hamiltonian. Representing this wavefunction as a TTNS and propagating it with TDVP yielded results consistent with conventional HEOM while substantially reducing computational cost. For the spin--boson model, the TTNS representation achieved approximately a fourfold speedup compared with an equivalent MPS decomposition~\cite{Ke2023}.
An example of how such a TTN representation of the vectorized or purified ADOs could be
\begin{equation}
    \begin{tikzpicture}
        [tensor/.style={circle,draw=black,fill=paletteBlue_Light3,
                     inner sep=0pt,minimum size=5mm},
        line/.style={draw, thick},
        leg/.style={draw, thick}]
        \node (c3) at (0.75,0) {$\left|\rho^\mathbf{n}\right)$};
        \node (eq3) at (1.25,0) {$=$} ;
        
        \node [tensor, fill=paletteBlue_Light3] (m1) at (2,-2) {$A_1$};
        \node (i13) at (2, -3) {$n_1$};
        \draw (i13) -- (m1.south);
        \node [tensor, fill=paletteBlue_Light3] (m2) at (3,-2) {$A_2$};
        \node (i23) at (3, -3) {$n_2$};
        \draw (i23) -- (m2.south);
        \node [tensor, fill=paletteBlue_Light3] (m3) at (4,-2) {$A_{3}$};
        \node (i33) at (4, -3) {$n_3$};
        \draw (i33) -- (m3.south);
        \node [tensor, fill=paletteBlue_Light3] (m4) at (5,-2) {$A_4$};
        \node (iN3) at (5, -3) {$n_4$};
        \draw (iN3) -- (m4.south);
        
        \node [tensor, fill=paletteBlue_Light1] (u1) at (2.5,-1) {};
        \node [tensor, fill=paletteBlue_Light1] (u3) at (4.5,-1) {};
        
        \node [tensor] (T13) at (3.5,-0) {$A_0$};
        \node (alpha3) at (3.5, -1) {$\alpha$};
        \draw (T13.south) -- (alpha3.north);
        
        \draw[line] (m1) -- (u1);
        \draw[line] (m2) -- (u1);
        \draw[line] (m3) -- (u3);
        \draw[line] (m4) -- (u3);
        \draw[line] (u1) --  (T13);
        \draw[line] (u3) -- (T13);
        \node at (6,0) {,};
    \end{tikzpicture}
\end{equation}
for a binary tree with $\Omega=4$.
A complementary formulation was introduced with the \textit{bexcitonic} approach to HEOM~\cite{Chen2024}, which interprets the hierarchy in terms of fictitious bosonic quasiparticles (\emph{bexcitons}) coupled through creation and annihilation operators. Building on this framework, a tree-tensor-network HEOM (TTN-HEOM) was developed by combining the bexcitonic formulation with tree tensor networks and propagating the resulting representation using Dirac--Frenkel variational dynamics~\cite{Chen2025}. This enabled the direct formulation of equations of motion for the tensors, eliminating the need for Trotterization and allowing higher-order integrators to be employed for TTN-HEOM propagation. As a result, arbitrary time-dependent system Hamiltonians can be treated efficiently, avoiding the potentially substantial computational cost associated with Trotterized implementations.

The development of TTN-HEOM has been strongly influenced by the multi-layer multi-configuration time-dependent Hartree (ML-MCTDH) method (see Sec.~\ref{sec:MLMCTDH}). The hierarchical decomposition and variational optimization strategies of ML-MCTDH provide the conceptual foundation for organizing auxiliary modes into tree structures, allowing TTN-HEOM to inherit many of the favorable compression and scaling properties of multilayer tensor methods.

The tensor network HEOM approach has the following convergence parameters
\begin{itemize}
    \item the required number of terms for the finite expansion of the correlation function $N_{R,I}$,
    \item the hierarchy depth $\Omega$,
    \item the bond dimension $\chi$ of the MPS or TTNS representation,
    \item the time-step $\Delta t$.
\end{itemize}

\subsubsection{Applications and implementations}
Due in part to the age of the approach, HEOM has been used to model a plethora of different physical systems. In its standard formulation, before the tensor network formulations, a canonical example in the field of quantum effects in biological systems was the study of coherence in the FMO complex using simplified environmental models~\cite{Ishiza_Theoretical_2009}. Other common problems tackled include effects of quantum tunneling in chemical processes~\cite{Ishizaki2005,Tanimura1992}, proton tunneling~\cite{Shi2011,Zhang2020}, electron transfer~\cite{Tanaka2009,Tanaka2010}, charge separation~\cite{Sakamoto2017}, photochemical processes~\cite{Ikeda2019}, and conical intersections~\cite{Qi2017}. Beyond chemical processes, HEOM has been used to explore quantum information and thermodynamics problems. These include quantum heat flow~\cite{Wang2018,Duan2020} and quantum heat engines~\cite{Kato2016}. Furthermore, HEOM has been used to study grand canonical electron baths, extending the approach to fermionic systems~\cite{Jin2007,Jin2008} and using barycentric methods for Anderson impurity models~\cite{Dan_Efficient_2023}. 
The extension of HEOM to tensor network methods has enabled the simulation of systems with considerably greater complexity, such as tens of sites, each with a Lorentzian environment~\cite{Takahashi_train_2024, takahashi_carrier_2025} or the simulation excitation energy transfer in the FMO complex, comprising seven sites\cite{Shi2018,Yan2021}, and in Photosystem I (PS I), comprising 96 sites~\cite{Yan2021}, considering environments of up to 74 undamped harmonic oscillators per site to model the low-frequency sector of experimentally estimated spectral densities, the PS I simulations used convergence parameters for a smaller FMO system.

Open source tensor network implementations of HEOM are available as part of the Python pyTTN package~\cite{Lindoy2025} and the Python TENSO package~\cite{Chen2025}.

\subsection{ML-MCTDH}\label{sec:MLMCTDH}

\subsubsection{Overview}

The Multi-Layer Multi-Configuration Time-Dependent Hartree (ML-MCTDH) method relies on the representation of the wave-function of the \{System + Environment\} as a tree tensor network. 
While not initially formulated in the language of tensor networks, ML-MCTDH is now recognized as belonging to the class of numerically exact tensor network methods.

Accounting for many-body correlations is a central challenge in quantum chemistry, where the interplay of electronic and vibrational degrees of freedom crucially determines the properties and dynamics of molecular systems. 
A common strategy is to start from a Hartree--Fock reference state (representing an uncorrelated solution) and include corrections from weakly correlated configurations. 
This approach underlies, for example, configuration interaction and coupled-cluster methods~\cite{cramer_essentials_2013}.
However, such methods are not guaranteed to converge to the exact dynamics and are computationally expensive.
An alternative method to capture correlations building up during the evolution of the quantum state consists of expanding the state on a time-dependent basis~\cite{meyer_multidimensional_2009}.
This is the approach at the heart of the Multi-Configuration Time-Dependent Hartree (MCTDH) method, which provides an ansatz for solving the Schrödinger equation for the full wave function of the system and its environment. 
It was developed during the final decade of the twentieth century within the field of molecular quantum dynamics~\cite{meyer_multi-configurational_1990, meyer_studying_2012, beck_multiconfiguration_2000}.
MCTDH rapidly became a standard tool in quantum chemistry, notably after it successfully reproduced the $S_2$ absorption spectrum of pyrazine while explicitly including all 24 vibrational modes of the molecule~\cite{worth_effect_1996, worth_relaxation_1998, raab_molecular_1999}.
However, as with other many-body approaches, its applicability was severely limited by the curse of dimensionality inherent to the structure of the Hilbert space.

A new formulation of MCTDH, the Multi-Layer Multi-Configuration Time-Dependent Hartree (ML-MCTDH) method~\cite{wang_multilayer_2003, manthe_multilayer_2008, vendrell_multilayer_2011, wang_multilayer_2015}, overcame this limitation by expressing the wave-functions of molecular vibrational quantum systems in terms of hierarchical, tree-like structures that are now recognized as tree tensor networks~\cite{larsson_tensor_2024}.
The ML-MCTDH method extended the tractable system size from a few dozen to thousands of degrees of freedom~\cite{larsson_accurate_2026}.
Nowadays, it stands as a reference approach for simulating molecular quantum dynamics, with further extensions enabling the treatment of finite-temperature effects and fermionic systems~\cite{manthe_wavepacket_2017, gatti_molecular_2026}.

\subsubsection{Derivation}

The ML-MCTDH method is designed to represent and evolve wave-functions that depend on discrete degrees of freedom.
However, many physical systems possess continuous variables, such as position or momentum.
To apply ML-MCTDH in these cases, the continuous variables must be discretized.
In the context of this review, the continuous degrees of freedom correspond to the normal modes of the environment.
When these modes are discretized, the resulting wave function is said, in the quantum chemistry literature, to be expressed in a discrete variable representation (DVR)~\cite{tannor_phase-space_2018}.
As for the ACE method (see Sec.~\ref{sec:ACE}), the choice of how to sample the environmental modes (such as the number of modes and their distribution) is typically left to the practitioner's discretion~\cite{Lindoy2025, wang_systematic_2001, chen_comparison_2026, walters_direct_2017}.

Once the problem is reduced to a set of $N$ degrees of freedom, the many-body wave function is the one given in \eqref{eq:fullHilbertstate}, which we recall here
\begin{equation}
    \ket{\psi} = \sum_{\{i_k\}}c_{i_1\ldots i_N}\ket{\phi_{i_1}}\ldots\ket{\phi_{i_N}}\ . \tag{\ref{eq:fullHilbertstate}}
\end{equation}
In quantum chemistry, this expression of the many-body wave-function is known as the full configuration interaction wave function, and the local basis $\{\ket{\phi_{i_k}}\}_{k=1}^N$ is known as the \emph{primitive basis}.

The ML-MCTDH ansatz consists of decomposing the tensor $c_{i_1\ldots i_N}$ into a hierarchical structure where the degrees of freedom have been grouped into $K<N$ clusters such that
\begin{equation}
    \ket{\psi} = \sum_{j_1\ldots j_K}A^{(1)}_{j_1\ldots j_K}\ket{\varphi^{(1;1)}_{j_1}}\ldots\ket{\varphi^{(1; K)}_{j_K}}\ , \label{eq:ML-MCTDH-1}
\end{equation}
and where the basis states of cluster $k\in\{1,\ldots K\}$, $\ket{\varphi^{(1; k)}_{j_k}}$ with ${j_k \in \{1,\ldots, D_k\}}$, are called \emph{single-particle functions}.
The single-particle functions are decomposed into lower-level single-particle functions following the same ansatz
\begin{equation}
    \ket{\varphi^{(1;k)}_{i}} = \sum_{j_1\ldots j_{K_k}}A^{(2; k)\ i}_{j_1\ldots j_{K_k}}\ket{\varphi^{(2;k;1)}_{j_1}}\ldots\ket{\varphi^{(2;k;K_k)}_{j_{K_k}}}\ ,\label{eq:ML-MCTDH-2}
\end{equation}
and so on.
Generically, we have decompositions of the following form, where $n$ labels layers in the hierarchy and we use the collective index ${(J) = (k;l;m;\ldots)}$,
\begin{equation}
    \ket{\varphi^{(n;J)}_i} = \sum_{j_1\ldots j_{K_J}}A^{(n+1;J)\ i}_{j_1\ldots j_{K_J}}\ket{\varphi^{(n+1;J;1)}_{j_1}}\ldots\ket{\varphi^{(n+1;J; K_J)}_{j_{K_J}}}\ ,\label{eq:ML-MCTDH-3}
\end{equation}
until reaching states belonging to the primitive basis
\begin{equation}
    \ket{\varphi^{(n;J; k)}_{i_k}} = \ket{\phi_{i_k}}\ .\label{eq:ML-MCTDH-4}
\end{equation}

\subsubsection{Tensor network representation}

The hierarchical decomposition in Eqs.~(\ref{eq:ML-MCTDH-1})-(\ref{eq:ML-MCTDH-4}) can naturally be understood as a tree tensor network structure where each $A^{(n;J)}$ is a rank-$K_J +1$ tensor connected with one `parent' tensor in layer $n-1$ and $K_J$ `child' tensors in layer $n+1$.
The numbers of single-particle functions $D_k$ are naturally identified as the bond dimensions of inner bonds.
The dimensions of the primitive basis are recognized as the local dimension of the physical legs of the TTN state.

The actual tree structure is left to one's choice, and different heuristics will be discussed in the next section.
For illustration purposes, let us consider an example with $N = 6$ degrees of freedom, for instance, the vibrational modes of ammonia~\cite{giri_full-dimensional_2011}.
Let us assume that we wish to decompose it into a four-layer ansatz as represented in \eqref{eq:example-MLMCTDH}
\begin{equation}
\begin{tikzpicture}[
    tensor/.style={
        circle,
        draw=black,
        fill=paletteBlue_Light3,
        minimum size=4mm,
        inner sep=0pt
    },
    line/.style={draw=black, thick},
    leg/.style={draw=black, thick}
]
\node (c) at (-0.25,3.6) {$\ket{\psi}$};
\node (eq) at (0.5,3.6) {$=$};

\draw[thin,dotted] (0.8,4) -- (6,4);
\draw[thin,dotted] (0.8,3) -- (6,3);
\draw[thin,dotted] (0.8,1.5) -- (6,1.5);
\draw[thin,dotted] (0.8,0.3) -- (6,0.3);
\draw[thin,dotted] (0.8,-0.9) -- (6,-0.9);

\node[tensor, fill=paletteBlue_Light3] (p5) at (3,0) {}; 
\node[left] at (p5.south west) {\tiny$A^{(4;2;1;1)}$};
\node[tensor, fill=paletteBlue_Light3] (p6) at (4,0) {};
\node[right] at (p6.south east) {\tiny$A^{(4;2;1;2)}$};

\node[tensor, fill=paletteBlue_Light2] (m1) at (1,1.2) {};
\node[left] at (m1.north west) {\tiny$A^{(3;1;1)}$};
\node[tensor, fill=paletteBlue_Light2] (m2) at (2,1.2) {};
\node[right] at (m2.north east) {\tiny$A^{(3;1;2)}$};
\node[tensor, regular polygon, regular polygon sides=3, fill=paletteBlue_Light2, minimum size=6mm] (m3) at (3.5,1.2) {};
\node[right] at (m3.east) {\tiny$A^{(3;2;1)}$};
\node[tensor, fill=paletteBlue_Light2] (m4) at (5,1.2) {};
\node[right] at (m4.east) {\tiny$A^{(3;2;2)}$};

\node[tensor, regular polygon, regular polygon sides=3, fill=paletteBlue_Light1, minimum size=6mm] (u1) at (1.5,2.4) {};
\node[right] at (u1.east) {\tiny$A^{(2;1)}$};
\node[tensor, regular polygon, regular polygon sides=3, fill=paletteBlue_Light1, minimum size=6mm] (u2) at (3.5,2.4) {};
\node[right] at (u2.east) {\tiny$A^{(2;2)}$};
\node[tensor, fill=paletteBlue_Light1] (u3) at (5.5,2.4) {};
\node[right] at (u3.east) {\tiny$A^{(2;3)}$};

\node[diamond, draw=black, fill=paletteYellow_Light1, minimum size=6mm, inner sep=0pt] (r) at (3.5,3.6) {};
\node[right] at (r.north east) {\tiny$A^{(1)}$};

\draw[line] (p5) -- (m3);
\draw[line] (p6) -- (m3);
\draw[line] (m1) -- (u1);
\draw[line] (m2) -- (u1);
\draw[line] (m3) -- (u2);
\draw[line] (m4) -- (u2);
\draw[line] (u1) -- (r);
\draw[line] (u2) -- (r);
\draw[line] (u3) -- (r);

\foreach \x in {p5,p6,m1,m2,m4,u3} {
    \draw[leg] (\x) -- ++(0,-0.8);
}

\node at (6.2,3) {.};
\end{tikzpicture}
\label{eq:example-MLMCTDH}
\end{equation}
Different shades of blue identify the different layers of the tree tensor network, which are also delineated by thin dotted lines.
In \eqref{eq:example-MLMCTDH}, the blue circles represent unitary tensors with a physical leg, blue triangles represent unitary tensors with inner legs only, while the head node (yellow diamond) corresponds to the core tensor obtained from the first Tucker decomposition.
In this representation, TTN is in a generalization of the mixed canonical form (see Sec.~\ref{sec:TN}) centered on the head node~\cite{evenbly_practical_2022}. 
In this ansatz, from left to right, the first two degrees of freedom are grouped in the third layer (from top to bottom); the next three degrees of freedom are grouped---two of them in the fourth layer, and the third one in the third layer---and the last degree of freedom is in the second layer.
If starting from an initially correlated state, this is achieved using a sequence of Tucker decompositions (see Sec.~\ref{sec:TN}); otherwise, if the initial state is a product state, this is directly written as the initial ansatz.

Time-evolution algorithms such as TDVP, introduced in Sec.~\ref{sec:TN}, can be directly applied to compute the dynamics of this state. 
In addition, time-propagation schemes specifically designed for ML-MCTDH have been proposed~\cite{lindoy_time_2021, lindoy_time_2021-1}.

The particular decomposition shown in \eqref{eq:example-MLMCTDH} is not unique; alternative tree structures with different numbers of layers or different groupings of the degrees of freedom could equally well be employed.

Given a discretization of the environment and a tree structure, the convergence parameters of ML-MCTDH are
\begin{itemize}
    \item the bond dimension $\chi$ of the tree tensor network,
    \item the truncation level of the environmental bosonic modes,
    \item the time step $\Delta t$ of the time-evolution.
\end{itemize}

ML-MCTDH is a pure-state method and therefore appears to be limited to zero-temperature dynamics.
Nevertheless, extensions of the method to address finite-temperature dynamics have been developed.
When the environment is Gaussian, an efficient approach is to exploit the purification of the density matrix via the thermo-field method~\cite{takahashi_thermo_1996, blasone_thermal_2011} to express the finite-temperature problem as a pure-state one~\cite{fischer_thermofield-based_2021}.
However, when Gaussianity cannot be assumed, one has to work in a density matrix framework.
Then, a straightforward, though numerically suboptimal, strategy consists of expressing the vectorized density operator (i.e., the density operator in Liouville space) as a TTN~\cite{van_haeften_propagating_2023}.
Another approach, amenable to parallelization but still computationally demanding, relies on statistical sampling of the density operator~\cite{weike_multi-configurational_2022}.

In the above discussion, we assumed that the degrees of freedom are bosonic; yet fermionic formulations of ML-MCTDH have also been developed~\cite{wang_numerically_2009, manthe_multi-layer_2017}.

Unlike some methods that aim at describing solely the influence of the environment on the system, the ML-MCTDH approach is not restricted to Gaussian environments.
It naturally accommodates anharmonicity.
The ML-MCTDH method has been applied, for instance, to the H\'enon-Heiles Hamiltonian, which contains cubic terms~\cite{vendrell_multilayer_2011}, electron transfer in an anharmonic environment with up to quartic terms in the free and interaction Hamiltonians~\cite{wang_quantum_2007}, or the Bose-Hubbard model~\cite{niermann_multi-layer_2024}.

\subsubsection{Tree structure}

Optimizing the tree structure, i.e., which modes to group together and how many layers should be considered, is a difficult task.
In most studies, the choice of the tree structure is mostly guided by intuition and trial and error.
Several structural optimization strategies have been proposed.
To show their diversity, we briefly summarize three examples of such structural optimization protocols.

\textcite{larsson_computing_2019} has formulated a `disentangling' approach.
Once some entangled state has been obtained with an initial tree that is based on an initial guess, the following steps are iterated several times:  
\begin{enumerate}
    \item Randomly choose two neighboring tensors $A$ and $B$ in the tree; 
    \item Make $A$ an orthogonality center;
    \item Contract the two tensors into a new tensor $C$;
    \item Randomly permute the legs of $C$;
    \item Perform an SVD for a randomly chosen bipartition of the indices of $C$;
    \item Keep the singular values that are larger than a chosen threshold. If the number of kept singular values is smaller than the previous bond dimension, replace $A$ and $B$ by $U$ and $SV^\dagger$ respectively.
\end{enumerate}  
For a given pair of neighboring tensors $A$ and $B$, this protocol should be repeated several times.

In a similar vein, \textcite{hikihara_automatic_2023} introduced a method where at each step of a DMRG sweep, SVDs are performed on the found local state for three different bipartitions of its indices. 
The retained bipartition is the one minimizing the entanglement entropy.

A different approach was proposed by \textcite{mendive-tapia_optimal_2023}, where correlations extracted from classical molecular dynamics simulations are used to guide the grouping of degrees of freedom. 
The procedure involves three main steps:
\begin{enumerate}
    \item Obtain a probability density using an approximate method, e.g., quasi-classical normal mode sampling, Metropolis sampling, or snapshots of molecular dynamics trajectories;
    \item From this discrete distribution, compute the corresponding correlation matrix for the degrees of freedom using a correlation measure of your choice;
    \item Estimate the `optimal' tree structure as the nearest tree-structured diagram of the correlation matrix through iterative clustering of the closest pairs of degrees of freedom.
\end{enumerate}



\begin{figure*}[t]
    \centering
    \includegraphics[width=\textwidth]{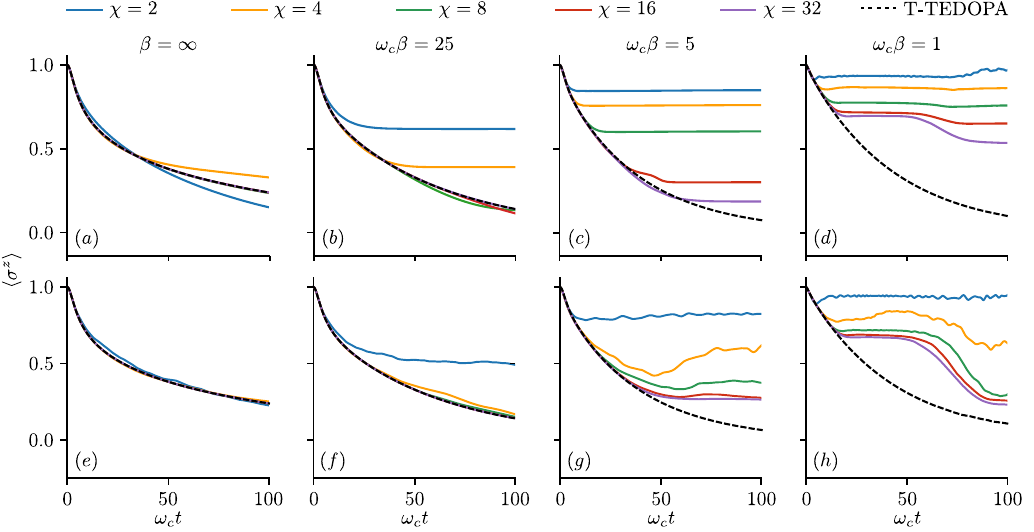}
    \caption{{Matrix product state (top row) and binary tree tensor network (bottom row) simulations of $\langle\sigma^z(t)\rangle$ for a spin--boson model with an Ohmic spectral density (exponential cutoff) at several inverse temperatures $\beta$. Each plot shows the simulation results for several bond dimensions $\chi$ (solid colored lines) in the case of (a)-(d) an MPS (star picture), or (e)-(h) a TTN. The reference converged results (black dashed lines) are obtained with T-TEDOPA simulations. The simulation parameters are $\epsilon = 0$, $\Delta = 0.2\omega_c$, and $\alpha=2.8$. Adapted from \textcite{Lindoy2025}.}}
    \label{fig:tree-convergence}
\end{figure*}
As these examples illustrate, a wide range of strategies can be devised for optimizing the tree structure of a TTN or ML-MCTDH state.
However, a common principle is that tensors in the network should ideally have at most three legs.
This follows from scaling considerations;
TTN states consist of two types of tensors: Tensors that only have inner bonds and tensors that have both physical and inner bonds.
The size of a tensor with $m$ inner legs of bond dimension $\chi$ scales as $\chi^m$, while that of a tensor with $m'$ inner bonds and one physical leg of dimension $d$ scales as $d \chi^{m'}$.
The tree structure minimizing the memory and computation costs, for fixed bond and physical dimensions, is one where $m = 3$ and $m'=1$.
These properties are satisfied in the TTN state in \eqref{eq:example-MLMCTDH}.
The memory cost of such a TTN state is $Nd \chi + M \chi^3$, where $M < N$ is the number of tensors without physical legs, whereas an MPS for the same state has a memory cost of $Nd \chi^2$.
It is assumed in the ML-MCTDH literature, from empirical numerical evidence, that compared to MPSs, TTN states often converge faster with respect to the bond dimension both in vibrational dynamics and in molecular electronic structure, i.e., $\chi_\text{MPS}> \chi_\text{tree}$.
This is illustrated in Fig.~\ref{fig:tree-convergence} where the convergence as a function of the bond dimension $\chi$ of MPS and TTN simulations of a spin--boson model (SBM) are compared.
The Hamiltonian ${H_S + H_I = \epsilon \sigma^z + \Delta\sigma^x + \sigma^z\int_{\mathbb{R}^+}d\omega \sqrt{J(\omega)}(a_\omega + a_\omega^\dagger)}$ and SD ${J(\omega) = 2\alpha\omega\exp(-\omega/\omega_c)}$ are considered, and a finite number of normal modes are sampled from the continuous bath (i.e. we consider the star geometry of Fig.~\ref{fig:star-geom}).
Figure~\ref{fig:tree-convergence} shows the expectation value $\langle\sigma^z(t)\rangle$, for several inverse temperatures $\beta$, simulated with a MPS -- Fig.~\ref{fig:tree-convergence}(a) to (d) -- or a binary TTN -- Fig.~\ref{fig:tree-convergence}(e) to (h)-- for several values of the tensor network state bond dimension $\chi$.
For this example, where the simulations are performed with the Hamiltonian in the star geometry, the TTN simulations converge faster than the MPS ones.
Note that this is no longer the case when using the chain-mapped geometry (see Sec.~\ref{sec:TEDOPA})~\cite{Lindoy2025}.

Furthermore, it is often assumed that the bond dimension required for a converged TTN state is similar to the physical dimension $\chi_\text{tree} \sim \mathcal{O}(d)$~\cite{larsson_tensor_2024}. 
However, it appears unclear when these conditions are fulfilled and under which circumstances; and counterexamples exist for model Hamiltonians such as the spin--boson model~\cite{lindoy_supplementary_2025}.
If such conditions are met, the typical scaling $d \chi_\text{MPS}^2$ of a tensor in an MPS is going to be larger than the typical scaling of a tensor in a binary tree, that is $\chi_\text{tree}^3$ and $d \chi_\text{tree}$.
Very few systematic comparisons of ML-MCTDH and MPS with the same numerical implementation have been performed, and they lead to contrasting results regarding the computational cost of MPS and TTN state~\cite{larsson_computing_2019, mainali_comparison_2021, dorfner_comparison_2024, Lindoy2025, li_further_2026}.

\subsubsection{Applications and implementations}
ML-MCTDH has been applied to a variety of problems such as, for example, photoinduced electron transfer in dye-semiconductor interfaces~\cite{kondov_theoretical_2006, kondov_quantum_2007, li_theoretical_2010, li_photoinduced_2012, li_quantum_2015}, and organic molecules~\cite{borrelli_quantum_2012}; electron transport in hetero-junctions~\cite{xie_full-dimensional_2015}, single molecule junctions~\cite{wang_multilayer_2013}, molecular quantum dots~\cite{albrecht_bistability_2012}, and cryptochromes~\cite{mendive-tapia_towards_2017, mendive-tapia_multidimensional_2018}.
Exciton dissociation in hetero-junctions has also been studied~\cite{dorfner_comparison_2024, li_further_2026, Lindoy2025}.
The method has also been applied to proton transfer in condensed-phase~\cite{craig_proton_2007}, in malonaldehyde~\cite{hammer_intramolecular_2011}, in salicylaldimine and porphycene~\cite{van_haeften_propagating_2023}; and to energy transfer in pigment-protein complexes~\cite{shibl_multilayer-mctdh_2017, schulze_multi-layer_2016}.
It has also been used to study heat transport in molecular junctions~\cite{velizhanin_heat_2008}.

Chemical reactivity has also been studied using ML-MCTDH, for instance to calculate chemical reaction rate constants~\cite{manthe_state_2008, laude_calculations_2020} and to study the influence of vibrational states on reactivity~\cite{ellerbrock_full-dimensional_2018}.

It has also been used to compute the absorption spectra of pyrazine~\cite{vendrell_multilayer_2011, larsson_dynamical_2017}  and of formaldehyde oxide~\cite{meng_full-dimensional_2014}; the photo-electron spectra of cations~\cite{meng_multilayer_2013}; and the infrared spectra of protonated water clusters~\cite{larsson_state-resolved_2022}.
Temperature effects on the ultrafast internal conversion dynamics in pyrazine have also been studied~\cite{fischer_thermofield-based_2021}.

Model Hamiltonians such as the Anderson impurity model~\cite{wang_multilayer_2018}; the spin--boson model~\cite{wang_multilayer_2017, Lindoy2025} and its phase transition~\cite{wang_quantum_2019}; a two-level system coupled to a spin bath~\cite{wang_dynamics_2012}; the H\'enon-Heiles model~\cite{vendrell_multilayer_2011}; or the Bose-Hubbard model~\cite{niermann_multi-layer_2024} have also been studied using ML-MCTDH.

In addition, we will mention that the method has been used to investigate singlet fission~\cite{zheng_ultrafast_2016}, the dynamics of ultra-cold atomic mixtures~\cite{kronke_non-equilibrium_2013, cao_multi-layer_2013}, the photodissociation of methyl iodide~\cite{westermann_photodissociation_2011}, the dynamics of quantum dots in a phononic environment~\cite{wilner_sub-ohmic_2015}, or the ground state of linear chains of rotors~\cite{mainali_comparison_2021}.

The most well-known packages for ML-MCTDH, the so-called `Heidelberg package'~\cite{worth_heidelberg_2022} and the \texttt{Quantics} package~\cite{worth_quantics_2020} are not open source.
The Python package \texttt{Renormalizer} implements tools for electron--phonon quantum dynamics with TTN states~\cite{renormalizer_2025}.
The Python Tree Tensor Network package (\texttt{pyTTN}) provides an implementation of ML-MCTDH in \texttt{Python}~\cite{Lindoy2025}.

\subsection{TEDOPA}\label{sec:TEDOPA}

\subsubsection{Overview}

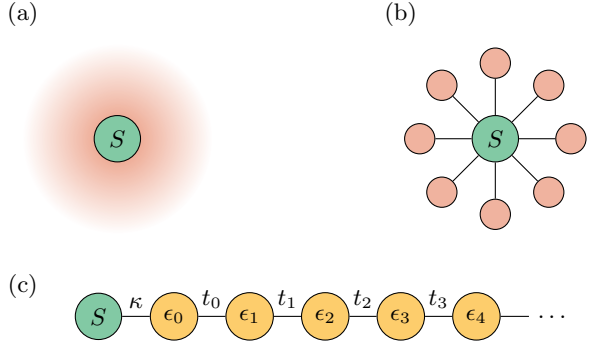
\begin{figure}
    \centering
    \begin{tikzpicture}
        \node at (-6,1.5) {(a)};
        \node at (-1,1.5) {(b)};
        \node at (-6,-2.1) {(c)}; 

        \begin{scope}[shift={(0.25,-0.15)}] 
            \shade[
                inner color=paletteRed_Light1,
                outer color=white,
                shading=radial
            ] (-5,0) circle[radius=1.25cm];

            \node[
                circle,
                minimum size=0.5cm,
                draw=black,
                fill=paletteGreen
            ] (S1) at (-5,0) {$S$};
        \end{scope}

        \begin{scope}[shift={(0.25,-0.15)}] 
            \node[
                circle,
                minimum size=0.5cm,
                draw=black,
                fill=paletteGreen
            ] (S2) at (0,0) {$S$};

            \foreach \x in {0,45,90,135,180,225,270,315} {
                \node (node\x) at (\x:1cm) [
                    circle,
                    minimum size=0.4cm,
                    draw=black,
                    fill=paletteRed_Light2
                ] {};
                \draw (S2) -- (node\x);
            }
        \end{scope}

        \node[
            circle,
            minimum size=0.5cm,
            draw=black,
            fill=paletteGreen
        ] (S3) at (-5,-2.5) {$S$};

        \node (dots) at (1,-2.5) {$\ldots$};

        \foreach \n in {0,...,4} {
            \node (\n) at (\n-4,-2.5) [
                circle,
                minimum size=0.4cm,
                draw=black,
                fill=paletteYellow_Light1
            ] {$\epsilon_\n$};
        }

        \draw (S3.east) -- node[above] {$\kappa$} (0.west);
        \foreach \n [evaluate=\n as \next using int(\n+1)] in {0,...,3} {
            \draw (\n.east) -- node[above] {$t_\n$} (\next.west);
        }
        \draw (4.east) -- (dots.west);

    \end{tikzpicture}
    \caption{Geometry of the system interacting with a bosonic or fermionic environment. Black lines represent interactions. (a) For a continuum of normal modes, depicted as a blurred red region. (b) For a finite set of modes, depicted as red discs. This structure, where the system couples to all the modes and the modes only to the system, is called the \emph{star geometry}. (c) After the chain mapping, the system interacts with a semi-infinite chain environment, shown as yellow discs. This \emph{chain geometry} is one-dimensional and well suited for an MPS representation.}
    \label{fig:star-geom}
\end{figure}

The \emph{Time Evolving Density operator with Orthogonal Polynomials Algorithm} (TEDOPA) uses a unitary transformation which exactly maps a continuous environment into a semi-infinite chain~\cite{chin_exact_2010, woods_mappings_2014}.
This unitary transformation, defined through a family of orthonormal polynomials and called `chain mapping', generates a discrete representation of the environment that naturally has a one-dimensional geometry and enables the simple retention of all the relevant bath modes.
These different geometries are shown in Fig.~\ref{fig:star-geom}.
The joint wave-function of the system and its environment can then be represented as an MPS (and the operators as MPOs)~\cite{prior_efficient_2010}, which has been shown to be typically more efficient than representing the star-picture environment as an MPS~\cite{hoeb_tensor-network_2019}.
The extension of the method to finite temperature is known as \emph{Thermalized} Time Evolving Density operator with Orthogonal Polynomials Algorithm (T-TEDOPA)~\cite{tamascelli_efficient_2019}.
The chain mapping procedure relies only on the fact that the harmonic bath modes are independent and that the interaction between the system and the environment is linear.
Therefore, the procedure can be applied to both bosonic environments and fermionic environments~\cite{nuseler_efficient_2020, kohn_efficient_2021} without additional complexity.

Historically, the idea of mapping a set of non-interacting degrees of freedom into a semi-infinite chain was introduced by \textcite{wilson_renormalization_1975} in the context of the \emph{Numerical Renormalization Group} (NRG).
It was later applied by \textcite{bulla_numerical_2003} to a bosonic environment in the context of the spin--boson model with SDs belonging to the Ohmic family~\cite{bulla_numerical_2005, bulla_numerical_2008, anders_equilibrium_2007}.
The NRG approach applied to OQS is to discretize the continuous environment and then perform an iterative numerical technique to bring the truncated finite Hamiltonian to the desired chain form.
However, this numerical transformation was unstable after a few dozen iterations.
The use of orthogonal polynomials~\cite{chin_exact_2010} simultaneously removes the need to discretize the environmental degrees of freedom and the numerical instabilities that occur in the NRG.

Because the orthogonal polynomial chain mapping itself and its extension to finite temperatures are unitary operations, the dynamics of the environment can also be directly investigated in the TEDOPA framework~\cite{riva_thermal_2023}.
Even though it is in general possible to indirectly reconstruct environment observables via the convolution of system observables, among the methods presented in this review, only TEDOPA, ML-MCTDH and ACE can \emph{directly} access the dynamics of environmental degrees of freedom.

Recent efforts have been made to connect the chain mapping approach to other methods commonly used to study OQS.
It has been shown that Markovian and non-Markovian quantum collision models can be derived rigorously using chain mapping techniques~\cite{lacroix_making_2025}.
Closer to the topics of this review, the outline of a process tensor construction (PT, see Sec.~\ref{sec:ACE}) starting from TEDOPA has been proposed by~\textcite{keeling_process_2026}.
The proposal for a PT-TEDOPA method opens many questions about the efficiency of such a procedure, the representation of the information encoded in the PT, and the separation of the environment into a non-Markovian and a Markovian part, leveraging the existence of the translationally invariant part of the chain (see below).
All these questions deserve to be investigated further.
Along similar lines, TEDOPA has been used to reconstruct non-Markovian dynamical maps in combination with the transfer tensor method in order to study long-time dynamics of open systems~\cite{rosenbach_efficient_2016}.\\

\subsubsection{Derivation}

\begin{figure}
    \centering
    \includegraphics[width=\columnwidth]{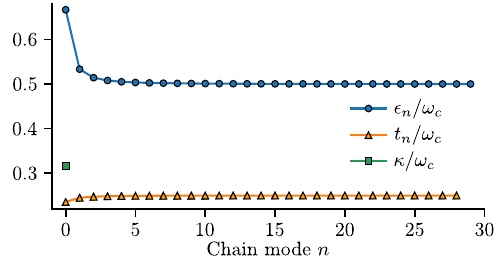}
    \caption{Chain coefficients for an Ohmic SD with a hard cutoff $J(\omega) = \alpha\omega H(\omega_c - \omega)$ with $\alpha=0.1$ and $H$ the Heaviside step function. $\{\epsilon_n\}_{n\in\mathbb{N}}$ are the chain on-site energies, $\{t_n\}_{n\in\mathbb{N}}$ are the chain hopping energies, and $\kappa$ is the coupling between the system and the first chain mode. After a few sites, the coefficients become homogeneous, making the chain translationally invariant.}
    \label{fig:chaincoeffs}
\end{figure}

Let us consider the Hamiltonian presented in Eqs.~(\ref{eq:Hbath}) and (\ref{eq:HSE}).
We can introduce a unitary transformation of the continuous normal modes $\aw$ to an infinite discrete set of interacting modes $\bn$~\cite{chin_exact_2010}
\begin{align}
    \aw &= \sum_{n=0}^{\infty} U_n(\omega)\bn = \sum_{n=0}^{\infty} \sqrt{J(\omega)}P_n(\omega)\bn\ , \label{eq:chain-mapping}
\end{align}
where $P_n(\omega)$ are real orthonormal polynomials defined such that
\begin{align}
    \int_{0}^{\infty} d\omega \,P_n(\omega)P_m(\omega)J(\omega) = \delta_{n,m}\ .
\end{align}
The orthonormality of the polynomials ensures the unitarity of the transformation defined in Eq.~(\ref{eq:chain-mapping}). The mapping from a continuous set of modes to a (still infinite) discrete set might seem counterintuitive; however, it is a direct consequence of the separability of the underlying Hilbert space.
Equation~(\ref{eq:chain-mapping}) can be seen as a change of basis between a generalized continuous basis of the Hilbert space (labeled with $\omega$) and a discrete one (labeled with $n$).
Performing the calculation for the interaction Hamiltonian yields
\begin{align}
    H_I &= O \int_{0}^\infty d\omega \sqrt{J(\omega)} \left(\aw +\awd\right)\\
    &= O \sum_{n=0}^\infty \left(\bn + \bnd\right) \kappa \int_{0}^\infty d\omega J(\omega) P_n(\omega)P_0 \\
    &= \kappa O \left({b}_0 + {b}_0^\dagger\right)\ ,
\end{align}
where $\kappa = \sqrt{\int_{0}^\infty\d\omega\ J(\omega)}$ and $P_0 = \kappa^{-1}$.
Applying the same transformation to $H_E$ leads to
\begin{align}
    H_E &= \sum_{n,m} \int_{0}^\infty d\omega\ \omega J(\omega) P_n(\omega) P_m(\omega) b_n^\dagger b_m\\
    &= \sum_{n,m} \int_{0}^\infty d\omega\ J(\omega) \Big(A_n P_{n-1}(\omega) + B_n P_n(\omega)\nonumber\\
    &\qquad\qquad\qquad+ C_n P_{n+1}(\omega) \Big) P_m(\omega) b_n^\dagger b_m\\
    &= \sum_{n} A_{n+1} b_{n+1}^\dagger b_{n} + B_n b_n^\dagger b_n + C_n b_n^\dagger b_{n+1}\ ,
\end{align}
where we used the recurrence relations between polynomials $\omega P_n = A_n P_{n-1} + B_n P_n + C_n P_{n+1}$ to be able to use the orthogonality condition.
Relations between recurrence coefficients can then be used to obtain a Hamiltonian with nearest-neighbor hopping.
The interested reader can find more details in~\textcite{chin_exact_2010}.
Under this transformation, the total Hamiltonian becomes
\begin{align}
    H = H_S &+ \sum_{n=0}^{\infty}\varepsilon_n\bnd\bn + t_n({b}_{n+1}^\dagger\bn + \hc)\nonumber\\
    &+ \kappa O({b}_0 + {b}_0^\dagger)\ . \label{eq:chain-Hamiltonian}
\end{align}
Hence, this mapping transforms the normal environment Hamiltonian into a Hamiltonian with on-site energies $\varepsilon_n$ and nearest-neighbor hopping energies $t_n$.
Another important consequence of this mapping is that now the system only interacts with the first mode $n=0$ of the chain-mapped environment.
We note that this is not the case when the chain mapping is applied in the interaction picture~\cite{nuomin_improving_2022}.
Figure~\ref{fig:star-geom}(c) shows a schematic drawing of the chain-mapped topology.
The chain coefficients $\varepsilon_n$, $t_n$, and the coupling $\kappa$ depend solely on the spectral density.
They can sometimes be calculated analytically~\cite{chin_exact_2010}, for instance, for the family of Ohmic spectral densities; otherwise, they are computed numerically~\cite{gautschi_algorithm_1994}.
For spectral densities belonging to the Szegő class~\cite{szego_orthogonal_1975}---physically speaking, SDs that are non-negative, without spectral gaps and with a finite support---the chain coefficients converge to asymptotic values for large modes $n$, thus making the chain translationally invariant for all practical purposes after a given number of sites.
These asymptotic values are universal as they depend solely on the difference between the maximum and minimum frequencies of the support of the spectral density
\begin{align}
    \lim_{n\to\infty} \epsilon_n &= \frac{\omega_\text{max} + \omega_\text{min}}{2}\ ,\\
    \lim_{n\to\infty} t_n &= \frac{\omega_\text{max} - \omega_\text{min}}{4}\ .\label{eq:t-infty}
\end{align}
Figure~\ref{fig:chaincoeffs} shows the chain coefficients of an Ohmic spectral density.
The dispersion relation of the asymptotic part of the chain is gapless and can support excitations over the full spectral range of the original environment.
Intuitively, this translationally invariant region of the chain is responsible for the irreversibility of the system dynamics, whereas the `disordered' early part of the chain is responsible for the non-Markovian dynamics.
Of course, in practice, the chain has to be truncated to a finite number of modes.
It has been shown that the truncation of the chain corresponds to an optimal discretization of the normal modes of the environment~\cite{de_vega_how_2015}.
How to do so will be discussed in Sec.~\ref{sec:TEPOPA_chain_length}.
In the case where the environment is actually comprised of a finite set of discrete modes, as depicted in Fig.~\ref{fig:star-geom}(b), the analytical determination of the polynomials is still possible for specific spectral densities~\cite{chin_exact_2010}; otherwise, the chain mapping can be performed using the Lanczos algorithm to tri-diagonalize the free environment Hamiltonian~\cite{golub_matrix_1996}.\\

\subsubsection{T-TEDOPA}\label{sec:T-TEDOPA}

The straightforward extension of the method to finite temperature would be to represent the density operator of the \{System + Chain\} as an MPO.
This description is naturally more complex than the zero-temperature case because of the squaring of the Hilbert space dimension when moving to density operators, and the non-zero initial number of environmental excitations in a Gibbs state at inverse temperature $\beta \neq \infty$.
Remarkably, TEDOPA can be extended to finite temperature while keeping a wave-function representation of the state of the \{System + Chain\}.
This can be done by combining the chain mapping approach with a thermo-field transformation~\cite{takahashi_thermo_1996, blasone_thermal_2011} of the finite-temperature environment to obtain \emph{two} chains at zero-temperature~\cite{de_vega_thermofield-based_2015}; or combining chain mapping with a transformation of the finite-temperature environment to an extended environment to obtain a \emph{single} chain at zero-temperature~\cite{tamascelli_efficient_2019}.
The latter approach is the one we will present here.
As we have seen in Sec.~\ref{sec:OQS}, for Gaussian environments the dynamics of the reduced system only depend on the bath correlation function $C(\tau)$.
Thus, the exact nature of the environment is not essential for calculating the system's observables, and the real microscopic bath can be replaced by a fictitious one that is more amenable to numerical approaches~\cite{TamascelliPRL2018}.
Using a zero-temperature bath with the same correlation function as the finite-temperature one, it is possible to keep an MPS representation of the quantum state and use TEDOPA.
For Gaussian baths, this can always be achieved by replacing the spectral density by the Fourier transform $C(\omega)$ of the environment correlation function $C(t)$ (as defined in \eqref{eq:Environment_Correlation_Function})
\begin{align}
    J_\beta(\omega)  &\overset{\text{def}}{=} C(\omega)\nonumber\\
    &= \big(1 + n_\beta(\omega)\big) J(\omega) H(\omega) \nonumber\\
    &\qquad+ n_\beta(|\omega|)J(|\omega|)H(-\omega)\label{eq:thermalizedSD}\ ,
\end{align}
where $\omega$ covers the entire real line and $H$ is the Heaviside step function.
We refer to \eqref{eq:thermalizedSD} as the \emph{thermalized} spectral densities~\cite{tamascelli_efficient_2019}.
This new fictitious bath at zero temperature replaces absorption of energy at frequency $\omega$ from the bath into the system by an emission in the bath mode of frequency $-\omega$.
The corresponding bath and interaction Hamiltonians with this `extended' environment reads as,
\begin{align}
    H_E^\text{ext} + H_I^\text{ext} &= \int_{-\infty}^\infty d\omega\ \omega\awd\aw \nonumber\\
    &+ O \int_{-\infty}^\infty d\omega\ \sqrt{J_\beta(\omega)}\left(\aw + \awd\right)\ .
\end{align}
The new effective environment then has a vacuum state as its initial condition.
From this extended Hamiltonian, the transformation in \eqref{eq:chain-mapping} can be applied, considering the thermalized spectral density.
By introducing negative-frequency modes, the method constructs a spectral density that embeds the thermal occupation of the environment directly into the chain's vacuum initial state. 
This allows the simulation to proceed entirely in a pure-state formalism, eliminating the need for sampling mixed initial conditions.
\\

\subsubsection{Tensor network representation}
The Hamiltonian in Eq.~(\ref{eq:chain-Hamiltonian}) has a one-dimensional chain topology naturally.
This makes the representation of the joint \{System + Environment\} wave-function as an MPS very natural.
After truncating to a finite number $N_m$ of modes, as discussed below, we can write the MPS
\begin{align}
    \label{eq:mps_tedopa}
    \begin{tikzpicture}
        [tensor/.style={circle,draw=black,fill=paletteBlue_Light3,
                     inner sep=0pt,minimum size=7mm}]
        \node (c) at (-0.25,0) {$\ket{\psi}$};
        \node (eq) at (0.25,0) {$=$} ;
        \node[circle,draw=black,fill=paletteGreen,inner sep=0pt,minimum size=7mm] (T1) at (1,0) {$T_S$} ;
        \node (i1) at (1, -1) {} ;
        \node[tensor] (T2) at (2,0) {$T_0$} ;
        \node (i2) at (2, -1) {} ;
        \node[tensor] (T3) at (3,0) {$T_1$} ;
        \node (i3) at (3, -1) {} ;
        \node (dots) at (4,0) {$\ldots$} ;
        \node[tensor] (TN) at (5.4,0) {$T_{N_m}$} ;
        \node (iN) at (5.4, -1) {} ;
        \draw (i1) -- (T1.south) ; \draw (T1.east) -- (T2.west) ; 
        \draw (i2) -- (T2.south) ; \draw (T2.east) -- (T3.west) ;
        \draw (i3) -- (T3.south) ; \draw (T3.east) -- (dots.west) ;
        \draw (iN) -- (TN.south) ; \draw (dots.east) -- (TN.west) ;
        \node at (6,0) {.} ;
    \end{tikzpicture}
\end{align}

When the system couples to several environments, a chain-mapping can be applied to each of them independently, and the resulting state can be represented as a TTN state where the system is the head node of the tree and each branch is one chain-mapped environment~\cite{schroder_tensor_2019, dunnett_influence_2021, lacroix_connectivity_2024}
\begin{align}
\begin{tikzpicture}
    [tensor1/.style={circle,draw=black,fill=paletteBlue_Light3,
                    inner sep=0pt,minimum size=7mm},
     tensor2/.style={circle,draw=black,fill=paletteBlue_Light3,
                    inner sep=0pt,minimum size=7mm},
     tensor3/.style={circle,draw=black,fill=paletteBlue_Light3,
                    inner sep=0pt,minimum size=7mm},
     system/.style={circle,draw=black,fill=paletteGreen,
                    inner sep=0pt,minimum size=7mm}]
    \node (c) at (-0.25,0) {$\ket{\psi}$};
    \node (eq) at (0.25,0) {$=$};
    \node[system] (TS) at (1,0) {$T_S$};
    \node (legS) at (1,-0.8) {};
    \draw (TS.south) -- (legS);
    \node[tensor1] (A1) at (2,1.5) {$T_0^{(1)}$};
    \node[tensor1] (A2) at (3,1.5) {$T_1^{(1)}$};
    \node (dotsA) at (4,1.5) {$\ldots$};
    \node[tensor1] (AN) at (5.4,1.5) {$T_{N_1}^{(1)}$};
    \foreach \x in {A1,A2,AN}
        \draw (\x.south) -- ++(0,-0.3);
    \draw (TS) -- (A1.west);
    \draw (A1.east) -- (A2.west);
    \draw (A2.east) -- (dotsA.west);
    \draw (dotsA.east) -- (AN.west);
    \node[tensor2] (B1) at (2,0) {$T_0^{(2)}$};
    \node[tensor2] (B2) at (3,0) {$T_1^{(2)}$};
    \node (dotsB) at (4,0) {$\ldots$};
    \node[tensor2] (BN) at (5.4,0) {$T_{N_2}^{(2)}$};
    \foreach \x in {B1,B2,BN}
        \draw (\x.south) -- ++(0,-0.3);
    \draw (TS.east) -- (B1.west);
    \draw (B1.east) -- (B2.west);
    \draw (B2.east) -- (dotsB.west);
    \draw (dotsB.east) -- (BN.west);
    \node[tensor3] (C1) at (2,-1.5) {$T_0^{(3)}$};
    \node[tensor3] (C2) at (3,-1.5) {$T_1^{(3)}$};
    \node (dotsC) at (4,-1.5) {$\ldots$};
    \node[tensor3] (CN) at (5.4,-1.5) {$T_{N_3}^{(3)}$};
    \foreach \x in {C1,C2,CN}
        \draw (\x.south) -- ++(0,-0.3);
    \draw (TS) -- (C1.west);
    \draw (C1.east) -- (C2.west);
    \draw (C2.east) -- (dotsC.west);
    \draw (dotsC.east) -- (CN.west);
    \node at (6,0) {.};
\end{tikzpicture}
\end{align}

The TEDOPA method has the following four convergence parameters:
\begin{itemize}
    \item the chain length $N_m$,
    \item the local Hilbert space dimension $d$ of the chain modes (in the bosonic case),
    \item the bond dimension $\chi$ of the MPS,
    \item the time-step $\Delta t$ of the time-evolution.
\end{itemize}

\subsubsection{Chain length}
\label{sec:TEPOPA_chain_length}
\begin{figure*}
    \centering
    \includegraphics[width=\textwidth]{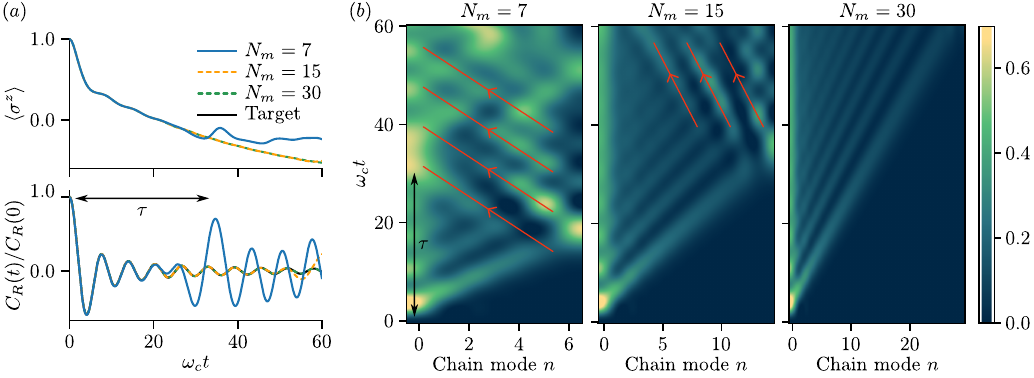}
    \caption{Dynamics of a two-level system for different chain length $N_m \in \{7, 15, 30\}$ and a fixed simulation time $\omega_cT = 60$. (a-top) Expectation value of $\sigma^z$. (a-bottom) Real part of the BCF generated by the chains compared to the analytical and target BCF. (b) Chain populations $b_n^\dagger b_n$. When the chain is too short, excitations are reflected toward the system. The red arrows are guides for the eyes. If this excitation has the time to reach the system, this unphysical interaction will lead to erroneous results. This behavior can also be interpreted as a consequence of using a chain-mapped environment whose BCF, from a certain time, deviates from the true BCF because of the truncation.\label{fig:chain-length}}
\end{figure*}
It is crucial in the TEDOPA approach to consider chains that are long enough to guarantee that the truncation does not impact the system dynamics because of unphysical boundary effects~\cite{trivedi_convergence_2021, woods_simulating_2015}.
The optimal choice is given by ${N_m = v T}$ where the characteristic speed $v = (\omega_\text{max} - \omega_\text{min})/4$ of the propagation of excitations in the chain is the asymptotic nearest-neighbor coupling in \eqref{eq:t-infty} and $T$ is the simulation time, with $\omega_\text{max/min}$ the largest/smallest frequency of the (thermalized) spectral density~\cite{woods_simulating_2015, woods_dynamical_2016, de_vega_how_2015}.

As an example, we here consider a two-level system interacting with a bosonic environment described by the following Hamiltonian
\begin{align}
    H = &\frac{\omega_0}{2}\sigma^z + \int_0^\infty d\omega \omega \awd\aw\nonumber\\
    &+ \frac{\sigma^x + \sigma^z}{\sqrt{2}}\int_0^\infty d\omega \sqrt{J(\omega)}(\aw + \awd)\ ,
\end{align}
where $\omega_0 = 0.2\omega_c$ and we considered an Ohmic bath spectral density with a hard cutoff $J(\omega) = 2\alpha\omega H(\omega_c - \omega)$ with $\omega_c$ the bath cutoff frequency, $\alpha = 0.1$ the Kondo parameter, and $H(\omega)$ the Heaviside step function.
With this set of parameters, the minimal length we presented yields ${N_m = vT = 15}$.

Figure~\ref{fig:chain-length}(a) shows how the system dynamics are affected by chains of different lengths, considering truncations smaller than ($N_m=7$), equal to ($N_m=15$), and larger than ($N_m=30$) the minimum required value. For $N_m=7$, the system dynamics begin to deviate from the converged result at approximately half the simulation time, whereas for $N_m=15$, they remain converged until the final few time steps. For $N_m=30$, the dynamics are fully converged.

This behavior can be rationalized by examining the populations of the chain modes, shown in Figure~\ref{fig:chain-length}(b). For $N_m=7$, the environmental excitations initially generated by the interaction with the system have sufficient time to reach the end of the chain, be reflected, and interact with the system again at approximately half the simulation time. For $N_m=15$, the reflected environmental excitations interact with the system only during the final few time steps, whereas for $N_m=30$, no reflections occur over the simulated timescale.

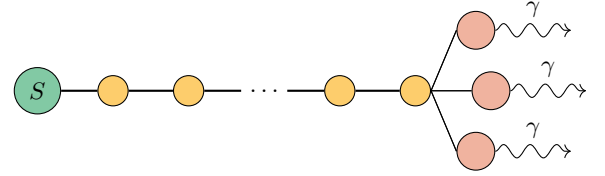
\begin{figure}
    \centering
        \begin{tikzpicture}
            \node[circle, minimum size = 0.5cm, draw=black, fill=paletteGreen] (S) at (-5,0) {$S$} ;
            \node (dots) at (-2,0) {$\ldots$} ;
            \foreach \n in {0,...,1} 
            {
             \node (\n) at (\n-4, 0) [draw=black, fill=paletteYellow_Light1, circle, minimum size=0.4cm] {} ;
            }
            \foreach \n in {3,...,4} 
            {
             \node (\n) at (\n-4, 0) [draw=black, fill=paletteYellow_Light1, circle, minimum size=0.4cm] {} ;
            }
            \draw[thick] (S.east) -- (0.west) ; 
            \draw[thick] (0.east) -- (1.west);
            \draw[thick] (1.east) -- (dots.west);
            \draw[thick] (dots.east) -- (3.west);
            \draw[thick] (3.east) -- (4.west);
            \node[circle, minimum size = 0.5cm, draw=black, fill=paletteRed_Light2] (MC1) at (1, 0) {};
            \node[circle, minimum size = 0.5cm, draw=black, fill=paletteRed_Light2] (MC2) at (0.8, 0.8) {};
            \node[circle, minimum size = 0.5cm, draw=black, fill=paletteRed_Light2] (MC3) at (0.8, -0.8) {};
            \draw (4.east) -- (MC1.west) ;
            \draw (4.east) -- (MC2.west) ;
            \draw (4.east) -- (MC3.west) ;
            \draw (4.east) -- (MC1.west);
            \draw (4.east) -- (MC2.west);
            \draw (4.east) -- (MC3.west);
            
            \draw[->,decorate,decoration=snake,draw=black]
                (MC1.east) -- node[above, yshift=2pt] {$\gamma$} ++(1.0,0);
            
            \draw[->,decorate,decoration=snake,draw=black]
                (MC2.east) -- node[above, yshift=2pt] {$\gamma$} ++(1.0,0.0);
            
            \draw[->,decorate,decoration=snake,draw=black]
                (MC3.east) -- node[above, yshift=2pt] {$\gamma$} ++(1.0,0);
        \end{tikzpicture}
    \caption{Schematic representation of the Markovian closure. The translationally invariant part of the semi-infinite chain is replaced by a finite set of sites undergoing Lindblad damping such that the overall spectral density of the environment remains unchanged. This procedure enables efficient long-time simulations using a finite chain.}
    \label{fig:MC}
\end{figure}

For long-time simulations, the required chain length and bond dimension typically increase, which may lead to computational bottlenecks.
However, as discussed previously, the chain coefficients asymptotically become translationally invariant, such that any excitation that propagates far enough into the chain cannot be back-scattered (except at the truncated end).
Thus, the translationally invariant part of the chain can be replaced by a set of sites undergoing Lindblad dissipation (i.e., sinks), if the dissipation rates are chosen such that the overall bath correlation function of the \{truncated chain + sinks\} is the same as the original one~\cite{TamascelliPRL2018}.
This construction, called Markovian closure~\cite{nuseler_fingerprint_2022, ferracin_spectral_2024}, allows the circumvention of the linear scaling of the chain length with simulation time at the cost of moving to a density matrix formalism to allow for the Lindblad dissipator. Furthermore, the correspondence to the original normal modes is lost.
The addition of a Markovian closure to the chain-mapped environment formally makes it a Markovian embedding, i.e., an extended system following a GKSL master equation.
This procedure is shown schematically in Fig.~\ref{fig:MC}.\\

\subsubsection{Applications and implementations}
TEDOPA has been applied, for instance, to the transport of electronic excitations in the presence of structured vibrational environments~\cite{prior_efficient_2010}, photonic crystals~\cite{prior_quantum_2013}, exciton transport in molecular systems~\cite{oviedo-casado_phase-dependent_2015} and vibration-induced coherence~\cite{Chin2013}.

TEDOPA has been used in chemical physics to study intramolecular charge transfer of a covalently linked tetracene dimer~\cite{alvertis_non-equilibrium_2019}, excited-state proton transfer~\cite{le_de_extending_2024, le_de_impact_2025} or photoinduced ultrafast vibration-coupled electron transfer reactions~\cite{wang_numerically_2025}.

The ability to take into account complex environments made it recently widely used for the calculation of absorption spectra of chromophores such as methylene blue~\cite{dunnett_influence_2021} and proflavine~\cite{hunter_environmentally_2024}; molecules in solutions such as pyrazine~\cite{lambertson_computing_2024} or indole~\cite{bashirova_importance_2026}; and pigment-protein complexes such as the water-soluble chlorophyll-binding protein of cauliflower~\cite{nuseler_fingerprint_2022, caycedo-soler_exact_2022}, the special pair of bacterial reaction centers~\cite{caycedo-soler_exact_2022}, and FMO dimers~\cite{LorenzoniPRL2024}.

Thanks to its versatility, TEDOPA has also been used in varied context such as quantum metrology~\cite{tamascelli_quantum_2020}, the formation of non-equilibrium steady states~\cite{dunnett_matrix_2021, riva_thermal_2023}, spatially correlated noise in nano-devices~\cite{lacroix_unveiling_2021, lacroix_non-markovian_2024}, relativistic light--matter interaction~\cite{jonsson_chain-mapping_2024}, singlet fission~\cite{nuomin_efficient_2024}, or the finite-power efficiency of quantum heat engines~\cite{weber_thermodynamic_2026}.

The fermionic version of TEDOPA has been applied to impurity models such as resonant level models~\cite{nuseler_efficient_2020} and the single-impurity Anderson model~\cite{kohn_efficient_2021, kohn_quench_2022, de_vega_thermofield-based_2015, ferracin_spectral_2024}.

TEDOPA has been used to benchmark the two-particle irreducible effective action approach in Schwinger-Keldysh field theory~\cite{reyes-osorio_schwingerkeldysh_2026}, and exact methods like HEOM~\cite{le_de_managing_2024, le_de_revisiting_2026}, DAMPF~\cite{lorenzoni_full_2025, Lemmer_2018}, and quantum collision models~\cite{lacroix_making_2025}.
It has also been combined with other methods such as the periodically refreshed bath approach~\cite{purkayastha_periodically_2021}, and transfer tensors~\cite{rosenbach_efficient_2016}.

Quantum algorithmic implementations of TEDOPA, which seek to digitally simulate open system dynamics on quantum hardware, have been proposed, namely Q-TEDOPA~\cite{guimaraes_digital_2024} and a combination with projected-variational quantum dynamics~\cite{nishi_simulation_2024}.

An implementation of TEDOPA is in the open-source \texttt{MPSDynamics.jl} Julia package~\cite{dunnett_mpsdynamicsjl_2025, lacroix_mpsdynamicsjl_2024}, which provides built-in methods to compute the chain coefficients, generate an MPS representation of the quantum state, and time-evolve them with several variants of the TDVP algorithm.
Another implementation in Python is also provided in the open-source \texttt{pyTTN} (Python Tree Tensor Network) package~\cite{Lindoy2025}.

\subsection{TEMPO}
\label{sec:TEMPO}
\subsubsection{Overview}
Instead of propagating the system via the construction of effective environment degrees of freedom, the influence of the environment on the reduced system can be expanded into a TN with two distinct \textit{temporal} directions: evolution and memory.
The starting point is the path-integral formulation of Feynman and Vernon~\cite{feynman_theory_1963} discussed in Sec.~\ref{sec:OQS}.
Expressing the influence functional as a causal TN enables an efficient representation of the flow of information in and out of the reduced system during the evolution and is the foundation of the Time-Evolving Matrix Product Operator (TEMPO) formalism~\cite{strathearn_efficient_2018}. Thus, the environment memory time naturally emerges as a convergence parameter that determines the size of the TN \cite{strathearn_efficient_2017}. For the sake of clarity, we will refer to this original method as ADT-TEMPO in this review, clarifying its core idea of reformulating the so-called \textit{augmented density tensor} (ADT) \cite{makri_tensor_1995, makri_tensor_1995-1} of the QUAPI method \cite{makarov_path_1994}, a representation of the system and its memory, as an MPO.

A memory-efficient implementation of this TN then relies on the compression of TN bonds that represent memory effects propagating through the network.
Multiple ways of compression lead to different resultant methods, some of which result in the construction of a process tensor, which will be presented in this section. The optimal choice of method might depend on the requirements of the specific problem that is to be addressed.

First, we describe the ADT-TEMPO algorithm \cite{ strathearn_efficient_2018} and thereby introduce the structure of the causal TN that forms the basis of all methods discussed in this section. We continue to show how different ways of contracting this network can be used to generate a process tensor (PT) in MPO form. The first approach for this has been termed PT-TEMPO \cite{jorgensen_exploiting_2019} and is presented in Sec.~\ref{sec:pt-tempo}. Notably, the causal TN exhibits repeating patterns and a time-translational symmetry which can be used to build the PT more efficiently, either using a divide-and-conquer scheme \cite{cygorek_sublinear_2024}, presented in Sec.~\ref{sec:divide-and-conquer}, or using infinite time-evolving block decimation (iTEBD) to build a time-translationally invariant PT as discussed in \cite{link_open_2024}, see Sec.~\ref{sec:link_struntz}.

\subsubsection{ADT-TEMPO}
The general idea of ADT-TEMPO is to construct a causal tensor network that can be used for the time-local propagation of the ADT, that stores the current system state and relevant previous states within the memory time, as an MPO.

To begin the derivation of this method, consider the time evolution of an open system, represented by the reduced density operator ${\rho}_\alpha$ in Liouville space.
We assume coupling to a continuous boson environment via the operator $O$, and we choose a basis $\{\ket{a}\}$ such that $O$ is diagonal, i.e., $O = \sum_{a}O_{aa} |a\rangle\langle a|$. This choice of basis is central to the standard construction of the causal tensor network. However, extensions to non-diagonal couplings do exist, for example in Refs.~\cite{zhang_time-evolving_2025, richter_enhanced_2022}, but their derivation lies outside the scope of this review. Dynamics of the system are captured by a Liouville operator $\mathcal{L}(t)$, which can depend on time. Using the Trotter-decomposed time evolution given by~\eqref{ReducedDensityMatrix} and the
Feynman-Vernon path integral formulation \eqref{eq:feynman-vernon-path-integral}, the influence tensor can be written as \cite{strathearn_efficient_2018}
\begin{align}
    \mathcal{I}^{(\alpha_n, \alpha^\prime_n)\dots (\alpha_1, \alpha^\prime_1) }  &= \left(\prod_{n^\prime=1}^n\mathcal{B}^{\alpha_{n^\prime}\dots\alpha_1}\right)\delta_{\alpha_n\alpha^\prime_n}\dots\delta_{\alpha_1\alpha^\prime_1}\label{eq:influence_funcitional_tempo},\\
    \mathcal{B}^{\alpha_n\dots\alpha_1} &= \prod^{n-1}_{k = 0}I_{k}(\alpha_n, \alpha_{n-k}),\label{eq:TEMPO_PathIntegral_row}
\end{align}
where the Kronecker deltas in \eqref{eq:influence_funcitional_tempo} result from the formulation in the basis that makes the environment coupling operator diagonal.
In this basis
\begin{align}
    I_{k}(\alpha, \beta) = \exp\left(-O_\alpha^{c}(O_\beta^{\jw{c}}\Re[\eta_k] + iO_\beta^{a}\Im[\eta_k])\right),\label{eq:TEMPO_InfluenceTensors}
\end{align}
with  $O^{c}_\alpha = O_{aa}- O_{bb}$ and $O^{a}_\alpha = O_{aa} + O_{bb}$, where ${\alpha = (a, b)}$ with the Hilbert space indices $a$ and $b$. Compared to \eqref{eq:feynman-vernon-path-integral}, this is an indexed version of the coupling operator in the interaction-picture formulation. Furthermore,
\begin{align}
    \eta_k = \begin{cases}
        \int_0^{\Delta t}dt^\prime\int_0^{\Delta t}dt^{\prime\prime}C(t^\prime-t^{\prime\prime} + k\Delta t),~ k > 0 \\
        \int_0^{\Delta t}dt^\prime\int_0^{t^\prime}dt^{\prime\prime}C(t^\prime-t^{\prime\prime}),~k=0
    \end{cases},
\end{align}
$C(t)$ being the environment correlation function defined in \eqref{eq:Environment_Correlation_Function}.

\begin{figure*}[t]
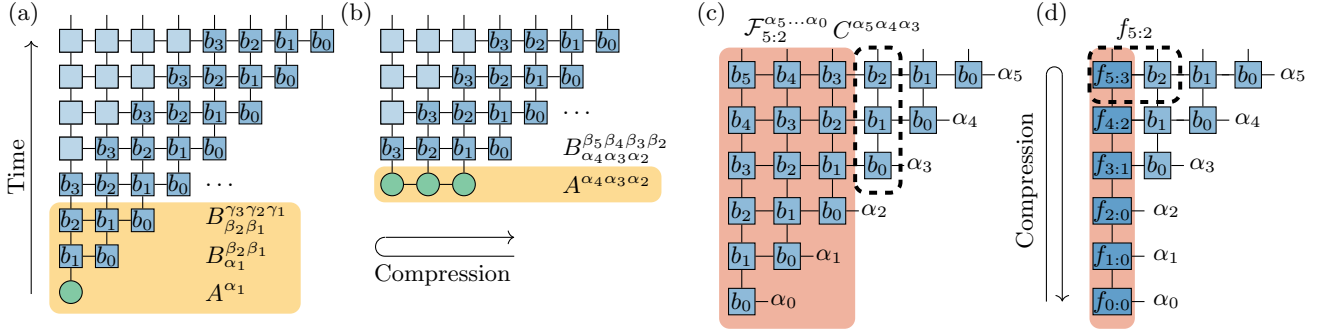

    \centering
    \include{Review/figs/Fig_TEMPO_1v3}
    \caption{(a) The full TEMPO network. Blue rectangles indicate the tensors as defined in Eqs.~(\ref{eq:TEMPO_InfluenceTensor1})-(\ref{eq:TEMPO_InfluenceTensor3}). The initial ADT is represented as a green circle. Grayed-out contributions indicate parts of the network that get neglected for a memory cutoff after three time steps. (b) The propagation phase of the ADT, which is an MPO represented by green circles. Inner bonds of the ADT (horizontal lines) are reduced using MPO compression techniques. (c) Construction of the PT in the PT-TEMPO scheme by the combination of vertical slices of the TEMPO network. Indices $\alpha_n$ correspond to the outer bonds of the resulting process tensor and connect to the reduced system density operator. (d) Contraction and compression in the MPO representation of the PT. }\label{fig:TEMPO_1}
\end{figure*}

While this formulation establishes the basis of QUAPI (as discussed in section \ref{sec:OQS}), the exponential growth of the retained paths with increasing memory time can be overcome by representing the influence functional as a tensor network. This was first realized by \textcite{strathearn_efficient_2018} and can be done by introducing suitable Kronecker deltas to expand the number of indices (legs) of each $I_k(\alpha, \beta),$~\eqref{eq:TEMPO_InfluenceTensors}. Thus, at every time step $t_n = n\Delta t$ one can define the rank-4 tensors 
\begin{align}
    [b_k]^{d, \alpha}_{d^\prime, \alpha^\prime} = \begin{cases}
        \delta_{d, d^\prime}\delta_{\alpha, \alpha^\prime}I_k(d, \alpha),~1<k < n-1\\
        \delta_{d, d^\prime}\delta_{\alpha, \alpha^\prime}I_k(d, \alpha)M^{d\alpha}, k = 1
    \end{cases}\label{eq:TEMPO_InfluenceTensor1}
\end{align}
within the network and the rank-2 and rank-3 tensors
\begin{align}
    [b_0]_{d}^\alpha &= \delta_{d, \alpha} I_{0}(\alpha, \alpha),\label{eq:TEMPO_InfluenceTensor2}\\
    [b_{n-1}]_{\alpha^\prime}^{d, \alpha} &= \delta_{\alpha, \alpha^\prime}I_{n-1}(d, \alpha)\label{eq:TEMPO_InfluenceTensor3}
\end{align}
at the edges of the network. The internal system dynamics are captured by $M^{d\alpha}$, as defined in \eqref{eq:internal_dynamics}.
Graphically, this can be represented as
\begin{align}
   \raisebox{-0.5em}{%
    \begin{tikzpicture}[scale=0.2]
      \node[draw, rectangle, fill=paletteBlue_Light2, minimum    size=0.5cm, inner sep=0pt] (n) at  ({-15cm},{0}) {$b_{n-1}$};
      \draw (n.north) -- ++(0,1cm);
      \draw (n.east) -- ++(1cm, 0);
      \draw (n.south) -- ++(0, -1cm);
      \node[] at (-15cm, 3cm) {$\alpha$};
      \node[] at (-11.75cm, 0cm) {$d$};
      \node[] at (-15cm, -3cm) {$\alpha^\prime$};
      \node[draw, rectangle, fill=paletteBlue_Light2, minimum    size=0.5cm, inner sep=0pt] (na) at  ({0},{0}) {$b_{k}$};
      \draw (na.north) -- ++(0,1cm);
      \draw (na.west) -- ++(-1cm, 0);
      \draw (na.east) -- ++(1cm, 0);
      \draw (na.south) -- ++(0, -1cm);
      \node[] at (-3cm, 0) {$d$};
      \node[] at (0, 3cm) {$\alpha$};
      \node[] at (3cm, 0) {$d^{\prime}$};
      \node[] at (0, -3cm) {$\alpha^\prime$};
      \node[draw, rectangle, fill=paletteBlue_Light2, minimum    size=0.5cm, inner sep=0pt] (nb) at  ({15cm},{0}) {$b_{0}$};
      \draw (nb.north) -- ++(0,1cm);
      \draw (nb.west) -- ++(-1cm, 0);
      \node[] at (15cm, 3cm) {$\alpha$};
      \node[] at (12cm, 0cm) {$d$};
    \end{tikzpicture}
   }.
   \nonumber
\end{align}
Then, the $\mathcal{B}^{\alpha_n\dots\alpha_1}$ defined in ~\eqref{eq:TEMPO_PathIntegral_row} can be written as the MPO 
\begin{align}
    B_{\alpha_{n-1}, \dots, \alpha_1}^{\alpha^\prime_n,\alpha^\prime_{n-1}, \dots, \alpha^\prime_1} &= [{b}_0]^{\alpha^\prime_{n}}_{d_1}\left(\prod_{k=1}^{n-2}[{b}_k]^{d_k, \alpha^\prime_{n-k}}_{d_{k+1}, \alpha_{n-k}}\right)[{b}_{n-1}]_{\alpha_1}^{d_{n-1} \alpha^\prime_1},
\end{align}
corresponding to rows of the network depicted in Fig.~\ref{fig:TEMPO_1}(a).
The $\alpha,\alpha^\prime$-indices correspond to vertical connections connecting one time step of the MPO propagation to the next, while the $d$ indices correspond to horizontal legs in the tensor network and capture the memory induced by the system--bath interaction. Then, time evolution corresponds to an iterative contraction of this tensor network along the time (vertical) axis (see Fig.~\ref{fig:TEMPO_1}(b)). This contraction can be interpreted as the time evolution of a time-nonlocal MPO, the ADT, giving rise to the method's name. 

For this, the initial ADT is defined as $A^{\alpha_1} = B^{\alpha_1}\rho^{\alpha_1}$ represented as a red circle in Fig.~\ref{fig:TEMPO_1}(a). Subsequently, one obtains the $n$-th time step ADT as 
\begin{align}
    A^{\alpha_n, \alpha_{n-1}, \dots, \alpha_1} = \sum_{\alpha^\prime_{n-1}\dots\alpha^\prime_1}B^{\alpha_n\alpha_{n-1} \dots \alpha_1}_{\alpha^\prime_{n-1}\dots \alpha^\prime_1}A^{\alpha^\prime_{n-1} \alpha^\prime_{n-2} \dots\alpha^\prime_1}.
\end{align}

The ADT then becomes an MPS with outer bonds (pointing upward) connecting to the next time step in the causal tensor network. Meanwhile, the inner bonds (horizontal) represent the non-Markovian influence of previous time steps on the current system state. The dimension of the inner bonds of the ADT can be reduced by sweeping forward and backwards along the horizontal direction and compressing using SVD (indicated by the loop in Fig.~\ref{fig:TEMPO_1}(b)). The rank of the ADT (i.e., the number of outer legs) increases by one at each time step. However, for a finite environment memory time, the growth of the ADT can be terminated once the MPO covers the memory time of the environment. For each time step, the tensors in the network encoding the influence of times larger than the memory time of the environment can be neglected because $\eta_k \approx 0 $ when $k\Delta t$ exceeds the memory time and correspondingly $I_k(d_\alpha) \approx 1$ in this case. This is represented with grayed-out boxes in Fig.~\ref{fig:TEMPO_1}. This is done by tracing over the earliest time index in the remaining tensor network
\begin{align}
    B_{\alpha_{n_c}\dots\alpha_1}^{\alpha^\prime_{n_c+1}\dots\alpha^\prime_2} =\sum_{\alpha^\prime_1}B_{\alpha_{n_c}\dots \alpha_1}^{\alpha^\prime_{n_c+1}\dots\alpha^\prime_1}\ ,
\end{align}
where the memory time of the environment is $n_c \Delta t$.
Thus, the contraction of the TEMPO network can be divided into two phases: The growth phase within the memory time, in which the length of the ADT increases by one each time step, and the propagation phase, during which the size of the ADT remains constant \cite{strathearn_efficient_2018}.    
The state of the reduced system can be obtained by contracting the current ADT from the left to the right
\begin{align}
    {\rho}^{j_n}(n\Delta t) = \sum_{j_1, \dots, j_{n-1}}A^{j_{n} \dots j_1}.
\end{align}
We conclude this section by commenting on the computational cost of the ADT-TEMPO method. For QUAPI, and without advanced path-filtering techniques \cite{SIM1997335}, the storage memory requirement grows exponentially in the environment memory time with $\mathcal{O}(M^{2n_c})$, where $n_c$ is the number of time steps necessary to cover the memory time and $M$ is the Hilbert space dimension. For ADT-TEMPO, in comparison, numerical tests have shown the memory requirement to effectively be between $\mathcal{O}(n_c)$ and, at worst, $\mathcal{O}(n_c^2)$ depending on the shape of the spectral density \cite{strathearn_efficient_2018}. This comes with the cost of having to perform $\mathcal{O}(n_c)$ SVDs of matrices of dimension $\chi^2 \times M^4$  at each time step, where $\chi$ is the inner bond dimension.

\subsubsection{Process tensor formulation: PT-TEMPO}\label{sec:pt-tempo}
As discussed in Sec.~\ref{sec:IF}, the process tensor (PT) can be a powerful tool for open quantum systems simulations; By using the causal TEMPO network, the process tensor can be built in MPO form with compressed inner bonds, which allows its use, even for the simulation of long-time dynamics.

This is achieved by recasting the TEMPO network in the form shown in Fig.~\ref{fig:TEMPO_1}(c)~\cite{jorgensen_exploiting_2019} by adding a leg to the $[b_0]$-tensors: 
\begin{align}
    [b_0]_{d_n}^{\alpha_{n+1}} &\to [b_0]_{\alpha_n}^{\alpha^\prime_nd_{n+1}} = \delta_{\alpha_n, \alpha^\prime_n}\delta_{\alpha_n, d_{n+1}} I_{0}(\alpha_n, \alpha_n),
\end{align}
and, according to the PT paradigm of creating a system-independent object, removing the system contribution from $[b_1]$
\begin{align}
    [b_1]^{d_{n} \alpha_{n}}_{d_{n-1}\alpha_{n-1}} = 
        \delta_{d_n, d_{n-1}}\delta_{\alpha_n, \alpha_{n-1}}I_1(d_n, \alpha_n),
\end{align}
where $\alpha_n$ connects to the system degrees of freedom and can therefore be considered an outer PT bond, while $d_n$ and $d_{n+1}$ are connections within the causal tensor network.
Graphically, this can be represented as, 
\begin{align}
    \raisebox{-0.5em}{%
    \begin{tikzpicture}[scale=0.2]
      \node[draw, rectangle, fill=paletteBlue_Light2, minimum    size=0.5cm, inner sep=0pt] (na) at 
            ({0},{0}) {$b_{0}$};
      \draw (na.north) -- ++(0,1cm);
      \draw (na.west) -- ++(-1cm, 0);
      \node[] at (-3cm, 0) {$d_n$};
      \node[] at (0, 3cm) {$\alpha_{n+1}$};
    \end{tikzpicture}
   }
    &\to
    \raisebox{-0.5em}{%
    \begin{tikzpicture}[scale=0.2]
      \node[draw, rectangle, fill=paletteBlue_Light2, minimum    size=0.5cm, inner sep=0pt] (na) at 
            ({0},{0}) {$b_{0}$};
      \draw (na.north) -- ++(0,1cm);
      \draw (na.west) -- ++(-1cm, 0);
      \draw (na.east) -- ++(1cm, 0);
      \node[] at (-3cm, 0) {$\alpha^\prime_n$};
      \node[] at (0, 3cm) {$d_{n+1}$};
      \node[] at (3.5cm, 0) {$\alpha_n$};
    \end{tikzpicture}
   }\nonumber.
\end{align}

Then, the causal tensor network can be contracted not row-by-row, but column-by-column. This means writing the PT $\mathcal{I}^{\alpha_k\dots\alpha_0}$ up to time $t_k = k\Delta t$ as \cite{jorgensen_exploiting_2019, cygorek_sublinear_2024}
\begin{align}
    \mathcal{I}^{\alpha_k\dots\alpha_0} &= \prod_{i=0}^{k}C^{\alpha_k\dots\alpha_i} \nonumber\\
    &= \prod_{i=0}^{k}[b_{k-i}]^{~~d_{k-i+1}}_{\alpha_kd_{k-i}}\prod_{j=1}^{k-i-1}[b_j]_{\alpha_{j+i}d_j}^{~~d_{j+1}}[b_0]_{\alpha_i}^{~~d_1} .
\end{align}
The $C^{\alpha_k\dots\alpha_0}$ now represent columns in Fig.~\ref{fig:TEMPO_1}(c) with legs pointing to the right. 
For notational clarity, and in analogy with Refs.~\cite{jorgensen_exploiting_2019, cygorek_sublinear_2024}, we have omitted here left-dangling legs that would connect to the neighboring column, thus removing one index from each $[b_j]$-tensor.
From this, the iterative scheme depicted in Fig.~\ref{fig:TEMPO_1}(c, d) can be derived. Defining $\mathcal{F}^{\alpha_k\dots\alpha_0}_{k:0} = C^{\alpha_k\dots\alpha_0}$ 
and
\begin{align}
\mathcal{F}_{k:j+1}^{\alpha_k\dots\alpha_0} = C^{\alpha_k\dots\alpha_{j+1}}\mathcal{F}_{k:j}^{\alpha_k\dots\alpha_0},
\end{align}
the final process tensor is obtained as $\mathcal{I}^{\alpha_k, \dots, \alpha_0} = \mathcal{F}^{\alpha_k, \dots, \alpha_0}_{k:k}$ as shown in Fig.~\ref{fig:TEMPO_1} (c).
However, no compression has been performed yet, and this network contraction is still unfeasible because of the exponential growth of the PT size. This is overcome by representing the PT as an MPO, which can be achieved as follows. Assume that $\mathcal{F}_{k:j}^{\alpha_k\dots\alpha_0}$ is in MPO form, i.e., 
\begin{align}
    \mathcal{F}_{k:j} 
    =\sum_{d_k, \dots, d_1}\left( \prod_{l=j+1}^{k}[f^{\alpha_{l}}_{l:l-j}]_{d_{l+1}d_{l}}\right)\left(\prod_{l=0}^{j} [f^{\alpha_{l}}_{l:0}]_{d_ld_{l+1}}\right),
\end{align}
where $[f_{k:j}]^{\alpha_l}_{d_{l}d_{l-1}}$ with physical leg $\alpha_l$ and inner bonds $d_l$, $d_{l-1}$ denotes the matrices making up the MPO. Note that in this notation we set $d_0=1$ to ensure that $f_{0:0}$ corresponds to a two-index tensor. The product is organized such that the right product signifies parts that already have been contracted with all $[b_j]-$tensors on their right (see Fig.~\ref{fig:TEMPO_1} (d)).
Now, the next column can be added to the MPO by updating
\begin{align}
    [f_{l:j}]^{\alpha_l}_{d_{l+1} d_{l}} = \begin{cases}[f_{l:j}]^{\alpha_l}_{d_{l+1} d_{l}},~&l\leq j  \\
    \sum_{\alpha_l^\prime}[f_{l:j+1}]^{\alpha^\prime_l}_{d^{\prime}_{l+1} d^\prime_{l}}[b_l]^{\alpha^\prime_ld^{\prime\prime}_{l+1}}_{\alpha_l d^{\prime\prime}_{l} },~&\text{otherwise}
    \end{cases},
\end{align}
such that $d_l = (d^\prime_l, d^{\prime\prime}_l)$. This procedure is depicted in Fig.~\ref{fig:TEMPO_1}(d). The right-dangling legs represent the outer indices of the PT, which connect to the internal (Markovian) dynamics of the system. Now, this MPO is compressed using sweeps of SVDs along the propagation time axis as depicted in Fig.~\ref{fig:TEMPO_1}(d). Overall, this leads to $\mathcal{O}(N^2)$ singular-value decompositions when building the PT, where $N$ is the number of simulation time steps. For a finite memory time, this complexity can be reduced to $\mathcal{O}(Nn_c)$.

\begin{figure}
    \centering
    \input{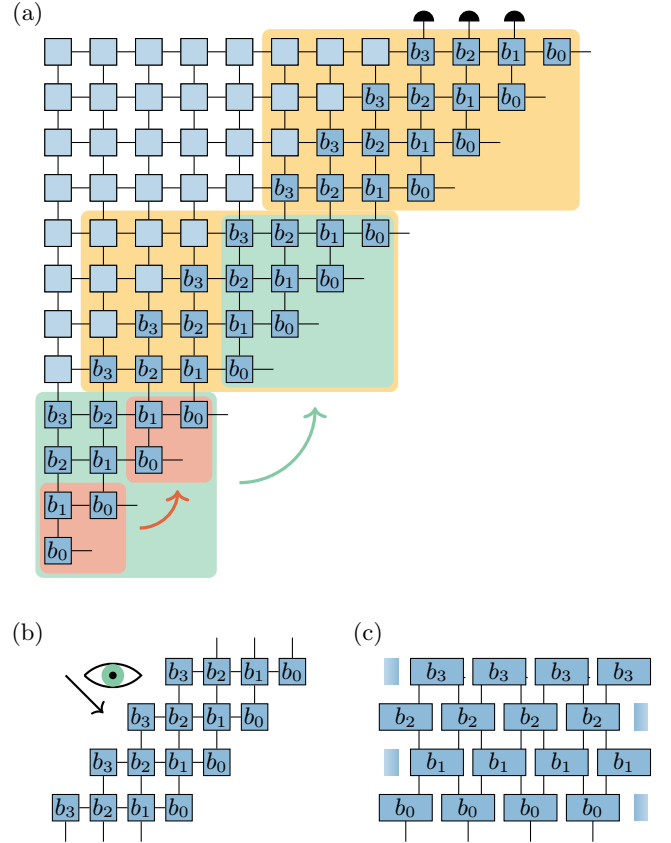}
    \caption{(a) Depiction of the symmetries within the TEMPO network enabling the divide-and-conquer method. Green and red boxes indicate parts of the network that can be reused for calculating the process tensor. The red, green, and orange boxes indicate the repeating pattern that can be reused to propagate to longer times. (b) Depiction of the key idea of Link \textit{et.~al.}: The TEMPO network with $n_\textrm{mem}=4$ somewhere in the middle of the time evolution. This can be looked at from the direction indicated by the arrow. This change of perspective leads to the network depicted in (c). This network can be treated with iTEBD when contracting along the memory time (from the top to the bottom). Adapted from Ref.~\cite{link_open_2024}.}
    \label{fig:TEMPO_2}
\end{figure}

\subsubsection{Divide-and-conquer}\label{sec:divide-and-conquer}
Some problems, such as the precise calculation of linear spectra, require simulations with tens to hundreds of thousands of time steps. In those cases, the PT created by the previously discussed sequential contraction scheme becomes difficult to calculate, as the number of singular-value decompositions is at a minimum linear in the number of simulation time steps. Indeed, as mentioned in Sec.~\ref{sec:TN}, the optimal contraction of a two-dimensional network is not a trivial task. Additionally, storage of the PT for this many time steps becomes unfeasible. 

This obstacle can be overcome by identifying that the causal tensor network possesses time-translational symmetry. Therefore, by using a divide-and-conquer scheme introduced in \textcite{cygorek_sublinear_2024} and discussed now, the computational cost can be reduced to $\mathcal{O}(N\log(N))$ for $N$ simulation time steps. Afterwards, we show how, given the memory cutoff $n_c$, the computational complexity can be further reduced to $\mathcal{O}(n_c\log(n_c))$, additionally reducing the storage cost of the PT by the use of repeating PTs.

\paragraph{Divide-and-conquer:}

Consider the finite-memory network depicted in Fig.~\ref{fig:TEMPO_2}(a). 
The algorithm is based on the observation that there exist repeating patterns within the causal tensor network. For example, the red encircled blocks are identical, just as the green encircled blocks are. Furthermore, the red block appears within the green block twice, indicating that, once calculated, it can be reused. Thus, the calculation of an $N = 2^m$ time step process tensor can be reduced to the combination of $m$ blocks, leading to the previously mentioned $\mathcal{O}(N\log(N))$ scaling. Formally, this procedure can be written as
\begin{align}
    \mathcal{F}^{\alpha_k \dots\alpha_1}_{k:2^{m+1}} = \mathcal{D}^{\alpha_k ... \alpha_{2^m+1}}\mathcal{F}^{\alpha_k \dots\alpha_1}_{k:2^{m}},
\end{align}
where $\mathcal{D}^{\alpha_k...\alpha_{2^m+1}}$ is contained in $\mathcal{F}^{\alpha_k \dots\alpha_1}_{k:2^{m}}$ but capped at the final time via the introduction of closures, as specified in~\textcite{cygorek_sublinear_2024}. 

Note that this process involves the combination and compression of operators with an increasingly large bond dimension, leading to inferior effective scaling. For sufficiently large time propagation, this can even become prohibitively difficult if standard SVD compression is used. This issue can be overcome by using a preselection of relevant singular values. For this consider the compression of two matrices $A = U^A\sigma^AV^{(A)\dagger}$ and $B = U^B\sigma^BV^{(B)\dagger}$. Then, the combined matrix can be written as
\begin{align}
    [AB]_{(a,b)(c,d)} = \sum_{jk} U_{a j}^AU_{b k}^B \sigma_j^A\sigma_k^B V^{(A)\dagger}_{jc}V^{(B)\dagger}_{kd}.
\end{align}
Therefore, only index combinations for which the product $\sigma_j^A\sigma_k^B$ exceeds a certain threshold $\epsilon_\textrm{select}$ will be relevant and need to be kept. This preselection is numerically efficient, as SVD scales unfavorably with matrix size. It is thus preferable to perform SVD on the smaller matrices which are to be combined. However, it has been found that this does not necessarily lead to the optimal low-rank representation of the PT, as the resulting object is not in canonical form. This can be optimized by choosing different forward, backward, and preselection thresholds \cite{cygorek_sublinear_2024}. 

\paragraph{Repeating process tensors:}

Consider again the network shown in Fig.~\ref{fig:TEMPO_2}(a). For a finite memory time, which again allows neglecting the grayed-out tensors, this network, stretching over $N=3$ memory times, possesses a repeating pattern of $N-1 = 2$ cores designated by the yellow rectangle. Thus, having calculated this core of the process tensor over the memory time, time evolution over $N^\prime$ memory times can be realized by repeating $N^\prime-1$ of these process tensor cores. 
Importantly, the bonds connecting the right-hand side of the $(N-1)$\textsuperscript{th} core with the left-hand side of the $N$\textsuperscript{th} core are identical to the bonds connecting the right-hand side of the $N$\textsuperscript{th} core with the left-hand side of the $(N-1)$\textsuperscript{th} core \cite{cygorek_sublinear_2024}. The time propagation is terminated by tracing over the dangling indices of the last core (indicated by black caps in Fig.~\ref{fig:TEMPO_2}(a). 
In addition to this core, one needs to calculate the initial part of the PT, which covers a single memory time of the environment (green square). The calculation of these parts is straightforward within the divide-and-conquer algorithm framework: For the initial part, one calculates the PT up to the memory time using divide and conquer. For the repeating part, this PT is contracted with the triangular network that makes up the rest of the orange box. 
\subsubsection{Infinite process tensors: UniTEMPO}\label{sec:link_struntz}
Under the assumption that the environment has finite memory time, the previous section has shown that it is sufficient to only create and store a repeating unit of the PT that spans that memory time. In fact, a fully time-translational PT, which is periodic not in the memory time, but in the time step, can be created from the causal network using the following method called UniTEMPO. This means that the PT can be written as
\begin{align}
    \mathcal{F}_{N}^{\alpha_N \dots \alpha_1} = \sum_{d_0\dots d_N}v^f_{d_N}\left[\prod_{n=1}^{N} f_{d_nd_{n-1}}^{\alpha_n} \right]v^i_{d_0}.\label{eq:PT_unitempo}
\end{align}
Denoting the inner bond dimension as $\chi$, $[f^{\alpha_n}]_{d_nd_{n-1}}$ is an $n$-independent $\chi \times \chi$ square matrix and $v^{i/f}_{d}$ are $\chi$-dimensional vectors needed to account for boundary conditions such as the initial state of the environment and tracing out the environment degrees of freedom at the end time \cite{link_open_2024}. 

The repeating operator $f^{\alpha_n}_{d_nd_{n-1}}$ can be obtained by considering the causal network after the memory time, depicted in Fig.~\ref{fig:TEMPO_2}(b). Rotating the view by $45^\circ$ reveals its similarity to a periodic quantum circuit model (cf.~Fig.~\ref{fig:TEMPO_2}(c)) for which efficient contraction schemes exist, for example, time-evolving block decimation (TEBD) \cite{orus_practical_2014}, which can be adapted for infinite tensor networks \cite{PhysRevB.78.155117}, immediately yielding $f^{\alpha_n}_{d_nd_{n-1}}$. 
To account for a finite simulation time, cap vectors $\mathbf{v}^{i/f}$ are introduced, which are obtained as the left and right eigenvectors of $f^0$ with eigenvalue one \cite{link_open_2024}. 
Importantly, \eqref{eq:PT_unitempo} together with \eqref{ReducedDensityMatrix} implies that $f^{\alpha_n}_{d_nd_{n-1}}$ can now be interpreted as a time-local and constant propagator for the system together with a non-physical auxiliary space of dimension $\chi$ \cite{garbellini2026}. The initial state is then given by $\rho_S \otimes \mathbf{v}_i$. Crucially, because this propagator is a time-translationally invariant square matrix, it describes evolution via a dynamical semigroup, allowing UniTEMPO to be adapted for Floquet theory \cite{mickiewicz_exact_2026}. 
This property also implies that, assuming that the Liouvillian governing the internal dynamics is time-constant, 
the matrix
\begin{align}
    Q_{(\mu, i)(\nu, j)} =  f^\mu_{ij}M^{\mu\nu}
\end{align}
can be spectrally decomposed, allowing for system steady states \cite{link_open_2024} and multi-time correlation functions \cite{garbellini2026} to be obtained without needing to propagate the dynamics.

TEMPO generally suffers from poor scaling with the size of the system Hilbert space, making the simulation of realistic quantum biological systems challenging. Recent work by \cite{cochin2026efficientconstructiontimeinvariantprocess} has shown that this issue can be addressed in UniTEMPO by introducing intermediate compression steps in the iTEBD.

\subsubsection{Applications and implementations}
All TEMPO-based methods have three main convergence parameters
\begin{itemize}
    \item $n_c$ the memory cutoff step 
    \item $\Delta t$ the timestep for the Trotterization
    \item $\epsilon$, the compression threshold used when performing SVDs or iTEBD.
\end{itemize}
As such, ADT-TEMPO and PT-TEMPO will perform better for environments with smaller memory times. Both divide-and-conquer and UniTEMPO are particularly useful in scenarios in which the propagation time far exceeds memory time, that is, situations in which the storage of the full PT provides a memory bottleneck. 
Divide-and-conquer additionally benefits from using different forward- and backward sweep thresholds, which introduce additional convergence parameters. 

A Python implementation of PT-TEMPO is available in the OQuPy package \cite{fux_oqupy_2024}.
The ACE package contains C++ implementations of TEMPO, PT-TEMPO, and the divide-and-conquer algorithm \cite{moritz_ACE_coding}.

Over the years, ADT-TEMPO has established itself as a versatile tool for capturing the impact of structured environments \cite{Gribben2020}, including $1/f$ noise with negative ohmicity \cite{t87m-7pwn}, and has served as a valuable tool for benchmarking approximate approaches \cite{10.1063/5.0255350}. Thus, applications range from foundational investigations in quantum thermodynamics \cite{PRXQuantum.2.020338} and non-additive effects between non-Markovian environments \cite{gribben_exact_2022, Hogg2024enhanced} to the description of driven condensed matter systems such as III-V quantum dots \cite{pgk3-b57l} and cavity QED with hBN nanostructures \cite{PhysRevResearch.5.L032037}.

The process tensor formulation, PT-TEMPO, has gained widespread attention due to its utility for performing parameter sweeps and multi-time correlation functions. These studies include simulation of 2D-spectroscopy \cite{doi:10.1021/acs.jpclett.5c00928, PhysRevResearch.7.013209}, control optimization in non-Markovian environments \cite{fux_efficient_2021, PhysRevLett.132.060401, meredith_efficient_2026}, and the description of driven condensed-matter systems, such as WSe$_2$ \cite{piccinini_high-purity_2025}, and GaAs QDs \cite{m1cc-l6ng}.
In addition, it has been combined with mean-field approaches to describe collective light--matter coupling \cite{Fowler-Wright2022, kjsb-h9s7}. 

 The favorable scaling of divide-and-conquer allows the calculation of two-time correlation functions spanning multiple timescales, allowing the simulation of multiple semiconductor QDs accounting for their phonon environment \cite{wiercinski_phonon_2023} and the analysis of phonon impact on collective light--matter coupling of up to five two-level systems \cite{wiercinski_phonon_2023, cygorek_sublinear_2024}. It also allows for the inclusion of photon degrees of freedom to access photonic observables in cavity QED settings \cite{Bracht:23}.

Due to their time-translational symmetry, infinite process tensors have been used in \cite{link_open_2024} to map out the phase transition of the famous spin--boson model and to reproduce results by \cite{PhysRevResearch.7.013209} on 2D spectroscopy \cite{garbellini2026}. Furthermore, Landau-Zener transitions for the numerically challenging sub-Ohmic regime in the spin--boson model have been analyzed in \cite{10.1063/5.0235741}. Its ability to push simulation time far beyond memory time has proven effective in the simulation of quantum system work statistics \cite{shubrook2025} and provided benchmarks for perturbative Floquet theory approaches \cite{mickiewicz_benchmarking_2026}.

With the use of Grassmann variables, a Feynman-Vernon path integral can be formulated for fermionic systems, leading to the development of fermionic TEMPO methods to treat Anderson impurity models \cite{xu_grassmann_2024, PhysRevB.109.045140, Ng2023}. Inspired by the infinite process tensor method Sec.~\ref{sec:link_struntz}, an infinite Grassmann-variable TEMPO method has recently been developed \cite{PhysRevB.110.045106}.

\section{CONCLUSION \& OUTLOOK}
\label{sec:outlook}
\begin{table*}[t]
    \raggedright
    \begin{tabular*}{\columnwidth}[t]{@{}c l l}
       \rowcolor{gray!50} \quad \normalsize\textbf{Method} \quad &  \quad \normalsize\textbf{Open source software packages} \quad & \quad\normalsize\textbf{Programming language} \qquad \\
        \rowcolor{gray!10} \quad ACE \quad & \quad ACE {\cite{moritz_ACE_coding}} & \quad C++, Python \\
        \rowcolor{gray!20} \quad DAMPF \quad & \quad  DAMPyF \cite{LorenzoniArxiv2026} & \quad Python\\
        \rowcolor{gray!10} \quad TN-HEOM \quad & \quad  TENSO {\cite{Tenso_Package_2026}} & \quad Python\\
        \rowcolor{gray!10} & \quad  mpsqd {\cite{guan_mpsqd_2024}} & \quad Python\\
        \rowcolor{gray!10} & \quad   pyTTN {\cite{Lindoy2025}} & \quad Python \\
        \rowcolor{gray!20} \quad ML-MCTDH \quad & \quad  TENSO {\cite{Tenso_Package_2026}} & \quad Python\\ 
        \rowcolor{gray!20} & \quad pyTTN~{\cite{Lindoy2025}} & \quad Python\\
        \rowcolor{gray!20} & \quad Renormalizer~{\cite{renormalizer_2025}} & \quad Python\\
        \rowcolor{gray!10} \quad TEDOPA \quad & \quad  MPSDynamics.jl~{\cite{lacroix_mpsdynamicsjl_2024}} & \quad Julia\\
        \rowcolor{gray!10} & \quad pyTTN {\cite{Lindoy2025}} & \quad Python\\
        \rowcolor{gray!20} \quad TEMPO \quad &  \quad OQuPy (PT-TEMPO) {\cite{fux_oqupy_2024}} & \quad Python\\
        \rowcolor{gray!20} & \quad  ACE (PT-TEMPO, Divide-and-Conquer) {\cite{moritz_ACE_coding}} & \quad C++, Python\\
        \rowcolor{gray!20} & \quad  UniTEMPO {\cite{link2024itebdtempo_code}} & \quad Python, Julia
    \end{tabular*}
    \caption{Summary of all the methods presented in this article with their open source implementations.}
    \label{tab:Summary_Methods}
\end{table*}

Non-Markovian effects in open quantum systems are ubiquitous; this underscores the vital need for methods capable of capturing them. 
Unraveling the consequences of non-Markovian effects is central to furthering our quantitative understanding across domains, in chemical physics, quantum technologies, quantum optics, quantum effects in biological systems, condensed-matter and solid-state physics, and many other areas.
Over the past few decades, the tensor-network formalism has enabled the numerically exact simulation of such open quantum systems, without relying on uncontrolled approximations. 

This progress has given rise to a broad range of methodologies, which differ primarily in (i) their representation of the environment, ranging from explicit physical modes to effective or auxiliary degrees of freedom, and (ii) the object being propagated, from the wave function or density operator of the enlarged system, obtained through direct solution of the Schrödinger or Liouville--von Neumann equation, to the reduced system being propagated through contractions with the influence functional. No single method is optimal in all settings; rather, their relative efficiency depends on the structure of the environment and on the quantities of interest. For Gaussian environments, path-integral-based methods provide a natural route to the reduced system dynamics. Among these, TEMPO is particularly effective when the environmental memory time is relatively short, whereas HEOM is advantageous when the bath correlation function can be decomposed into a finite number of exponentials. Methods based on an explicit or effective representation of the environment offer complementary advantages. TEDOPA is naturally suited to continuous environments with linear system--environment coupling, and provides direct access to environmental observables. DAMPF is particularly effective for highly structured environments that admit finite pseudomode representations and can efficiently tackle multi-site systems. ML-MCTDH is particularly valuable when correlations between the relevant degrees of freedom can be efficiently organized in a tree topology. When the same environmental influence must be reused, such as for different control protocols or for computing multi-time observables, process-tensor approaches become particularly appealing. This includes PT-TEMPO and ACE; the latter is well suited to environments that can be decomposed into independent discrete modes.

For all these methods, there are open-source software packages---often accompanied by extensive documentation and examples---that further enhance their accessibility, moving these methods beyond the domain of specialists and broadening their reach to researchers seeking to model realistic physical systems. Table~\ref{tab:Summary_Methods} summarizes the existing open-source implementations of the methods reviewed here.

\subsection{A unifying view of open quantum systems}

\begin{figure*}
    \centering
        \resizebox{\textwidth}{!}{%
        \input{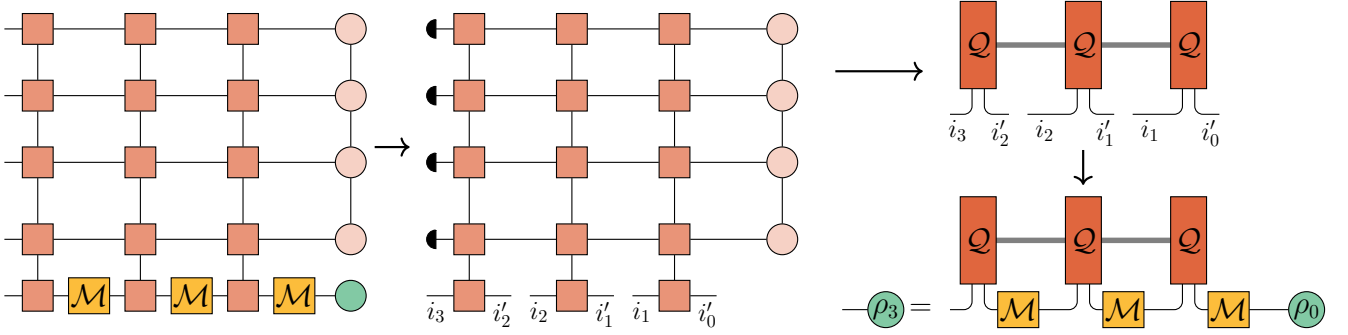}
    }
    \caption{(Left) The time evolution of TEDOPA, DAMPF, or MPS-HEOM can be represented as a two-dimensional tensor network where $\mathcal{M}$ is the free evolution of the system for a time step $\Delta t$ and the evolution of the environment---including interactions with the system---is generically described by an MPO (red tensors). (center) Tracing the environmental modes at the final time (black caps) and contracting the network `vertically', one can construct a process tensor (PT, see right-top). This PT could be used to evolve any initial state of the system (right-bottom). This construction generalizes to tree tensor network states as used in TTN-HEOM or ML-MCTDH. As discussed in Secs.~\ref{sec:ACE} and \ref{sec:TEMPO}, ACE and TEMPO are already formulated as two-dimensional networks from which a PT can be obtained.}
    \label{fig:Fig_outlook}
\end{figure*}

Although the methods discussed in this review originate from various theoretical perspectives, their expression within a shared tensor-network framework reveals conceptual and computational similarities.
Formulating these approaches in a common language has a twofold advantage. 
First, it helps to clarify connections between seemingly disparate theoretical frameworks. 
Second, it facilitates synergistic developments and the design of hybrid numerical methods. 
As emphasized by \textcite{keeling_process_2026,Ortega2024,pollock_non-markovian_2018,jorgensen_exploiting_2019}, the process-tensor (PT) concept (see Sec.~\ref{sec:IF}) has considerable unifying power and may provide insights into more efficient representations of the OQS problem~\cite{Dowling2024_Capturing_Long_Range}. 
Indeed, any method that explicitly represents environmental degrees of freedom within a single tensor network---such as ACE, TEDOPA, or ML-MCTDH---can be used to construct a PT by forming a two-dimensional tensor network that extends in both time and `space'.
Rather than evolving the state one time step at a time, the sequence of MPO or few-body gates (e.g. the two-site gates in TEBD) can be stacked---and not contracted---until the desired final time is reached.
In this representation, one dimension of the network corresponds to physical time, while the other encodes entanglement between the modes.
This procedure is schematically depicted in Fig.~\ref{fig:Fig_outlook}.
The same construction can be extended to the auxiliary modes that appear in HEOM and the pseudomodes of DAMPF; and, as explained in Sec.~\ref{sec:TEMPO}, the two-dimensional representation of the influence functional in TEMPO can be contracted into a PT.

However, the mere existence of a PT representation does not by itself guarantee numerical efficiency, and identifying when such constructions yield practical advantages remains an open question.

Adopting a broader view, the methods reviewed here may be regarded as different contraction strategies for an underlying two-dimensional tensor network, with one dimension corresponding to physical time and the other to environmental memory, or equivalently system--environment correlations. 
However, unlike one-dimensional tensor networks, where optimal compression and contraction strategies are available, the contraction of general two-dimensional tensor networks is computationally hard (see Sec.~\ref{sec:TN}), and no universally optimal contraction scheme is known~\cite{Markov2008,Dumitrescu2018-qn,Stoian2024}. 
Though recent developments on approximate contraction schemes, such as belief propagation methods~\cite{tindall_gauging_2023, park_simulating_2025, alkabetz_tensor_2021}, could offer a potential route to circumvent this issue if the approximations incurred are controlled.
Similarly, the recent possibilities of employing artificial intelligence (AI) for finding solutions to the contraction problem~\cite{meirom_optimizing_2022, sharir_neural_2022} are particularly stimulating, as are the possible connections between emerging `neural network' approaches to open systems and the tensor methods described in this review~\cite{hartmann_neural-network_2019, banchi_modelling_2018, PhysRevLett.122.250501, PhysRevA.109.062215, cao_simulating_2026}.
It therefore remains doubtful that a single dominant method will emerge; instead, we expect that progress will be driven by complementary algorithms tailored to different physical regimes.

At the same time, the tensor-network \emph{lingua franca} naturally encourages hybrid algorithmic strategies that combine the strengths of different approaches. 
A clear example is the pseudomode construction central to DAMPF, which can be incorporated into HEOM and ACE to provide a compact, physically transparent representation of environmental structure and memory effects while reducing effective dimensionality.
Related ideas have inspired effective-mode extensions of HEOM, suggesting the possibility of a systematic reformulation of environmental representations across methods~\cite{muller2026onetoonecorrespondencehierarchicalequations,xu_colloquium_2026}. 
Additional salient connections arise between TEDOPA and both ML-MCTDH and ACE. These latter two require discretizing the environment, and the TEDOPA truncated orthogonal-polynomial chain mappings are known to provide optimal discretizations for continuous environments~\cite{de_vega_how_2015}.
In the case of a Gaussian environment, ML-MCTDH could also make use of the concept of thermalized spectral density~\cite{tamascelli_efficient_2019} to improve the efficiency of finite-temperature simulations.
Moreover, the recently established equivalence between TEDOPA and collision models~\cite{lacroix_making_2025} could offer an alternative ACE formulation in which environmental modes interact with the system only for a finite duration, enabling a natural time-dependent compression of the environmental Hilbert space and further improving computational efficiency.
Additional conceptual links emerge between HEOM and TEMPO when the former is derived via a time discretization of the bath correlation function, revealing formal analogies between hierarchical indices and memory-time indices~\cite{link_open_2024}. 
Making this correspondence explicit could offer a unified understanding of how non-Markovian effects are encoded in time-propagation schemes and may enable hybrid formulations of these techniques.

\subsection{Hardware improvement}

The rapid advancement of GPU technology has transformed computational capabilities across scientific fields. In particular, it enables efficient computation of linear algebraic operations that naturally appear in quantum system dynamics, particularly across the methods discussed. Developing algorithms that exploit the massively parallel architecture of GPUs offers significant potential for accelerating many of the methods described here.
Properly harnessing the power of GPUs will, however, require the wider diffusion of the most recent GPU-native implementations of SVD algorithms or the use of alternative decomposition schemes, for instance, tensor cross-interpolations~\cite{nunez_fernandez_learning_2025}.

Tensor operations are in fact so important in many modern applications, for instance, data analysis and AI, that they are now even driving the development of hardware, namely so-called Tensor Processing Units (TPUs)~\cite{jouppi2017indatacenterperformanceanalysistensor}.
Just as the explosion of computing power in recent times has allowed the AI revolution, perhaps long-standing issues with higher-dimensional tensor network schemes will also become tractable with the growing power of emerging computational capabilities, making them clearly a useful tool for tensor network methods in open quantum systems. However, it should be noted that this does not circumvent the curse of dimensionality, which will require additional methodological advances beyond hardware improvements.

The continued propitious development of quantum computing devices along with emerging quantum algorithms offers a complementary route for simulating open quantum systems~\cite{delgadogranados2024quantumalgorithmsapplicationsopen}. Quantum computers offer a natural framework for simulating open quantum systems, as quantum information is encoded directly in logical quantum states. This one-to-one representation circumvents the exponential memory overhead inherent to classical simulations, substantially alleviating the curse of dimensionality. This advantage is particularly evident in analog quantum computing, where a controllable quantum system is engineered to emulate the dynamics of a target quantum system through a direct mapping of their Hamiltonians~\cite{Pastawski_2011_Quantum_memory,Zanardi_2016_Dissipative,Ding_2024_Simulating_Open,Kayshap_2025_Accuracy_Analog_Open_Quantum_Sim}; this has recently been experimentally validated for modeling chemical dynamics~\cite{Navickas_2025_experimental_quantum_sim,MacDonell_2021_Analog_quantum_sim}. Specifically for the methods discussed here, a key insight from \cite{link_open_2024} was that the UniTEMPO process tensor can be interpreted as a shallow-depth quantum circuit comprised of non-unitary gates. Recent work has further developed explicit quantum implementations of established methods, including HEOM~\cite{dan_simulating_2025} and TEDOPA~\cite{guimaraes_digital_2024}. In addition, substantial progress has been made in the quantum simulation of Lindblad dynamics~\cite{Oh2024_SVD_Quantum,Schlimgen2022,Watad_2024,Sun_2024,Guimaraes2023}, as well as quantum algorithmic implementations of pseudomode-based approaches such as DAMPF~\cite{Lemmer_2018,So2025,Sun2025}. Although current implementations remain largely exploratory, these developments suggest that quantum processors could, in the long term, provide native access to environmental memory, complementing classical tensor-network and GPU-based strategies.

\subsection{Standardizing numerical methods}

To facilitate synergistic developments and cross-fertilization between methods, setting standards in the community---as has been done, for instance, in computational chemistry~\cite{Goerigk2011,Goerigk2017}---is crucial.
Such standardization could include, for example, a common definition of Hamiltonian models and the spectral density, or the definition of a data format for storing process tensors.
Another benefit of standardization would be to enable the development of user-friendly front-end software able to call different methods in the back-end to perform simulations.
Standardization would also facilitate the benchmarking of these numerical methods.
Partial comparisons have already been performed, for instance, between MPS-HEOM and TEDOPA~\cite{le_de_managing_2024, le_de_revisiting_2026}, DAMPF and TEDOPA~\cite{lorenzoni_full_2025}, ML-MCTDH and TTN-HEOM~\cite{chen_comparison_2026}, or between ACE and TEMPO~(see the supplementary material of \cite{mortiz_2022_ACE}). A limited comparison between the different TEMPO-based PT methods can be found in \cite{cygorek_sublinear_2024}.
Yet, systematic comparisons on shared models, ranging from simple spin--boson systems to complex biomolecular aggregates, are still lacking. 
Such benchmarks would help quantify trade-offs between computational efficiency, memory scaling, and achievable accuracy, and offer valuable guidance for practitioners selecting methods suited to specific regimes of system size, coupling, temperature, and environment structure.
Such a benchmarking would also illuminate how the strengths of different methods can be combined in a complementary manner. 
Establishing rigorous correspondences among the methods presented in this review and developing standardized benchmark tests will be key milestones for the field in the coming years.
Through a deeper understanding of the synergies between the different approaches, we anticipate the development of new methods that leverage the strengths and concepts of each technique.

\subsection{Toward many-body open quantum systems}

Looking ahead, the further development of scalable non-perturbative tensor-network methods for open quantum systems may be viewed as bringing the field full circle. Originally introduced for the study of strongly correlated closed many-body systems, tensor-network approaches have since been extended to describe the dynamics of open quantum systems. The role of memory effects in many-body open quantum systems has been seen to be increasingly important in understanding driven-dissipative phase transitions, non-equilibrium many-body states, or dissipative state preparation~\cite{huelga_non-markovianity-assisted_2012, westhoff_tensor_2026,debecker_controlling_2024, debecker_role_2025,Fowler-Wright2022, kjsb-h9s7}.
In contrast to the examples presented in this review, where the system was always described by only a few degrees of freedom and the entire complexity originated from the many-body nature of the environment, the accurate simulation of many-body open quantum systems is considerably more challenging. 
In this case, the curse of dimensionality applies both to the many-body system itself and its environment.
This challenge is especially severe when each constituent of the many-body system is coupled to its own local non-Markovian environment~\cite{flannigan_many-body_2022}.
However, even in the presence of a single global environment, the problem remains difficult because the environment can mediate and encode non-trivial spatiotemporal correlations.
Whilst recent work on polaritonic chemistry has employed TN-HEOM to tackle collective light-matter interactions in a regime where several sites can be treated non-perturbatively \cite{ke_cooperative_2026, ke_stochastic_2025, ke_probing_2026}, the methods presented in this review are not yet capable of tackling genuinely large many-body systems because of the sheer size of the resulting tensor networks.
Extending those capabilities to treat systems that contain tens or hundreds of sites can be seen as the ultimate goal of the field.
We anticipate that continued progress in OQS methods will ultimately enable the simulation of many-body systems embedded in realistic environments, providing a unified framework in which correlations, dissipation, and non-Markovian effects are treated on equal footing.

\begin{acknowledgments}
We would like to thank all the people with whom we have interacted and collaborated on the topics covered in this review. We are also grateful to everyone who commented and provided feedback on this manuscript.

NL, TL, JL, SFH and MBP acknowledge support from the EU project SPINUS (Grant No.~101135699), the BMFTR project PhoQuant (Grant No.~13N16110), the Volkswagen Foundation (Grant No.~0200187), the ERC Synergy Grant HyperQ (Grant No.~856432), the State of Baden-Württemberg through bwHPC, and the German Research Foundation (DFG) through Grant No.~INST 40/575-1 FUGG for the JUSTUS 2 cluster.
AB, JW, KD and EMG acknowledge the Leverhulme Trust (Grant No.~RPG-2022-335), the Volkswagen Foundation (Grant No.~0200195), and Innovate UK (Grant No. 10120741) under the Horizon Europe Pathfinder Challenge and EPSRC (Grant No. EP/W524669/1).
BWL and JMJK acknowledge funding from EPSRC (Grant No.~UKRI3778).
\end{acknowledgments}

\bibliography{Review/bibliography_Review}

\end{document}